# Continuum lowering - a new perspective

**B J B Crowley**[1,2,3]

[1]Department of Physics, University of Oxford, Parks Road, Oxford OX1 3PU, UK
[2]AWE PLC, Reading RG7 4PR, UK

[3]Email: basil.crowley@physics.ox.ac.uk

What is meant by continuum lowering and ionization potential depression (IPD) in a Coulomb system depends very much upon precisely what question is being asked. It is shown that equilibrium (equation-of-state) phenomena and non-equilibrium dynamical processes like photoionization are characterised by different values of the IPD. In the former, the ionization potential of an atom embedded in matter is the difference in the free energy of the many-body system between states of thermodynamic equilibrium differing by the ionization state of just one atom. Typically, this energy is less than that required to ionize the same atom *in vacuo*. Probably, the best known example of this is the IPD given by Stewart and Pyatt (SP). However, it is a common misconception that this formula should apply directly to the energy of a photon causing photoionization, since this is a local adiabatic process that occurs in the absence of a response from the surrounding plasma. To achieve the prescribed final equilibrium state, in general, additional energy, in the form of heat and work, is transferred between the atom and its surroundings. This additional relaxation energy is sufficient to explain the discrepancy between recent spectroscopic measurements of IPD in dense plasmas and the predictions of the SP formula. This paper provides a detailed account of an analytical approach, based on SP, to calculating thermodynamic and spectroscopic (adiabatic) IPDs in multicomponent Coulomb systems of arbitrary coupling strength with $T_e \neq T_i$. The ramifications for equilibrium Coulomb systems are examined in order to elucidate the roles of the various forms of the IPD and any possible connection with the plasma microfield. The formulation embodies an analytical equation of state (EoS) that is thermodynamically self- consistent, provided that the bound and free electrons are dynamically separable, meaning that the system is not undergoing pressure ionization. Apart from this restriction, the model is applicable in all coupling regimes. The Saha equation, which is generally considered to apply to weakly-coupled non-pressure-ionising systems, is found to depend on the Thermodynamic IPD (TIPD), a form of the IPD which takes account of entropy changes. The average Static Continuum Lowering (SCL) of SP relates to changes in potential energy alone and features in EoS formulae that depend on the variation of the mean ionization state with respect to changes in volume or temperature. Of the various proposed formulae, the Spectroscopic (adiabatic) IPD (SIPD) gives the most consistent agreement with spectroscopic measurements.

**PACS**: 52.25.-b, 52.25.Jm, 52.25.Kn, 32.80.-t, 32.80.Fb

**Keywords**: Coulomb systems, Ionization Potential Depression, Continuum lowering, Photoionization, Plasma equation of state.

CONTENTS

# 1 Introduction

## *1.1 Background*

The fact that electrons bound to atoms in plasmas and metals require less energy to liberate them into the continuum than from equivalent states in isolated atoms was, until recently, generally thought to be reasonably well understood to the extent that it could be described in terms of a simple model, despite a lack of sound experimental validation of any such model. Direct spectroscopic observation of ionisation potential depression, or continuum lowering as it is sometimes called, is generally frustrated by the Inglis-Teller effect [1] whereby the ”true” bound-free edge is obscured through becoming merged with nearby bound-bound transitions. Indirect methods have generally been too imprecise to discriminate between possible alternative models.

Interest in the phenomenon has been revived by some recent spectroscopic measurements [2], [3], [4] exploiting new facilities, of dense plasmas, that claim to have circumvented the Inglis-Teller effect to yield good quantitative data. However, rather than confirming the generally accepted thinking, as embodied in the well-known Stewart-Pyatt model [5], for example, they have exposed inconsistencies and deficiencies in some well-established current models, and thereby in prior understanding of this phenomenon, while raising deeper questions about the underlying concepts.

In one type of experiment [2], [3], a tuneable x-ray laser (FEL) is used to ionize the K-shell in solid state aluminium. Whether ionization occurs or not is a direct function of the laser energy and is diagnosed by measuring the subsequent Kα emission. The experiment is thus a clean measurement of the spectroscopic ionization potential that does not depend on any underlying model of the subject system. The results of this experiment are illustrated in **Figure 2**, in which the observed ionisation depression for various ionisation states of aluminium are compared with different theoretical predictions. It turns out that the results of this experiment significantly disagree with the predictions of Stewart and Pyatt [5] and are best described by an old model proposed by Ecker and Kröll [6] . This conclusion has raised concerns that the hitherto widely favoured model of Stewart and Pyatt is at fault raising concerns over the validity of the large amount of data derived using it.

In another recent experiment [4] spectroscopic measurements are carried out on laser-shocked Aluminium and the presence or absence of the 1-3 lines as a function of temperature and density used as a diagnostic of the continuum lowering. The results of this experiment, and comparisons with various theories, are given in **Table 2** . While the interpretation of this experiment does depend, to some extent, on modelling of the *in situ* $n=3$ atomic levels to represent the effect of the various continuum lowering models, the results appear conclusive and are consistent with a simple ion-sphere model, which is much closer to Stewart and Pyatt than Ecker and Kröll.

Both experiments claim to be able to discriminate between different models of the ionization potential depression with the FEL direct ionization measurement apparently supporting Ecker and Kröll while the laser driven shock measurements are presented as being more consistent with Stewart and Pyatt. Neither model is capable of fitting both experiments.

The Stewart-Pyatt has the virtue of possessing a physics-based derivation, albeit a far from exact one, and incorporates the ion-sphere and Debye-Hückel models in its limits. Simple alternatives, such as Ecker-Kröll, are more *ad hoc* in nature, and/or are of more limited validity, so it is logical that Stewart-Pyatt should carry favour over them. So why experiment should take a contrary view and, in certain circumstances, favour a less well justifiable alternative models seems difficult to understand. Ecker-Kröll depends upon an *ad hoc* assumption, which, even in hindsight, remains unsupported. The application of the ion-sphere model to the laser driven shock experiment does not appear to be justified either, due to the ion coupling being insufficiently strong. Moreover, since all of these models, Ecker-Kröll, Stewart-Pyatt and ion-sphere, claim to model the same thing, any inconsistencies are indicative only of deficiencies in one or more of them. Which model should be used is certainly not a matter of arbitrary choice or preference. While it may be that, of the various models considered, only Stewart-Pyatt appears to be rationally supportable, it is undeniable that *both* sets of experiments clearly demonstrate that the spectroscopically-determined ionization potential depression in dense matter is significantly greater than that predicted by this model.

This is unfortunate. It is not just that a simple formula, like Stewart and Pyatt's, is too useful to discard lightly. While it is true that a detailed atomic physics calculation, using a many-body implementation of density functional theory, for example, that captures the essential physics, might be expected to reproduce observational data, this is not always feasible. This capability is recent and, even now, not all plasma regimes are accessible to such calculations. The formula is incorporated or is implicit in many atomic physics codes still in use or which have been sources of currently available atomic data. So the failure of experiment to support this model is of considerable concern and raises two immediate questions: What is wrong with the model? and Can it be fixed?

This is our starting position. A first step is to review the basis of the Stewart-Pyatt and closely-related formulae to ascertain why they may not yield the results expected of them. Theoretical treatments of continuum lowering typically approach the problem from the point of view of thermodynamic equilibrium. It is true that neither of the experiments is characterised by full thermodynamic equilibrium, but this does not in itself offer a satisfactory or useful explanation for the discrepancies. Continuum lowering features in non-equilibrium situations. In strongly coupled plasmas, it is largely determined by the potential energy, which is dependent on the spatial configuration of the system independently of whether the system is in thermal equilibrium. Nevertheless, it is the presumed connection with equilibrium that turns out to be very much at the heart of the matter.

The modelling of plasmas in equilibrium typically treats ionization as a quasistatic transition between states of thermodynamic equilibrium. In equations that model equilibrium, such as the Saha equation or the Gibbs distribution of the Canonical Ensemble, the continuum lowering appears as a correction to the free energy. We shall refer to this as the *thermodynamic ionization potential depression* (TIPD). Stewart and Pyatt, however, in the derivation of their formula, consider the effect in terms of the average electrostatic potential experienced by the electrons in an atom, or, equivalently, the self-energy of the ion-electron system, which is consistent with the approach taken by average-atom models, while disregarding the effect of fluctuations that would be associated with the entropy term in the free energy. We refer to the depression of the ionization potential in the average electrostatic potential as the *static continuum lowering* (SCL). However spectroscopy probes dynamical process occurring between plasma microstates, in which the changes of state of individual electrons/atoms are observed on timescales that might not allow a response from the surrounding plasma to each individual transition.

In the linear regime of single-photon interactions, the active electron remains close to the atom during the spectroscopic process, specifically within the range of its initial wavefunction. The electron hole created by the ionisation process, which occurs on a timescale $\gtrsim 1/\omega$ where $\omega$ is the photon frequency, does not become visible to the surrounding plasma until the electron moves a distance comparable to the scale length of the plasma (such as may be represented by the Debye length or the mean ionic separation distance). This happens *after* the spectroscopic interaction has occurred on timescales determined by the inverse of the electron plasma frequency. The response of the ions is much slower still, occurring on timescales determined by the inverse of the ion plasma frequency. Spectroscopic observations therefore see atomic transitions as being effectively uncorrelated with changes in the plasma microstate. No energy, in the form of either heat or work, is exchanged with the surrounding plasma during the spectroscopic process itself, the only energy exchange being that between the atom and the probe photon. Such a process is adiabatic, in the sense of preserving entropy, as well as occurring at constant volume. The energy of the ionizing photon depends only upon its ability to excite a bound electron into a continuum state, one in which the electron is able to migrate away from the ion, while the surrounding plasma does not suffer any immediate change. This contimuum threshold may differ from that which applies after the plasma has relaxed in response to the changed charge state of the ion. The continuum lowering seen in spectroscopic measurements is therefore not the same as either the SCL or TIPD, a point which seems to have originally been made by Ecker and Weizel in 1956 [7] and reiterated by Ecker and Kröll [6].

The difference between the plasma environment for a microstate of the plasma and that due to the fully (space and time) averaged equilibrium state is commonly known as the *microfield* [8]. As the microfield represents departures from the idealised equilibrium state, it is, by definition, zero for the equilibrium state. However, since the application of any electric field to an atom seemingly lowers the ionization potential irrespective of the direction of the field, it would appear that the average ionization potential should be lowered by the microfield. That is to say, for adiabatic processes, the microfield has the potential to increase the spectroscopic ionization potential depression relative to the thermodynamic value. This question is examined in detail in section 6.2 where it is concluded that this argument is spurious and that the microfield effects are distinct from the continuum lowering and should be treated separately.

In the following, we examine these assertions from the general perspective of a Coulomb system in which the ions behave classically. As a framework for this, we use the static continuum lowering model developed previously by the author. This reproduces the Stewart-Pyatt model while incorporating the effect of a non-uniform free electron distribution induced by their polarisation in the field of the ion. For fast (adiabatic) processes, an additional term is postulated, representing the subsequent relaxation energy that needs to be subtracted out. In conjunction with the SCL, this yields a more correct form of the spectroscopic IPD (SIPD), one which provides reasonably good agreement with both sets of experiments.

## 2 Theory

### *2.1 The ionization potential and static configuration lowering*

We consider an electrically neutral plasma comprising a Coulomb system of electrons and atomic ions in thermodynamic equilibrium. The ions comprise a fixed number $N_\mathrm{i}$ of immutable atomic nuclei and variable numbers of bound electrons, which can be exchanged with the surrounding plasma. For added generality (and in order to model the experiments) the ions and electrons (including bound electrons) are considered to have different temperatures, $T_\mathrm{i}$ and $T_\mathrm{e}$ respectively. (Such a temperature separation can be expected occur when the ion and electron subsystems are weakly coupled on the dynamical timescales controlling the ionization, to which a contributory factor is the smallness of the electron mass compared to the masses of the ions.) We start by *defining* the *ionization potential* of a plasma in equilibrium to be the total energy required to change the charge state of a single ion in the plasma, through excitation of an electron from a bound state in that ion to the plasma continuum while requiring that the electronic chemical potential $\mu_\mathrm{e}$ and the temperature(s) $T_\mathrm{e}$, $T_\mathrm{i}$ of the plasma should be maintained precisely at their initial equilibrium values. We shall presume, at this stage, that we know what is meant by a "bound state" and the "continuum". Because ionizations are considered as occurring one ion at a time, individual ionization processes are conceived as being dynamically independent of each other, and, importantly, that each process does not directly influence the equilibrium ionization state of other ions, or of the plasma as a whole, which would be the case if $\mu_\mathrm{e}$ and $T_\mathrm{e}$ were allowed to change. The ionization process is thereby considered as a quasistatic transition between two equilibrium states of the plasma during which the plasma is considered to be in contact with the appropriate heat baths (at temperatures $T_\mathrm{e}$, $T_\mathrm{i}$) and electron reservoir (at chemical potential $\mu_\mathrm{e}$). Maintaining $\mu_\mathrm{e}$ and $T_\mathrm{e}$, for fixed numbers of the nuclear species, is equivalent to maintaining $n_\mathrm{e}$ and $T_\mathrm{e}$, where $n_\mathrm{e} = \bar{Z} n_\mathrm{i}$ is the free electron density. It will be shown that these constraints are equivalent to considering the ionization process to be occurring, in the closed system, at constant pressure and temperature. These conditions are therefore sufficient for the process to maintain thermal and mechanical equilibrium and hence be considered to be quasistatic.

The ionization potential defined this way is the total energy change of the plasma given by

$$\Delta E_{j\alpha} = \phi_{j\alpha} + \Delta U_{j\alpha} + \bar{\varepsilon} \tag{1}$$

where $\phi_{j\alpha} > 0$ is the ionization potential of the electron state $\alpha$ in an isolated ion, $j$; $\bar{\varepsilon} = \bar{\varepsilon}(\mu_e, T_e)$ is the mean (kinetic) energy of a free electron in the surrounding plasma (which appears by virtue of $T_e$ being maintained) and $\Delta U_{j\alpha} < 0$ is the contribution to the potential energy of the bound electron from the surrounding plasma, taken as the static average. For sufficiently localised, ie deeply bound, states $\Delta U_{j\alpha}$ is independent of the state, $\alpha$ and so $\Delta U_{j\alpha} = \Delta U_j$, which is the *static continuum lowering*.

The static continuum lowering is that which is associated with the static average-atom potential as seen by a test charge and is what is generally considered to be given by the Stewart-Pyatt formula [5] and the limiting (ion-sphere and Debye-Hückel) forms in their respective regimes of applicability, which is in accord with the derivations (see also [9]). In reality, this potential is modified by the microfield, which represents the spatial- and time-dependent fluctuations, around the average, of the electrostatic field experienced by individual electrons. A consequence of this is that some electronic states that are bound in the static average potential may, when subject to the microfield, not be bound to a single ion but rather exist in *transient localised states*.

It is significant perhaps that these are the same states that persistently oscillate between the continuum and the bound levels during explicit iteration of an average atom calculation. Calculations that force convergence by placing these states in the continuum (*coarse convergence*) define a different continuum to those that determine the true static potential by carefully controlling the convergence process (*fine convergence*), while it is possible that some detailed configuration accounting (DCA) calculations include the microfield at the outset. These factors need to be borne in mind when comparing calculations between different codes since they determine where these codes place the continuum as well as influencing how well they might agree with experiments.

In a formal many-body theory description, the total potential energy associated with an ion is usually referred to as its *self energy*, for which formal expressions are provided in terms of response functions [10]. Let the ionization process be considered to be a change of state of an ion-electron system embedded in the plasma and let the initial self-energy of the bound electron-ion system, considered to be effectively at rest, be $\Sigma_0$. In the final equilibrium state, in which the emitted electron is absorbed by the plasma so as to maintain a constant electron density and temperature, neglecting any change in motion of the ion, the total energy is $\Sigma_0 + \Delta\Sigma + \phi_{j\alpha} + \bar{\varepsilon}$ in which $\Delta\Sigma$ is the change in the self energy of the ion due to the quasi-static response of the plasma to the change of the charge state of the ion. Comparison with (1) shows that $\Delta U = \Delta\Sigma$ so that the static continuum lowering is synonymous with the change in the self-energy of the ion during the complete ionization process. This connection with formal many-body theory will be revisited in section 6.

As will be shown, the static continuum lowering provides a complete description of the physical ionization depression only in the strongly-coupled limit (high densities, low temperatures), which is also when ion microfield fluctuations become negligible. At finite coupling strengths, where the interaction energy depends upon temperature, the role of entropy needs to be considered as well.

## *2.2 Static continuum lowering*

A suitably general reference model of the static continuum lowering in a multicomponent Coulomb system is that presented elsewhere by this author [9]. This uses the same ion pair correlation function $g_j(r)$ incorporated in Stewart and Pyatt's method, where, for some $r_j$, which we call the ion core radius,

$$\begin{aligned} g_j(r) &= -1 \qquad && (r < r_j) \\ g_j(r) &= -\frac{r_j}{r}\exp\left(\left(r_j - r\right)/D\right) \qquad && (r > r_j) \end{aligned} \tag{2}$$

where $D$ is the total plasma screening length, while allowing for electron screening, which is the polarization of the electron density in the electrostatic potential of the ion. The static continuum lowering provided by this model is represented by the following formula [9],

$$\Delta U_j = \frac{kT_{\rm i}}{Z_{\rm p}X_j}\left[X_j - \Gamma_j\left(1 + \tfrac{1}{2}X_j^3\right)\right] - \tfrac{3}{2}\frac{kT_{\rm i}}{Z_{\rm p}}\Gamma_j X_j^2\left(\alpha^2 - 1\right) \tag{3}$$

in which the first term is due to the ions and the second term gives the electron contribution, and where

$$Z_{\rm p} = \frac{\sum_j Z_j^2}{\sum_j Z_j} = \frac{\langle Z^2 \rangle}{\langle Z \rangle} \tag{4}$$

is the effective plasma perturber charge, which turns out to be essentially the same as $z^*$ as originally defined by Stewart and Pyatt [5], which note is a property of the plasma as a whole ; $Z_j$ is the charge state of a particular ion $j$ ; $X_j$ is the positive real root of

$$X_j^3 + 3\alpha X_j^2 Y_j + 3X_j Y_j^2 - 1 = 0 \tag{5}$$

which yields the ion core radius,

$$r_j = X_j R_j \tag{6}$$

in terms of the *ion-sphere radius*,

$$R_j = \left( \frac{3Z_j}{4\pi n_e} \right)^{1/3} \tag{7}$$

and $\alpha$ is a number $\geq 1$, which represents the screening effect of the electrons according to

$$\alpha = \frac{D_i}{D} \tag{8}$$

which is the ratio of the ion screening length $D_i$ that is deemed applicable in the regime $r > r_j$, to the plasma screening length, $D$; and

$$Y_j = \frac{D_i}{R_j} = \frac{1}{\sqrt{3\Gamma_j}}$$

$$\Gamma_j = \frac{Z_j Z_p e^2}{4\pi \epsilon_0 R_j k T_i} \tag{9}$$

The ion screening length is taken to be given generally by the classical Debye formula

$$\frac{1}{D_i^2} = \frac{Z_p n_e e^2}{\epsilon_0 k T_i} \tag{10}$$

in terms of which the total screening length is given, in the Thomas-Fermi approximation, by

$$\frac{1}{D^2} = \frac{1}{D_i^2} + \frac{1}{D_e^2} = \frac{1}{D_i^2} + \frac{n_e e^2}{\epsilon_0 k T_e} \frac{I'_{1/2}(\mu_e/kT_e)}{I_{1/2}(\mu_e/kT_e)} \tag{11}$$

in which $I_j(x) = \int_0^\infty \frac{y^j}{1+\exp(y-x)} \mathrm{d}y$ denotes the Fermi function, and $I'_j$ its derivative, where, for sake of argument, the free electron screening is taken to be given by the finite-temperature Thomas-Fermi model.

The ion core radius $r_j$ defines the radius of the core region that characterises the local environment of the ion $j$, within which, according to (2), there are no other ions. In this region, the electrons are dominated by the strong central field of the ion. The region $R_j \gtrsim r > r_j$ is an intermediate region containing both ions and electrons in which the correlations with the central ion are respected according to the second part of (2). The region $r >> R_j$ is the external "collective" region occupied by the rest of the plasma, within which no individual ion is considered to have a dominant influence.

Equation (9) introduces the ion coupling parameter $\Gamma_j$, which, in the form given, is a measure of the relative strength of the electrostatic potential energy of the ion to its thermal kinetic energy. In the strong coupling limit, $\Gamma_j \gg 1$, which implies $r_j \sim R_j$; while, in the weak coupling limit, $\Gamma_j \ll 1$ implying $r_j \sim e^2 Z_j Z_\mathrm{p} / 4\pi\epsilon_0 kT_\mathrm{i}$ , which corresponds to the Landau length for a plasma perturber ion in the vicinity of the subject ion. Equation (5) expresses the condition that the derivative of the potential (the radial electric field) is continuous at $r = r_j$ . Conceptually, $r_j$ corresponds to the separation radius of Ecker and Kröll [6] who give a different, overtly *ad hoc,* formula for it, and the core radius of Stewart and Pyatt [5], whose treatment is basically similar to the above.

With the aid of (5), the formula (3) can be rendered in the much more elegant form,

$$\Delta U_j = -\frac{kT_\mathrm{i}}{2Z_\mathrm{p}}\left(\left(1+\tilde{\Lambda}_j\right)^{2/3} - 1\right) \tag{12}$$

where

$$\tilde{\Lambda}_j = \xi_j \Lambda_j$$

$$\Lambda_j = \frac{1}{Y_j^3} = \left(3\Gamma_j\right)^{3/2} \tag{13}$$

$$\xi_j = \alpha\left(1 + \left(\alpha^2 - 1\right) X_j^3\right) > 1$$

In the special case when $\alpha = 1$, which applies when the electrons are uniformly distributed, the above reduces to Stewart and Pyatt's well-known analytical formula [5], [9], [11], [12],

$$\Delta U^\mathrm{SP} = -\frac{kT_\mathrm{i}}{2Z_\mathrm{p}}\left(\left(1+\Lambda\right)^{2/3} - 1\right) \tag{14}$$

which, note, depends on the properties of the ions alone. (Note also that here it is $Z_\mathrm{p}$ that appears in the denominator, rather than $Z_\mathrm{p} + 1$, as in Stewart and Pyatt's original formula [5][i].) Equation (12) is however applicable to multicomponent plasmas, in which electron screening can also contribute to the continuum lowering. In the strong-coupling limit ($\Gamma_j \gg 1 \;\Rightarrow\; \Lambda_j \gg 1,\;\; X_j \simeq 1,\;\; \xi \simeq \alpha^3$ ):

$$\frac{\Delta U_j}{kT_\mathrm{i}} = -\frac{3\alpha^2}{2}\frac{Z_j e^2}{4\pi\epsilon_0 R_j kT_\mathrm{i}} = -\frac{3}{2}\frac{Z_j e^2}{4\pi\epsilon_0 R_j}\left(\frac{1}{kT_\mathrm{i}} + \frac{1}{Z_\mathrm{p} kT_\mathrm{B}}\right) \tag{15}$$

---

[i] This is due to an inconsistency in Stewart and Pyatt's argument in relation to the assumption of uniformly distributed free electrons. The origin of the $Z_\mathrm{p} + 1$ in Stewart and Pyatt's formula is the embedded relation between the classical Debye lengths, $Z_\mathrm{p} + 1 = Z_\mathrm{p} D_\mathrm{i}^2 / D^2 = Z_\mathrm{p}\left(1 + D_\mathrm{i}^2 / D_\mathrm{e}^2\right)$, which does not apply in this case. See also section 2.3.

where

$$T_{\mathrm{B}} = T_{\mathrm{e}} \frac{I_{1/2}\left(\mu_{\mathrm{e}}/kT_{\mathrm{e}}\right)}{I'_{1/2}\left(\mu_{\mathrm{e}}/kT_{\mathrm{e}}\right)} \tag{16}$$

If the electrons are non-degenerate and $T_{\mathrm{e}} = T_{\mathrm{i}} = T$ , this yields

$$\frac{\Delta U_j}{kT} = -\frac{3}{2}\frac{\left(Z_{\mathrm{p}}+1\right)}{Z_{\mathrm{p}}}\frac{Z_j e^2}{4\pi\epsilon_0 R_j kT} \tag{17}$$

Note well that the ion-sphere radius is distinct from the *Wigner-Seitz radius*,

$$R_{\mathrm{WS}} = \left(\frac{3}{4\pi n_{\mathrm{i}}}\right)^{1/3} \tag{18}$$

to which it is related by

$$R_j = \left(\frac{Z_j}{\bar{Z}}\right)^{1/3} R_{\mathrm{WS}} \tag{19}$$

The former is a property of the plasma, while the latter is a property of an individual ion. In terms of the Wigner-Seitz radius, equation (17) is

$$\Delta U_j = -\frac{3}{2}\frac{\left(Z_{\mathrm{p}}+1\right)}{Z_{\mathrm{p}}} Z_j^{\,2/3}\bar{Z}^{1/3} u_{\mathrm{WS}}^0 \tag{20}$$

which expresses the different scalings with respect to $Z_{\mathrm{p}} = \left\langle Z^2\right\rangle / \left\langle Z\right\rangle$, $\bar{Z} = \left\langle Z\right\rangle$ and $Z_j$, and where

$$u_{\mathrm{WS}}^0 = \frac{e^2}{4\pi\epsilon_0 R_{\mathrm{WS}}} \tag{21}$$

is the *Wigner-Seitz energ*y, which is defined independently of the ion charges.

In the weak coupling limit ($\Gamma_j \ll 1 \;\Rightarrow\; \Lambda_j \ll 1,\;\; X_j \ll 1,\;\; \xi \simeq \alpha$) equation (12) gives

$$\Delta U_j \simeq -\frac{kT_{\mathrm{i}}}{3Z_{\mathrm{p}}}\xi_j \Lambda_j \simeq -\alpha\frac{Z_j e^2}{4\pi\epsilon_0 D_{\mathrm{i}}} = -\frac{Z_j e^2}{4\pi\epsilon_0 D} = -Z_j u_{\mathrm{DH}}^0 \tag{22}$$

which is the Debye limit of the static continuum lowering, with the electronic screening included and where

$$u_{\mathrm{DH}}^0 = \frac{e^2}{4\pi\epsilon_0 D} \tag{23}$$

is the *Debye-Hϋckel electrostatic interaction energy.*

*2.3 Electrostatic potential energy*

Let us proceed by first considering the above in situations when the electron screening is negligible compared to the effect of the ions, ie when $\alpha \simeq 1$ . In this case the electrostatic potential energy, or self-energy, of the ion is given by

$$\frac{\Delta U_j}{kT_\mathrm{i}} = -\frac{1}{2Z_\mathrm{p}} h\left(\Lambda_j\right)$$

$$h(\Lambda) = (1+\Lambda)^{2/3} - 1 \tag{24}$$

For fixed $Z_j,\ Z_\mathrm{p}$ ,

$$n_\mathrm{e}\left(\frac{\partial \Lambda_j}{\partial n_\mathrm{e}}\right)_{T_\mathrm{i}} = -V\left(\frac{\partial \Lambda_j}{\partial V}\right)_{T_\mathrm{i}} = \tfrac{1}{2}\Lambda_j$$

$$T_\mathrm{i}\left(\frac{\partial \Lambda_j}{\partial T_\mathrm{i}}\right)_{n_\mathrm{e}} = T_\mathrm{i}\left(\frac{\partial \Lambda_j}{\partial T_\mathrm{i}}\right)_{V} = -\tfrac{3}{2}\Lambda_j \tag{25}$$

while, for fixed $Z_\mathrm{p},\ n_\mathrm{e},\ T_\mathrm{i}$ , which are regarded as properties of the surrounding plasma, the derivative with respect to the ion charge is

$$Z_j \frac{\partial \Lambda_j}{\partial Z_j} = \Lambda_j \tag{26}$$

The static continuum lowering represents the change in the electrostatic potential energy per unit charge $\delta = 1$, of the whole plasma, when a single ion undergoes a transition $Z_j^+ \to \left(Z_j + \delta\right)^+ + \delta \mathrm{e}^-$ , without affecting $Z_\mathrm{p},\ n_\mathrm{e},\ T_\mathrm{i}$. So, if the total Coulomb energy $U$ is assumed to be given to be in the form

$$\frac{U}{kT_\mathrm{i}} = -\frac{1}{2Z_\mathrm{p}} \sum_j Z_j g\left(\Lambda_j\right) \tag{27}$$

the static continuum lowering is yielded as

$$\frac{\Delta U_j}{kT_\mathrm{i}} = -\frac{1}{2Z_\mathrm{p}} \frac{\partial}{\partial Z_j}\left(Z_j g\left(\Lambda_j\right)\right) = -\frac{1}{2Z_\mathrm{p}}\left(g\left(\Lambda_j\right) + \Lambda_j g'\left(\Lambda_j\right)\right) \tag{28}$$

Hence, upon comparing with (24),

$$h(\Lambda) = \frac{\partial}{\partial \Lambda}\left(\Lambda g\left(\Lambda\right)\right) \tag{29}$$

which can be integrated to yield

$$g(\Lambda) = \frac{1}{\Lambda}\int_0^\Lambda h(\lambda)\,\mathrm{d}\lambda = \frac{1}{\Lambda}\left[\tfrac{3}{5}\left((1+\Lambda)^{5/3}-1\right)-\Lambda\right] \tag{30}$$

Combining equations (27) and (30) yields the electrostatic potential energy as

$$U = -\frac{kT_\mathrm{i}}{2Z_\mathrm{p}}\sum_j Z_j \frac{1}{\Lambda_j}\left[\tfrac{3}{5}\left((1+\Lambda_j)^{5/3}-1\right)-\Lambda_j\right] \tag{31}$$

which is the Coulomb energy in the Stewart-Pyatt (aka Generalized Ion Cell [9], [12], [13]) approximation, and which yields $U < 0$ for all possible $Z_j \geq 0$, $Z_\mathrm{p} > 0$. In the strong coupling limit, $\Gamma \gg 1 \Rightarrow \Lambda \gg 1$, $g(\Lambda) \sim \tfrac{3}{5}\Lambda^{2/3} = \tfrac{9}{5}\Gamma$ and so

$$U \sim U_\mathrm{is} = -\tfrac{9}{10}kT_\mathrm{i}\sum_j \frac{Z_j}{Z_\mathrm{p}}\Gamma_j = -\tfrac{9}{10}\sum_j \frac{Z_j^{\,2}e^2}{4\pi\epsilon_0 R_j} \tag{32}$$

which is recognised as the Coulomb energy in the well-known ion-sphere approximation [14].

In the weak coupling limit, $\Gamma \ll 1 \Rightarrow \Lambda \ll 1$, $g(\Lambda_j) \sim \tfrac{1}{3}\Lambda_j = \dfrac{\Gamma_j R_j}{D_\mathrm{i}}$ and so

$$U \sim U_\mathrm{DH} = -\tfrac{1}{2}\sum_j \frac{Z_j^{\,2}e^2}{4\pi\epsilon_0 D_\mathrm{i}} = -N_\mathrm{i}\frac{\bar{Z}Z_\mathrm{p}e^2}{8\pi\epsilon_0 D_\mathrm{i}} \tag{33}$$

which is the Debye-Hückel electrostatic energy due to the ions [10]. We observe that the electrostatic energy (31) generally satisfies both the Lieb-Narnhofer [15], [16] ($U \geq U_\mathrm{is}$) and Mermin [17] ($U \geq U_\mathrm{DH}$) bounds. In terms of the elementary Wigner-Seitz and Debye-Hückel energies, (21) and (23) respectively,

$$\begin{aligned} U_\mathrm{is} &= -\tfrac{9}{10}N_\mathrm{i}\left\langle Z^{5/3}\right\rangle\left\langle Z\right\rangle^{1/3}u_\mathrm{WS}^0 \\ U_\mathrm{DH} &= -\frac{N_\mathrm{i}}{2\alpha}\left\langle Z^2\right\rangle u_\mathrm{DH}^0 \end{aligned} \tag{34}$$

which are indicative of the different dependences on the charge state distribution (CSD).

The Stewart-Pyatt and Generalized Ion Cell models are presented as being applicable, at some level of approximation, in regimes of arbitrary plasma coupling. Importantly, equations (27)ff can be generalized to accommodate different, potentially more accurate, parameterizations of the potential energy, examples of which are the fits to hypernetted chain (HNC) and Monte-Carlo calculations of the one component plasma (OCP) fluid [18] as well as parameterizations more applicable to metallic solids.

According to the virial theorem, the pressure contribution from the Coulomb energy is $\tfrac{1}{3}U/V$.

*2.4 Generalizations*

We now extend above model to account for the electron contribution to the self-energy, which is the result of screening due to polarization of the electrons in the monopole electric field of the ion. We are able to do this, while also taking account of different possible geometric arrangements of the ions, as in a crystal lattice. The continuum lowering, with electron polarization included, is given by (12), in which, in the limit of strong coupling, $\xi_j \sim \alpha^3$ while, in the weak coupling limit, $\xi_j \sim \alpha$ , neither of which, it should be noted, depend upon the properties of the individual ion. In general,

$$Z_j \frac{\partial \tilde{\Lambda}_j}{\partial Z_j} = \tilde{\Lambda}_j \left( 1 + \frac{Z_j}{\xi_j} \frac{\partial \xi_j}{\partial Z_j} \right) = \tilde{\Lambda}_j + \Lambda_j \left( \alpha^2 - 1 \right) Z_j \frac{\partial X_j^3}{\partial Z_j} \tag{35}$$

in which, according to (8),

$$\alpha^2 - 1 = \frac{D_\mathrm{i}^2}{D_\mathrm{e}^2} \simeq \frac{T_\mathrm{i}}{Z_\mathrm{p} T_\mathrm{B}} \tag{36}$$

which is, in highly ionized and/or degenerate plasmas, typically small. Solving (5) for $X_j$ with $\alpha = 1$ yields $X_j = (1+x)^{1/3} - x^{1/3}$ where $x = 1/\Lambda_j$ , from which the logarithmic derivative with respect to $Z_j$ can be calculated according to

$$Z_j \frac{\partial X_j^3}{\partial Z_j} = Z_j \frac{\partial X_j^3}{\partial x} \frac{\partial x}{\partial \Lambda_j} \frac{\partial \Lambda_j}{\partial Z_j} = -x \frac{\partial X_j^3}{\partial x} = \frac{y(1-y)^2(1+y)}{1+y+y^2} \tag{37}$$

where use has been made of (26) and where,

$$y = \left( \frac{x}{1+x} \right)^{1/3} = \left( 1 + \Lambda_j \right)^{-1/3} \tag{38}$$

The expression on the right hand side of (37) has a maximum value of 0.137521 in the range $0 \le y \le 1$ corresponding to $y = 0.306$ (which corresponds to $\Lambda_j = 33.9$ and $\Gamma_j = 3.49$ ). Therefore, since $\xi_j > 1$ ,

$$1 + 0.13752\left( \alpha^2 - 1 \right) > \frac{Z_j}{\tilde{\Lambda}_j} \frac{\partial \tilde{\Lambda}_j}{\partial Z_j} > 1 \tag{39}$$

The approximation

$$\frac{Z_j}{\tilde{\Lambda}_j} \frac{\partial \tilde{\Lambda}_j}{\partial Z_j} \simeq 1 \tag{40}$$

is therefore reasonable in virtually all circumstances and means that the electronic contribution to the static continuum lowering and the Coulomb energy can be represented by replacing $\Lambda_j$ by $\tilde{\Lambda}_j$ as defined by (13) in which $\xi_j$ is regarded as possessing a negligible derivative with respect to $Z_j$. Hence, as well as the continuum lowering being given by (12), the total electrostatic energy given by (27) becomes

$$U = -\frac{kT_\mathrm{i}}{2Z_\mathrm{p}} \sum_j Z_j g\left(\tilde{\Lambda}_j\right) \tag{41}$$

This now offers the possibility of further generalization, whereby the formula for $\xi_j$ is extended to account for the ions being arranged on a regular close-packed lattice, by means of the introduction of a constant, $C_j$, which is assumed to be of $O(1)$, according to

$$\xi_j = \alpha\left(1+\left(\alpha^2\left(\tfrac{10}{9}C_j\right)^{3/2}-1\right)X_j^{\ 3}\right) \tag{42}$$

The Coulomb energy, in the strong-coupling limit, then becomes

$$U \sim -\alpha^2 \sum_j C_j \frac{Z_j^{\ 2} e^2}{4\pi\epsilon_0 R_j} = N_\mathrm{i} C \alpha^2 \bar{Z}^{1/3} \left\langle Z^{5/3} \right\rangle u_\mathrm{WS}^0 \tag{43}$$

where $\alpha^2$ is given by (36), $u_\mathrm{WS}^0$ is the Wigner-Seitz energy (21) given in terms of the Wigner-Seitz radius (18) and

$$C = \sum_j C_j Z_j^{\ 5/3} \Big/ \sum_j Z_j^{\ 5/3} \tag{44}$$

In the weak coupling however, the energy becomes independent of $C$, which is consistent with any regular close-packed structure disappearing in this limit. In the ion sphere approximation, $C = \frac{9}{10}$, which is considered to apply to fluid-like systems with no discrete symmetry. For close packing, $C = \frac{1}{2}\alpha_\mathrm{M} / \varphi^{1/3}$ where $\varphi$ is the packing fraction and $\alpha_\mathrm{M}$ is the appropriate Madelung constant. Taking values of the Madelung constants from [19] yields values of $C$ for various close-packed lattices, as given in the following table:

| ion sphere | fcc/ hcp | bcc | sc |
|---|---|---|---|
| 9/10 | 0.99025 | 1.01875 | 1.09189 |

**Table 1.** Values of the force constant $C$ for various lattices.

In all these cases, $C \geq 9/10$, which preserves the property $\xi_j > 1$. (The dependence of $C_j$ on the ion species incorporates the possibility of different ion species being arranged on different interpenetrating lattices.)

# 3 Photoionization

## *3.1 Spectroscopic ionization potential depression*

So far, we have treated ionization as a quasi-static process connecting two states of thermodynamic equilibrium. We now consider ionization to be a dynamical process, in particular *photoionization,* in which radiation in the form of a single photon ionizes an atom in a discrete event, during which no changes to the surrounding plasma are induced. This can be because the electron remains within or very close to the atom during the photon interaction process or because the process occurs on a timescale short enough to be considered instantaneous. Either way, the plasma does not respond to the changed state of the ion until the electron has moved a significant distance into the plasma, by which time the photon interaction has ceased, and the immediately resulting state of the system cannot be considered to be in local equilibrium. If the system is constrained at fixed $n_{\mathrm{e}}$, $T_{\mathrm{e}}$ (by contact with electron and thermal reservoirs) it subsequently relaxes to equilibrium during which process energy, hereinafter referred to as the *relaxation energy*, is implicitly exchanged with these surroundings. The total energy supplied to the system in attaining the final equilibrium state is then $\hbar\omega + \Delta\chi$ , where $\hbar\omega$ is the photon energy and $\Delta\chi$ is the *relaxation energy*, and is, by definition, equal to the ionization potential (1) , whereupon $\phi_{j\alpha} + \Delta U_j + \bar{\varepsilon} = \hbar\omega + \Delta\chi$ . Writing $\hbar\omega = \hbar\omega_0 + \hbar\Delta\omega_j + \Delta_{\mathrm{e}}$ where $\hbar\omega_0 = \phi_{j\alpha}$ is the ionization potential for the isolated ion, yields the *spectroscopic ionization potential depression* (SIPD),

$$\hbar\Delta\omega_j = \Delta U_j - \Delta\chi + \bar{\varepsilon} - \Delta_{\mathrm{e}} \tag{45}$$

In section 5.4, it is shown that the quantity $\Delta U_j - \Delta\chi + \bar{\varepsilon}$ is the *adiabatic IPD*. The extra term, $\Delta_{\mathrm{e}}$ in (45), is an offset introduced so that the SIPD corresponds to the photoionization threshold, which is how it is generally conceived, and will be explained later. For the moment, it is sufficient to note that such a term, with $\Delta_{\mathrm{e}} = \frac{3}{2}kT_{\mathrm{e}}$, is necessary to cancel $\bar{\varepsilon}$ in the low-density limit.

Equation (45) shows that the SIPD and the static continuum lowering are generally different. Moreover, the SIPD relates to all adiabatic dynamical processes that change the ionization state of individual atoms, including collisional ionization and recombination. A purely kinetic model that describes the time evolution of a plasma in terms of microscopic physical processes at the atomic level will thus involve only the SIPD. According to this picture, the static continuum lowering is an emergent property of the plasma as a whole that does not relate to any individual atom or electronic state.

Following a discrete ionization process whereby a single electron is promoted to the continuum with sufficient energy to put it in thermal equilibrium with the other free electrons, the plasma is considered to undergo relaxation, through contact with the surroundings, to a new thermodynamic state in which the temperature(s) of the electrons and ions and the free electron density remain at their original values. Since $\Delta U_j$ has been defined to give the energy change at constant $n_e$, the extra continuum electron means that the system must expand by an amount $\Delta V = 1/n_e$ .

The energy $\Delta E$ transferred to a general system during an isothermal incremental volume change $\Delta V$ is given by the *First Energy Equation of Thermodynamics* [20], according to which

$$\Delta E = \left( T \left( \frac{\partial P}{\partial T} \right)_V - P \right) \Delta V \tag{46}$$

where $P$ is the pressure, and in which the first term represents heat transfer and the second, the work done. In this case, we have $\Delta E = \Delta \chi$ and $\Delta V = V/N_i \bar{Z}$ , where $N_i$ is the total number of atomic nuclei (ions). The energy deficit following an adiabatic ionization process is therefore,

$$\Delta \chi = \frac{T}{\bar{Z}} \left( T \frac{\partial}{\partial T} \left( \frac{PV}{N_i T} \right) \right) \tag{47}$$

which must be evaluated for fixed $N_i$, since the number of ions is fixed, and for fixed ionization, $\bar{Z} = N_e/N_i$ , since both the unrelaxed and relaxed states of the system are defined to differ from the initial state by the ionization state of a single ion. Equation (47) shows that this relaxation energy results from departures of the equation of state (of a fixed number of ions and free electrons) from a perfect gas. These departures are predominantly due to the Coulomb energy and electron degeneracy. The plasma equation of state can be written as

$$PV = N_i k T_i + \tfrac{2}{3} N_i \bar{Z} k T_e \frac{I_{3/2}(\eta_e)}{I_{1/2}(\eta_e)} + \tfrac{1}{3} U \tag{48}$$

where $U$ is the electrostatic potential energy and $\eta_e = \mu_e / kT_e$ . In situations when the direct electron contribution to the continuum lowering can be ignored ($\alpha \simeq 1$), $U$ is given by (31) and depends only upon the ion temperature. Equations (47) and (48) then give the relaxation energy as the sum of two terms,

$$\Delta \chi = \Delta \chi_i + \Delta \chi_e \tag{49}$$

which comprise the contribution from the ion subsystem,

$$\Delta \chi_i = \frac{kT_i}{3N_i \bar{Z}} \left( T_i \frac{\partial}{\partial T_i} \left( \frac{U}{kT_i} \right) \right) \tag{50}$$

and the electron contribution

$$\Delta \chi_e = \tfrac{2}{3} k T_e^2 \frac{\partial}{\partial T_e} \left( \frac{I_{3/2}(\eta_e)}{I_{1/2}(\eta_e)} \right) \tag{51}$$

We now consider these two contributions in more detail.

### *3.2 Electrostatic contribution to the relaxation energy*

Substituting for $U$ according to (27) into (50) and making use of (25) and (29), yields

$$\begin{aligned}\frac{\Delta\chi_{\mathrm{i}}}{kT_{\mathrm{i}}} &= -\frac{1}{6N_{\mathrm{i}}\bar{\bar{Z}}}T_{\mathrm{i}}\frac{\partial}{\partial T_{\mathrm{i}}}\left(\sum_j \frac{Z_j}{Z_{\mathrm{p}}}g\left(\Lambda_j\right)\right)\\ &= -\frac{1}{6N_{\mathrm{i}}\bar{\bar{Z}}}\sum_j \frac{Z_j}{Z_{\mathrm{p}}}g'\left(\Lambda_j\right)T_{\mathrm{i}}\frac{\partial\Lambda_j}{\partial T_{\mathrm{i}}}\\ &= \frac{1}{4N_{\mathrm{i}}\bar{\bar{Z}}}\sum_j \frac{Z_j}{Z_{\mathrm{p}}}g'\left(\Lambda_j\right)\Lambda_j\\ &= \frac{1}{4N_{\mathrm{i}}\bar{\bar{Z}}}\sum_j \frac{Z_j}{Z_{\mathrm{p}}}\left(h\left(\Lambda_j\right)-g\left(\Lambda_j\right)\right)\end{aligned} \tag{52}$$

where the functions $h(\Lambda)$ and $g(\Lambda)$ are given by (24) and (30) respectively.

Equation (52) applies only when the electron component of the plasma is negligibly polarised, ie $\alpha \simeq 1$. When $\alpha \neq 1$, the potential energy depends non-trivially on both electron and ion temperatures and the electron and ion subsystems are not thermodynamically decoupled vis à vis equation (46). However we can still apply this equation if the temperature ratio $T_{\mathrm{e}}/T_{\mathrm{i}}$ is non-vanishing and a function of volume (or density) alone (the special situation of $T_{\mathrm{e}} = 0$ having been already been addressed above) which embraces thermal equilibrium ($T_{\mathrm{e}}/T_{\mathrm{i}} = 1$ ). Then, referring to (36) and making use of equations in APPENDIX A,

$$T_{\mathrm{i}}\frac{\partial\alpha^2}{\partial T_{\mathrm{i}}} = T_{\mathrm{e}}\frac{\partial\alpha^2}{\partial T_{\mathrm{e}}} = \tfrac{3}{2}\left(\alpha^2-1\right)\left(\lambda_T-1\right) \tag{53}$$

where $\lambda_T$ is the electronic isothermal bulk modulus pressure coefficient (as defined in APPENDIX A). In the case of non-degenerate electrons, $\lambda_T = 1$, while, in the limit of extreme degeneracy, $\lambda_T = \tfrac{5}{3}$ . Therefore, in the non-degenerate limit, $\partial\alpha^2/\partial T = 0$ while, at extreme degeneracy ($T_{\mathrm{i}} << T_{\mathrm{B}}$ ), $T\left(\partial\alpha^2/\partial T\right) = \alpha^2 - 1 \sim 0$. The temperature derivative of $\alpha$ therefore vanishes in both limits and, since $1 \le \lambda_T \le \tfrac{5}{3}$ , remains close to unity, is small enough to ignore in all degeneracy regimes. (For $\eta_{\mathrm{e}} \simeq 0$ , $\tfrac{3}{2}\left(\alpha^2-1\right)\left(\lambda_T-1\right) \simeq \tfrac{1}{4}$.) Considering the temperature derivative of $X_j^3$ , using the argument given in section 2.4, gives, analogously to equation (37)ff, making use of (25),

$$T_{\mathrm{i}}\frac{\partial X_j^{\;3}}{\partial T_{\mathrm{i}}}=T_{\mathrm{i}}\frac{\partial X_j^{\;3}}{\partial x}\frac{\partial x}{\partial \Lambda_j}\frac{\partial \Lambda_j}{\partial T_{\mathrm{i}}}=\tfrac{3}{2}x\frac{\partial X_j^{\;3}}{\partial x}=-\tfrac{3}{2}\frac{y(1-y)^2(1+y)}{1+y+y^2} \tag{54}$$

where $0\le y\le 1$ is given by (38). This yields

$$-0.20628\le T_{\mathrm{i}}\frac{\partial X_j^{\;3}}{\partial T_{\mathrm{i}}}\le 0 \tag{55}$$

Combining these arguments, it follows that the temperature derivative of $\xi_j$, as defined by either (13) or (42), is virtually always negligible, which then allows the electron polarization screening to be treated by the simple device of replacing $\Lambda_j$ everywhere with $\tilde{\Lambda}_j=\xi_j\Lambda_j$, and treating $\xi_j$ as if it were constant. The result is that the relaxation energy (52) is generalized to

$$\frac{\Delta\chi_{\mathrm{i}}}{kT_{\mathrm{i}}}=\frac{1}{4N_{\mathrm{i}}\bar{Z}}\sum_j\frac{Z_j}{Z_{\mathrm{p}}}\left(h\left(\tilde{\Lambda}_j\right)-g\left(\tilde{\Lambda}_j\right)\right) \tag{56}$$

which yields, in the strong coupling limit, $\Gamma\gg 1\;\Rightarrow\;\Lambda\gg 1,\;h\left(\tilde{\Lambda}_j\right)-g\left(\tilde{\Lambda}_j\right)\sim\tfrac{4}{3}C\alpha^2\Gamma_j$,

$$\begin{aligned}\Delta\chi_{\mathrm{i}}&=\tfrac{1}{3}C\alpha^2\frac{\langle Z^{5/3}\rangle}{\bar{Z}^{2/3}}\left(\frac{e^2}{4\pi\epsilon_0 R_{\mathrm{WS}}}\right)\\&=\tfrac{1}{3}C\frac{\langle Z^{5/3}\rangle}{\bar{Z}^{2/3}}\left(1+\frac{T_{\mathrm{i}}}{Z_{\mathrm{p}}T_{\mathrm{B}}}\right)u_{\mathrm{WS}}^0\end{aligned} \tag{57}$$

where $R_{\mathrm{WS}}$ is the Wigner-Seitz radius (18), and $u_{\mathrm{WS}}^0$ is the Wigner-Seitz energy (21); while, for weak coupling, $\Gamma\ll 1\;\Rightarrow\;\Lambda\ll 1,\;h\left(\tilde{\Lambda}_j\right)-g\left(\tilde{\Lambda}_j\right)\sim\tfrac{1}{3}\tilde{\Lambda}_j=\dfrac{\Gamma_j R_j}{D}$ whereupon

$$\Delta\chi_{\mathrm{i}}\sim\frac{1}{4N_{\mathrm{i}}\bar{Z}}\sum_j\frac{Z_j^{\;2}e^2}{4\pi\epsilon_0 D}=\frac{Z_{\mathrm{p}}e^2}{16\pi\epsilon_0 D}=\tfrac{1}{4}Z_{\mathrm{p}}u_{\mathrm{DH}}^0 \tag{58}$$

where $u_{\mathrm{DH}}^0$ is the Debye-Hückel interaction energy (23).

*3.3 Electron degeneracy contribution to the relaxation energy*

According to (51), the contribution to the relaxation energy made by electron degeneracy is, for a free-electron gas, (APPENDIX A),

$$\Delta\chi_{\rm e} = kT_{\rm e}^{\,2}\left(1-\tfrac{2}{3}\frac{I_{3/2}I'_{1/2}}{I_{1/2}^{\;2}}\right)\frac{\partial\eta_{\rm e}}{\partial T_{\rm e}}$$

$$= -\tfrac{3}{2}kT_{\rm e}^{\,2}\left(1-\tfrac{2}{3}\frac{\bar{\varepsilon}}{kT_{\rm B}}\right)\left(\frac{T_{\rm B}}{T_{\rm e}^{\,2}}\right) \qquad (59)$$

$$= \bar{\varepsilon} - \tfrac{3}{2}kT_{\rm B}$$

which is equivalent to

$$\Delta\chi_{\rm e} = -n_{\rm e}\left(\frac{\partial\bar{\varepsilon}}{\partial n_{\rm e}}\right)_{T_{\rm e}} \qquad (60)$$

According to (45), the total electronic contribution to the SIPD is then,

$$\Delta\chi_{\rm e} - \bar{\varepsilon} = -\tfrac{3}{2}kT_{\rm B} \qquad (61)$$

which, in the limit of extreme degeneracy, is $-kT_{\rm F}$, according to which the spectroscopic ionization potential is raised by precisely the Fermi energy. This expresses the known result that , in an adiabatic isochoric ionization process in a fully degenerate system, the electron must be elevated, in accordance with the Pauli principle, to at least the energy of the Fermi surface, there being no available states of lower energy. The fact that the theory makes this adjustment automatically is reassuring and means that, for degenerate systems, a separate adjustment for the Fermi energy does not need to be made. It is also an example of a previously well understood circumstance when the static continuum lowering, which gives the bottom of the Fermi continuum, differs from the spectroscopic ionization potential, which corresponds to the Fermi surface. In the non-degenerate limit, (61) reduces to $\Delta\chi_{\rm e} - \bar{\varepsilon} = -\tfrac{3}{2}kT_{\rm e}$ which is just the average energy of a free electron.

Ionization potential depression is often thought of in terms of a change in the threshold energy, that being the minimum photon energy deemed to be required to cause ionization. In partially degenerate systems, this is not well-defined, because the photoionization edge is blurred by the thermal distribution. This has not been an issue thus far, because the ionization potential has been defined in terms of well-defined initial and final thermodynamic states of the plasma. However spectroscopic observation looks for thresholds, such as those relating to bound-free edges or the existence or non-existence of lines. These thresholds may not be sharply defined in terms of photon energy, resulting in some indefiniteness in how the SIPD is defined. In the non-degenerate and fully degenerate limits, this is not a problem: the ionization threshold energies are $\hbar\omega_0 + \Delta U_j - \Delta\chi_\mathrm{i}$ and $\hbar\omega_0 + \Delta U_j - \Delta\chi_\mathrm{i} + kT_\mathrm{F}$ respectively. At arbitrary electron temperatures (partial degeneracy) a reasonable definition of an effective photoionization threshold that interpolates between these limits is given by (45) with the reference energy offset given by

$$\Delta_\mathrm{e} = \tfrac{3}{2}kT_\mathrm{B} - kT_\mathrm{e}w(\eta_\mathrm{e}) = \tfrac{3}{2}kT_\mathrm{e}\left(I_{1/2}(\eta_\mathrm{e})/I'_{1/2}(\eta_\mathrm{e}) - \tfrac{2}{3}w(\eta_\mathrm{e})\right) \tag{62}$$

which is everywhere $\mathcal{O}(kT_\mathrm{e})$ and is a constant in the context of the problem (since $T_\mathrm{e}$ and $\eta_\mathrm{e}$ both relate to the defined initial state of the plasma). This leads to the effective photoionization threshold

$$\hbar\omega_j = \hbar\omega_0 + \Delta U_j - \Delta\chi_\mathrm{i} + kT_\mathrm{e}w(\eta_\mathrm{e}) \tag{63}$$

which defines the electron degeneracy-related contribution to the SIPD entirely in terms of the function $w(\eta)$ whose properties are that it is monotone, positive definite and possesses the following behaviour

$$\begin{aligned} w(\eta) &\sim \eta - 2 \qquad && \eta \gg 2 \\ w(\eta) &\sim 0 && \eta < 2 \end{aligned} \tag{64}$$

where the effective half-width of the Fermi surface is taken to be $2T_\mathrm{e}$ (corresponding to the intercepts of the tangent at $\varepsilon = \mu$ ). The first of equations (64) places the threshold at $\mu_\mathrm{e} - 2kT_\mathrm{e}$ in the regime of $\eta_\mathrm{e} \gg 2$ ; When $\eta < 2$ , the Fermi surface lies, fully or partially, below the continuum sufficient for there to be no degeneracy shift in the threshold energy. A suitable simple function complying with these limits is

$$w(\eta) = (\eta - 2)\,\mathrm{H}(\eta - 2) \tag{65}$$

where $\mathrm{H}(x)$ is the Heaviside function. Note that, defining $\Delta_\mathrm{e} = \tfrac{3}{2}kT_\mathrm{e}$ , as formerly proposed, leads to $kT_\mathrm{e}w(\eta_\mathrm{e}) = \tfrac{3}{2}k(T_\mathrm{B} - T_\mathrm{e})$ , which although possessing the correct extreme limits, vanishes too slowly with temperature in the non-degenerate regime at high densities, when the electrons are compressed to within separation distances of about a Bohr radius. The leading term in the high-temperature expansion ( $T >> T_\mathrm{F}$ ) gives

$$\tfrac{3}{2}k(T_\mathrm{B} - T_\mathrm{e}) \sim \frac{1}{2}\left(\frac{2}{\pi}\right)^{1/2}\frac{kT_\mathrm{F}^{3/2}}{T_\mathrm{e}^{1/2}} = \frac{3\pi}{8}\sqrt{a_\infty^{\;3}n_\mathrm{e}}\,\frac{e^2}{4\pi\epsilon_0 D_\mathrm{e}} \tag{66}$$

where $a_\infty$ is the Bohr radius. At solid density (see section 3.4) $a_\infty^{\;3}n_\mathrm{e} \sim \bar{Z}^2$ which leads to $\tfrac{3}{2}k(T_\mathrm{B} - T_\mathrm{e}) \sim \bar{Z}e^2/4\pi\epsilon_0 D_\mathrm{e}$ which is of classical proportions, despite being quantum-mechanicalin origin, showing that quantum effects in dense non-ideal plasmas can persist even at quite high temperatures. The persistence of this offset term in (63), would have led to unreasonably large corrections to the IPD at moderately high temperatures.

### *3.4 Relaxation energy in cold condensed matter*

In the special case of cold condensed matter, the total pressure is zero and (47) yields $\Delta\chi = 0$ . However, in a metal, the spectroscopic continuum must nevertheless start at the top of the Fermi surface, which implies $\Delta\chi_{\rm e} = \bar{\varepsilon} - kT_{\rm F}$. This means

$$\Delta\chi_{\rm i} = \Delta\chi - \Delta\chi_{\rm e} = \tfrac{3}{2}kT_{\rm B} - \bar{\varepsilon} \simeq \tfrac{2}{5}kT_{\rm F} \tag{67}$$

which is the electron degeneracy pressure. The vanishing of the relaxation energy in this regime is just an expression of the fact that the repulsive electron degeneracy pressure is balanced by the attractive Coulomb bonding forces. The SIPD is then given, according to (45), by

$$\hbar\Delta\omega_j = \Delta U_j - \Delta\chi_{\rm i} + \tfrac{3}{2}kT_{\rm B} - \Delta_{\rm e} = \Delta U_j + \bar{\varepsilon} - \Delta_{\rm e} \tag{68}$$

which includes a net lowering, equal to the electron degeneracy pressure, compared to static equilibrium theories, as shown in **Figure 1**.

Equating the Coulomb relaxation energy $\Delta\chi_{\rm i}$, (57), in the low temperature limit, to the electron degeneracy pressure given by (67) and using the standard relationship between the Fermi temperature and the electron density, yields the following estimate of the Brueckner parameter for solid metal at ambient,

$$r_s \equiv \frac{R_{\rm WS}}{a_\infty} = \frac{3}{5C\bar{Z}^{1/3}}\left(\frac{9\pi}{4}\right)^{2/3} = \frac{2.21}{C\bar{Z}^{1/3}} \tag{69}$$

and hence

$$a_\infty n_{\rm i}^{1/3} = \left(\frac{\bar{Z}}{2}\right)^{1/3}\left(\frac{10}{9\pi}C\right) \gtrsim 0.25\,\bar{Z}^{1/3} \tag{70}$$

in which $n_{\rm i}$ is both comparable with and greater than the Mott density [10], in line with expectation.

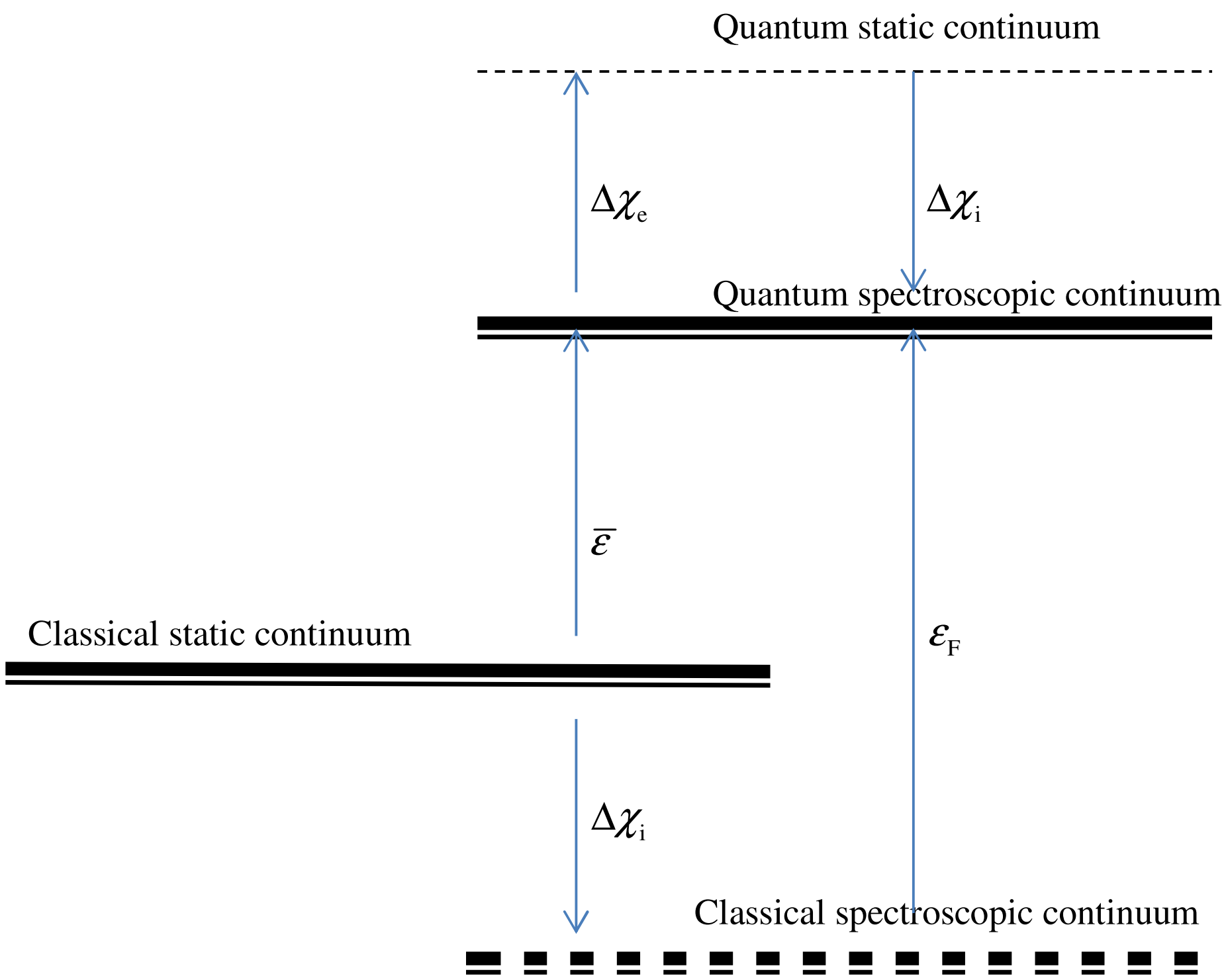

**Figure 1.** Diagram showing the various contributions to the shift in the continuum threshold for highly degenerate electrons in cold condensed matter, in which the net relaxation energy $\Delta\chi = \Delta\chi_e + \Delta\chi_i$ vanishes, showing how this results in a net downward shift of $\Delta\chi_i$ . The Classical static continuum is the equilibrium continuum level for classical electrons (No Pauli blocking due to electron degeneracy). The Classical spectroscopic continuum is the spectroscopic continuum level for classical electrons. The Quantum spectroscopic continuum is the resultant continuum threshold for degenerate electrons, for which $\varepsilon_F = kT_F$ is the Fermi energy.

## 4 Continuum lowering and ionization potential depression for discrete processes

### *4.1 Static continuum lowering*

The static continuum lowering, defined by (28) for example (in common with Stewart-Pyatt and related formulae) relates to the reversible excitation of an infinitesimally charged electron. More precisely, the continuum lowering contribution to the ionization potential depression is given by

$$\begin{aligned}\Delta U_j &= U\left(Z_j+1\right)-U\left(Z_j\right)\\ &\simeq \left.\frac{\partial U}{\partial Z_j}\right|_{Z_j+\frac{1}{2}} = \Delta U\left(Z_j+\tfrac{1}{2}\right)\end{aligned} \tag{71}$$

which states that the continuum lowering contribution to the IPD is, to an approximation, the static continuum lowering evaluated for the average charge state.

### *4.2 Continuum lowering in the strong coupling limit*

For strongly coupled systems, the electrostatic energy is given by (43), which implies the static continuum lowering,

$$\begin{aligned}\Delta U_j &= -C\alpha^2\overline{Z}^{1/3}\left(\left(Z_j+1\right)^{5/3}-Z_j^{\,5/3}\right)u_{\mathrm{WS}}^0\\ &\simeq -\tfrac{5}{3}C\alpha^2\overline{Z}^{1/3}\left(Z_j+\tfrac{1}{2}\right)^{2/3}u_{\mathrm{WS}}^0\end{aligned} \tag{72}$$

in which the error resulting from the final-stage approximation is $<5\%$ even in the worst case of $Z_j = 0$. From (57), the corresponding relaxation energy is

$$\Delta\chi_{\mathrm{i}} = \frac{1}{3}C\alpha^2\frac{\left\langle Z^{5/3}\right\rangle}{\overline{Z}^{2/3}}u_{\mathrm{WS}}^0 \tag{73}$$

Upon combining the above results, the total SIPD (45) is given by

$$\hbar\Delta\omega_j = \Delta U_j - \Delta\chi_{\mathrm{i}} + kT_{\mathrm{e}} w(\eta_{\mathrm{e}})$$

$$= -Cu_{\mathrm{WS}}^0 \bar{Z}^{1/3} \left( (Z_j+1)^{5/3} - Z_j^{5/3} + \frac{\langle Z^{5/3} \rangle}{3\bar{Z}} \right) \left( 1 + \frac{T_{\mathrm{i}}}{Z_{\mathrm{p}} T_{\mathrm{B}}} \right) + kT_{\mathrm{e}} w(\eta_{\mathrm{e}}) \tag{74}$$

$$\simeq -\tfrac{1}{3} Cu_{\mathrm{WS}}^0 \bar{Z}^{1/3} \left( 5(Z_j+\tfrac{1}{2})^{2/3} + \frac{\langle Z^{5/3} \rangle}{\bar{Z}} \right) \left( 1 + \frac{T_{\mathrm{i}}}{Z_{\mathrm{p}} T_{\mathrm{B}}} \right) + kT_{\mathrm{e}} w(\eta_{\mathrm{e}})$$

*4.3 Continuum lowering in the weak coupling limit*

In the limit of weak coupling, the plasma energy is given by

$$U = -\tfrac{1}{2}\alpha \sum_j \frac{Z_j^2 e^2}{4\pi\epsilon_0 D_{\mathrm{i}}} = \alpha U_{\mathrm{DH}} = -\tfrac{1}{2} N_{\mathrm{i}} \bar{Z} Z_{\mathrm{p}} u_{\mathrm{DH}}^0 \tag{75}$$

The static continuum lowering implied by (33) is now

$$\Delta U_j = -\tfrac{1}{2}\left( (Z_j+1)^2 - Z_j^2 \right) u_{\mathrm{DH}}^0$$

$$= -(Z_j+\tfrac{1}{2}) u_{\mathrm{DH}}^0 \tag{76}$$

$$= -\frac{(Z_j+\tfrac{1}{2})e^2}{4\pi\epsilon_0 D}$$

in obtaining which no (further) approximation is necessary. From (58), the corresponding relaxation energy is

$$\Delta\chi_{\mathrm{i}} \sim \frac{\alpha}{4N_{\mathrm{i}}\bar{Z}} \sum_j \frac{Z_j^2 e^2}{4\pi\epsilon_0 D_{\mathrm{i}}} = -\tfrac{1}{4} Z_{\mathrm{p}} u_{\mathrm{DH}}^0 \tag{77}$$

Upon combining the above results, in the weak-coupling limit, assuming non-degenerate electrons ( $w(\eta_{\mathrm{e}}) = 0$ ) the total SIPD (45) is given by

$$\hbar\Delta\omega_j = \Delta U_j - \Delta\chi_{\mathrm{i}}$$

$$= -(Z_j + \tfrac{1}{4} Z_{\mathrm{p}} + \tfrac{1}{2}) u_{\mathrm{DH}}^0 \tag{78}$$

*4.4 General formula for the spectroscopic IPD*

The formula for the SIPD of plasmas under regimes of arbitrary coupling and electron degeneracy results from a combination of equations (63), (49), (61), (56), (71) and (12) , which yields

$$\hbar\Delta\omega_j = -\frac{kT_\mathrm{i}}{2Z_\mathrm{p}}\left( h\left(\tilde{\Lambda}_j^+\right) + \frac{1}{2N_\mathrm{i}\bar{Z}}\sum_{j'} Z_{j'}\left(h\left(\tilde{\Lambda}_{j'}^0\right) - g\left(\tilde{\Lambda}_{j'}^0\right)\right)\right) + kT_\mathrm{e}w\left(\eta_\mathrm{e}\right) \tag{79}$$

$$\tilde{\Lambda}_j^\nu = \xi_j^\nu \Lambda_j^\nu$$

$$\Lambda_j^0 = \frac{Z_j}{\bar{Z}}\Lambda_0^0, \quad \Lambda_j^+ = \frac{Z_j + \frac{1}{2}}{\bar{Z}}\Lambda_0^0, \quad \Lambda_0^0 = \left(3\Gamma\right)^{3/2} \tag{80}$$

$$\Gamma = \frac{Z_\mathrm{p}\bar{Z}e^2}{4\pi\epsilon_0 R_\mathrm{WS} kT_\mathrm{i}}$$

in which $\nu$ denotes the index 0 or + as defined above and $\xi_j^\nu$ is given by (42) in terms of $X_j$, which is the positive real solution of (5) with $Y_j = \left(\Lambda_j^\nu\right)^{-1/3}$ ; $C_j$, which is the force constant ($= 9/10$ for fluid systems) as discussed in sect. 2.4 ; and $\alpha$, which is the ratio of the screening lengths as expressed by (8). Equation (79) depends on the CSD. If this is not precisely known, or when a simpler result is required, (79) can be replaced with the more approximate formula

$$\hbar\Delta\omega_j = -kT_\mathrm{i}\left(\frac{1}{2Z_\mathrm{p}} h\left(\tilde{\Lambda}_j^+\right) + \frac{1}{4\bar{Z}}\left(h\left(\tilde{\Lambda}_0^0\right) - g\left(\tilde{\Lambda}_0^0\right)\right)\right) + kT_\mathrm{e}w\left(\eta_\mathrm{e}\right) \tag{81}$$

In the strong-coupling limit, (81) yields

$$\hbar\Delta\omega_j = -\tfrac{1}{3}C\alpha^2 u_\mathrm{WS}^0\left(5\left(Z_j + \tfrac{1}{2}\right)^{2/3}\bar{Z}^{1/3} + Z_\mathrm{p}\right) + kT_\mathrm{e}w\left(\eta_\mathrm{e}\right) \tag{82}$$

which agrees with (74) , subject to the approximation $\left\langle Z^{5/3}\right\rangle \sim \left\langle Z^2\right\rangle / \left\langle Z\right\rangle^{1/3} = Z_\mathrm{p}\bar{Z}^{2/3}$. For sharply peaked CSDs likely to be encountered in strongly-coupled dense matter, the error is of the order of a few percent or less, and the approximation is consistently, albeit marginally, better than $\left\langle Z^{5/3}\right\rangle \sim \left\langle Z\right\rangle^{5/3}$ .

In the limit of weak-coupling and weak-degeneracy, (81) leads directly to the result (78) without further approximation.

## 5 Thermodynamic treatment of ionization

### 5.1 Entropy and free energy

Important insight into the problem is gained by considering ionization fully from the perspective of a thermodynamic process in an electrically neutral plasma comprising a fixed number, $N_{\mathrm{i}}$ , of atomic nuclei of a single species. (The generalization to multiple nuclear species, while straightforward, is omitted here in order to maintain clarity.) Let $E$ be the plasma energy, which is the energy residing in the degrees of freedom involving the component particles (ions and free electrons) including their mutual (Coulomb) interactions but excluding the energy contained in internal states of the atomic ions (electrons in bound states). Let $z$ represent one or more internal microscopic configuration variables describing these internal states, and suppose that it is possible to vary $z$ through the application of external influences, such as electromagnetic fields or radiation. In such circumstances, it is reasonable to promote $z$ to the status of a thermodynamic variable, in which case we can define $I$ to be the thermodynamic potential associated with $z$ whereby a change $\mathrm{d}z > 0$ in $z$ is associated with some energy $I\,\mathrm{d}z$ being made available. For an infinitesimal reversible process in such a system,

$$\mathrm{d}E + P\,\mathrm{d}V - T\,\mathrm{d}S - I\,\mathrm{d}z = 0 \tag{83}$$

where $S(V,T,z)$ is the entropy function, in terms of which the probability of $z$ in a closed system in equilibrium at fixed $V,T$ is given by the usual Gibbs distribution,

$$p(z) = \mathrm{e}^{-S(V,T,z)} \tag{84}$$

which satisfies the condition $\sum_z p(z) = 1$ or $\int p(z)\rho(z)\,\mathrm{d}z = 1$ , where $\rho(z)$ is the density of states represented by $z$ , depending on whether $z$ takes on discrete or continuous values. The distinction is unimportant, and, for sake of argument, we shall start by assuming the latter. The expectation value of $z$ is

$$\overline{z} = \langle z \rangle \equiv \int p(z)\, z\rho(z)\,\mathrm{d}z \tag{85}$$

corresponding to the macroscopic entropy

$$S_0(V,T) = -\int p \ln(p)\,\rho(z)\,\mathrm{d}z = \langle S(V,T,z)\rangle \tag{86}$$

which follows from (84). Expanding $S(V,T,z)$ about $z = \overline{z}$ and taking the average yields

$$\langle S(V,T,z)\rangle = S(V,T,\overline{z}) + \tfrac{1}{2}S''(V,T,\overline{z})\langle \Delta z^2\rangle + \ldots \tag{87}$$

where $S''(V,T,\overline{z}) = \left(\partial^2 S(V,T,z)/\partial z^2\right)_{V,T}\Big|_{z=\overline{z}}$ and $\Delta z = z - \overline{z}$ . If $z$ is a normally distributed variate, then

$$S(V,T,z) = S_0(V,T) + \tfrac{1}{2}\left((z-\bar{z})^2 \big/ \langle \Delta z^2 \rangle - 1\right) \tag{88}$$

in which case

$$S_0(V,T) = \langle S(V,T,z) \rangle = S(V,T,\bar{z}) + \tfrac{1}{2} \tag{89}$$

and, in particular,

$$S'(V,T,\bar{z}) \equiv \left( \frac{\partial}{\partial z} S(V,T,z) \right)_{V,T} \Bigg|_{z=\bar{z}} = 0 \tag{90}$$

which expresses the important property, which will be shown to hold generally, that the equilibrium values of the macroscopic thermodynamic coordinates are stationary with respect to the microscopic variables $z$. Equation (83) implies the following additional Maxwell relations,

$$\begin{gathered} \left(\frac{\partial T}{\partial z}\right)_S = \left(\frac{\partial I}{\partial S}\right)_z \quad , \quad \left(\frac{\partial S}{\partial z}\right)_T = -\left(\frac{\partial I}{\partial T}\right)_z \\ \left(\frac{\partial P}{\partial z}\right)_V = -\left(\frac{\partial I}{\partial V}\right)_z \quad , \quad \left(\frac{\partial V}{\partial z}\right)_P = -\left(\frac{\partial I}{\partial T}\right)_z \end{gathered} \tag{91}$$

If $Z_K$ is the charge on an ion in the state $K$, and $N_K$ is the mean number of ions in that state, charge neutrality is expressed by

$$N_{\mathrm{e}} = \sum_K Z_K N_K \tag{92}$$

which, note, is also a statement about the average charge state of the plasma, for fixed

$$N_{\mathrm{i}} = \sum_K N_K \tag{93}$$

Maximising the entropy subject to the constraints of particle numbers and total energy yields

$$S = \frac{E}{T} - \eta_{\mathrm{e}} N_{\mathrm{e}} - \sum_K \eta_K N_K + \mathfrak{Z} \tag{94}$$

where $1/T$, $\eta_{\mathrm{e}}$, $\{\eta_K\}$, $\mathfrak{Z}$ are the Lagrange multipliers, with $\ln \mathfrak{Z}$ being the *partition function* and $T\mathfrak{Z}$ the *grand potential*. We now make an important departure from the standard theory of equilibrium systems by generalizing to systems exhibiting weak electron-ion coupling by treating these as separate subsystems, with different temperatures $T_{\mathrm{e}} \neq T_{\mathrm{i}}$. Writing

$$S = S_{\mathrm{i}}(V,T_{\mathrm{e}},T_{\mathrm{i}},V,z) + S_{\mathrm{e}}(V,T_{\mathrm{e}},z) \tag{95}$$

and maximising the entropies independently yields

$$S_{\mathrm{i}} = \frac{E_{\mathrm{i}}}{T_{\mathrm{i}}} - \sum_K \eta_K N_K + \mathfrak{Z}_{\mathrm{i}}$$

$$S_{\mathrm{e}} = \frac{E_{\mathrm{e}}}{T_{\mathrm{e}}} - \eta_{\mathrm{e}} N_{\mathrm{e}} + \mathfrak{Z}_{\mathrm{e}} \tag{96}$$

in which we have made the assumption that the (free) electron dynamics are negligibly affected by the ions being at a different temperature (should this be so). Equation (83) then generalizes to

$$\mathrm{d}E + P\,\mathrm{d}V - T_{\mathrm{i}}\,\mathrm{d}S_{\mathrm{i}} - T_{\mathrm{e}}\,\mathrm{d}S_{\mathrm{e}} - I\,\mathrm{d}z = 0 \tag{97}$$

where $P = P_{\mathrm{e}} + P_{\mathrm{i}}$ and $E = E_{\mathrm{i}} + E_{\mathrm{e}}$ , with the respective temperatures given by

$$\left(\frac{\partial E}{\partial S_{\mathrm{i}}}\right)_{V,z} = T_{\mathrm{i}}, \quad \left(\frac{\partial E}{\partial S_{\mathrm{e}}}\right)_{V,z} = T_{\mathrm{e}} \tag{98}$$

These equations describe the ion and electron subsystems as being independently in equilibrium.

The Gibbs free energies for the ion and electron subsystems are defined in the usual way

$$G_{\mathrm{i}} = \sum_K \mu_K N_K$$

$$G_{\mathrm{e}} = \mu_{\mathrm{e}} N_{\mathrm{e}} \tag{99}$$

where $\mu_{\mathrm{e}} = T_{\mathrm{e}}\eta_{\mathrm{e}}$ and $\mu_K = T_{\mathrm{i}}\eta_K$ are the electron and ion chemical potentials respectively. The chemical potentials are intensive quantities that are, for a given plasma composition, functions of the respective temperatures and pressures. The plasma composition is determined by *chemical equilibrium* between the electrons and the various ion states.

### *5.2 Chemical equilibrium*

The general changes in the respective Gibbs free energies of the system are given by [20]

$$\mathrm{d}G_{\mathrm{e}} = -S_{\mathrm{e}}\,\mathrm{d}T_{\mathrm{e}} + V\,\mathrm{d}P_{\mathrm{e}} + \mu_{\mathrm{e}}\,\mathrm{d}N_{\mathrm{e}}$$

$$\mathrm{d}G_{\mathrm{i}} = -S_{\mathrm{i}}\,\mathrm{d}T_{\mathrm{i}} + V\,\mathrm{d}P_{\mathrm{i}} + \sum_K \mu_K\,\mathrm{d}N_K \tag{100}$$

which hold for *any* infinitesimal process involving a change in the plasma charge state. Chemical equilibrium, at constant pressure and temperature(s) depends upon the total Gibbs free energy $G = G_{\mathrm{e}} + G_{\mathrm{i}}$ being minimised with respect to variations $\mathrm{d}N_K$ in the composition, subject to the number of ions being fixed, whereupon

$$\sum_K \mathrm{d}N_K = 0 \tag{101}$$

and charge neutrality

$$\mathrm{d}N_\mathrm{e} = \sum_K Z_K \,\mathrm{d}N_K \tag{102}$$

The minimization condition

$$\left(\frac{\partial G}{\partial N_K}\right)_{P,T} = 0 \tag{103}$$

where hereafter, until otherwise indicated, $T$ denotes $T_\mathrm{e}$ and $T_\mathrm{i}$ severally, where these are distinct, implies that the chemical potentials of those species present in the system must satisfy

$$\mu_K + Z_K \mu_\mathrm{e} = \mu_0 \tag{104}$$

for some fixed $\mu_0$ that does not depend on the atomic configuration, and which corresponds, by inspection, to the chemical potential of a neutral atom. It is important to recognise that (104) is a condition for equilibrium and is not a constitutive relation. Consideration from the point of view of equilibrium at constant volume follows equivalent lines, except that it is then the Helmholtz free energy $F = G - PV$ that is minimised. Since, for any internal configuration variable $z$ ,

$$\begin{aligned}
\left(\frac{\partial F}{\partial z}\right)_{V,T} &\equiv \left(\frac{\partial G}{\partial z}\right)_{V,T} - V\left(\frac{\partial P}{\partial z}\right)_{V,T} \\
&\equiv \left(\frac{\partial G}{\partial z}\right)_{P,T} + \left(\frac{\partial G}{\partial P}\right)_{z,T}\left(\frac{\partial P}{\partial z}\right)_{V,T} - V\left(\frac{\partial P}{\partial z}\right)_{V,T} \\
&\equiv \left(\frac{\partial G}{\partial z}\right)_{P,T}
\end{aligned} \tag{105}$$

this also leads to (104).

Thus chemical equilibrium between the electron and ion subsystems generally depends upon the chemical potential differences,

$$\Delta_{JK} = \mu_J - \mu_K + \left(Z_J - Z_K\right)\mu_\mathrm{e} \tag{106}$$

vanishing for $\forall J, K$. Whenever any $\Delta_{JK} \neq 0$, the system is not in equilibrium, with $\Delta_{JK} > 0$ (or $\Delta_{JK} < 0$) implying a tendency for the reaction between the states $J$ and $K$ to proceed spontaneously in the direction $J \Rightarrow K$ (or $K \Rightarrow J$).

Ionization of a single atom or ion in a plasma in thermodynamic equilibrium, under conditions when the electronic chemical potential $\mu_{\mathrm{e}}$, and the temperatures are fixed, via the reaction $J \Rightarrow K + \mathrm{e}$ corresponds, using (99), to $\mathrm{d}G_{\mathrm{i}} = \mu_K - \mu_J$, $\mathrm{d}G_{\mathrm{e}} = \mu_{\mathrm{e}}$ and hence $\mathrm{d}G = 0$, which, in a closed system, corresponds to an isobaric, isothermal process.

The general change in the total Gibbs free energy during an infinitesimal process of a system in chemical equilibrium is, making use of (102), (104) and (101),

$$\begin{aligned} \mathrm{d}G = \mathrm{d}G_{\mathrm{e}} + \mathrm{d}G_{\mathrm{i}} &= -S_{\mathrm{e}}\,\mathrm{d}T_{\mathrm{e}} - S_{\mathrm{i}}\,\mathrm{d}T_{\mathrm{i}} + V\left(\mathrm{d}P_{\mathrm{e}} + \mathrm{d}P_{\mathrm{i}}\right) + \mu_{\mathrm{e}}\,\mathrm{d}N_{\mathrm{e}} + \sum_K \mu_K\,\mathrm{d}N_K \\ &= -S_{\mathrm{e}}\,\mathrm{d}T_{\mathrm{e}} - S_{\mathrm{i}}\,\mathrm{d}T_{\mathrm{i}} + V\,\mathrm{d}P + \sum_K \left(\mu_K + Z_K\mu_{\mathrm{e}}\right)\mathrm{d}N_K \\ &= -S_{\mathrm{e}}\,\mathrm{d}T_{\mathrm{e}} - S_{\mathrm{i}}\,\mathrm{d}T_{\mathrm{i}} + V\,\mathrm{d}P \end{aligned} \tag{107}$$

which, when combined with (97), yields that, for a reversible process of the closed system ,

$$\begin{aligned} \mathrm{d}G &= \mathrm{d}E - S_{\mathrm{e}}\,\mathrm{d}T_{\mathrm{e}} - T_{\mathrm{e}}\,\mathrm{d}S_{\mathrm{e}} - S_{\mathrm{i}}\,\mathrm{d}T_{\mathrm{i}} - T_{\mathrm{i}}\,\mathrm{d}S_{\mathrm{i}} + V\,\mathrm{d}P + P\,\mathrm{d}V - I\,\mathrm{d}z \\ &= \mathrm{d}E - \mathrm{d}\left(T_{\mathrm{e}}S_{\mathrm{e}}\right) - \mathrm{d}\left(T_{\mathrm{i}}S_{\mathrm{i}}\right) + \mathrm{d}\left(PV\right) - I\,\mathrm{d}z \end{aligned} \tag{108}$$

which reveals that $I\,\mathrm{d}z$ is a total differential , ie $I\,\mathrm{d}z = \mathrm{d}\Phi$ where $\Phi$ is a function of the state variables, and moreover that $I$ must depend on $z$ alone. Equations (91) then imply that the first derivatives of $P, V, T, S$ with respect to z all vanish, which indicates that, *at equilibrium*, these variables are all at extrema with respect to $z$, so, for any independent set of coordinates, eg $S, V, T$,

$$\left(\frac{\partial S}{\partial z}\right)_{V,T} = 0; \quad \left(\frac{\partial V}{\partial z}\right)_{S,T} = 0 \tag{109}$$

Integrating (108) then yields

$$G = E - T_{\mathrm{e}}S_{\mathrm{e}} - T_{\mathrm{i}}S_{\mathrm{i}} + PV - \Phi \tag{110}$$

and, upon referring to (96), the grand potential is

$$T_{\mathrm{i}}\mathfrak{Z}_{\mathrm{i}} + T_{\mathrm{e}}\mathfrak{Z}_{\mathrm{e}} = PV - \Phi \tag{111}$$

Since, at equilibrium, $I$ does not depend on the macroscopic coordinates, it can depend only the internal coordinates, and the energy variation is given by $\mathrm{d}E(z)=\mathrm{d}E(\overline{z})-\mathrm{d}\Phi$, which yields $\Phi$ as the deviation from equilibrium of the total binding energy of the electrons in the atomic system configurations, when the atoms are completely isolated from each other:

$$\Phi(\mathbf{z})=\sum_j\left(E_{\mathrm{c}}\left(\overline{\mathbf{z}}_j\right)-E_{\mathrm{c}}\left(\mathbf{z}_j\right)\right)=\sum_J\left(\overline{N}_J-N_J\right)E_{\mathrm{c}}\left(\mathbf{z}_J\right) \tag{112}$$

with $\mathbf{z}_j=\left(z_{j\alpha},z_{j\beta},...\right)$ denoting the electronic configuration of the atom $j$, and where $E_c\left(\mathbf{z}_J\right)\leq 0$ is the total energy of the configuration $J$ defined by $\mathbf{z}_J=\left(z_{J\alpha},z_{J\beta},\ldots\right)$. Equation (112) is exact, being an irrefutable consequence of the thermodynamics. It means that any changes to the configuration energies due to interactions between ions are contained in the other thermodynamic terms. Note that, while the average of $\Phi$ vanishes identically, the fluctuations of this quantity nevertheless have an important role to play.

Let $z$ denote some $z_{j\alpha}$, which is the occupancy of an energy level $\alpha$ in the atom, $j$, whose initial configuration is $J$. The ionization reaction $J\Rightarrow K+\mathrm{e}$, where $Z_K=Z_J+1$, and $\mathbf{z}_K=\left(z_{J\alpha}-1,z_{J\beta},\ldots\right)\equiv\mathbf{z}_J-\hat{\boldsymbol{v}}_\alpha$ then corresponds to $\Delta z=-1$. The reaction can then be expressed by the differential relations

$$\begin{aligned}&\frac{\partial N_{\mathrm{e}}}{\partial z}=-1,\quad\frac{\partial N_J}{\partial z}=+1,\quad\frac{\partial N_K}{\partial z}=-1\\&\frac{\partial N_L}{\partial z}=0,\quad L\neq J,K\end{aligned} \tag{113}$$

$$\frac{\partial Z_j}{\partial z}=-1 \tag{114}$$

and hence, from (112) , using (113),

$$I=\frac{\partial\Phi}{\partial z}=\frac{\partial\Phi}{\partial z_{J\alpha}}=E_c\left(\mathbf{z}_K\right)-E_c\left(\mathbf{z}_J\right)\equiv\phi_{J\to K}=\phi_{J\alpha} \tag{115}$$

where, for $\hat{\boldsymbol{v}}_\alpha=\mathbf{z}_J-\mathbf{z}_K$ , $\phi_{J\alpha}=\phi_{J\to K}>0$ is the ionization potential from the level $\alpha$ , in an isolated ion in the configuration $J$ , leading to the configuration $K$ .

*5.3 The thermodynamic ionization potential*

Equations (100) and (105) in conjunction with (113) yield

$$\left(\frac{\partial G}{\partial z}\right)_{P,T} \equiv \left(\frac{\partial F}{\partial z}\right)_{V,T} = \mu_K - \mu_J - \mu_e = \Delta_{KJ} \tag{116}$$

The condition for ionization equilibrium, $\Delta_{KJ} = 0$ , $\forall\ K,J$ , is therefore expressed by

$$\left(\frac{\partial G}{\partial z}\right)_{P,T} = \left(\frac{\partial F}{\partial z}\right)_{V,T} = 0 \tag{117}$$

where the chemical potentials are given by

$$\begin{aligned} \left(\frac{\partial G}{\partial N_e}\right)_{P,T} &= \left(\frac{\partial F}{\partial N_e}\right)_{V,T} = \mu_e \\ \left(\frac{\partial G}{\partial N_K}\right)_{P,T} &= \left(\frac{\partial F}{\partial N_K}\right)_{V,T} = \mu_K \end{aligned} \tag{118}$$

where, it should be noted, $N_K$ , $N_e$ are the actual particle numbers, which are independent of the macroscopic thermodynamic variables $P,V,T\ldots$ , in contradistinction to their averages, which are generally presumed to be functions of the macroscopic thermodynamic variables.

Now, let the total Helmholtz free energy be expressed in the form

$$F = F^0 + \Delta F \tag{119}$$

where $F^0 = F_i^0 + F_e^0$ is the free energy of a system comprising the same mixture (expressed in terms of $\{N_K\}, N_e$ ) of *non-interacting* particles at the same volume and temperature. The condition for equilibrium (117) then becomes

$$\left(\frac{\partial F}{\partial z}\right)_{V,T} = \mu_K^0 - \mu_J^0 - \mu_e^0 + \left(\frac{\partial \Delta F}{\partial z}\right)_{V,T} = 0 \tag{120}$$

where $\mu_K^0\left(n_K, T_i\right)$, $\mu_e^0\left(n_e, T_e\right)$ are the *non-interacting* ion and electron chemical potentials at the respective particle densities and temperatures, and where

$$\begin{aligned} F_i^0 + P_i^0 V &= \sum_K \mu_K^0 N_K \\ F_e^0 + P_e^0 V &= \mu_e^0 N_e \end{aligned} \tag{121}$$

are the Gibbs free energies of the non-interacting system at the pressure $P^0 = P_i^0 + P_e^0$ corresponding to *the same particle densities*. The equivalent decomposition of the Gibbs free energy, $G = G^0 + \Delta G$ , on the other hand, leads to

$$\left(\frac{\partial G}{\partial z}\right)_{P,T} = \tilde{\mu}_K^0 - \tilde{\mu}_J^0 - \tilde{\mu}_e^0 + \left(\frac{\partial \Delta G}{\partial z}\right)_{P,T} = 0 \tag{122}$$

which, by virtue of (117), is equivalent to (120), in that $G^0$ , and the associated chemical potentials $\tilde{\mu}_K^0$ , $\tilde{\mu}_e^0$ are now those that correspond to the non-interacting system at the same total pressure, $P$ and temperature. The relationship between $\Delta F$ and $\Delta G$ is expressed by

$$F = F^0 + \Delta F = G - PV = G^0 + \Delta G - PV \tag{123}$$

where

$$\begin{aligned} G^0 &= G_0(P,T) = F_0\left(V^0,T\right) + PV^0 \\ F^0 &= F_0(V,T) = G_0\left(P^0,T\right) - P^0 V \end{aligned} \tag{124}$$

and where $G_0(P,T)$ and $F_0(V,T)$ are the Gibbs and Helmholtz functions respectively for the non-interacting particle systems having the same particle concentrations as the interacting system, and

$$V^0 = \left(\frac{\partial G_0(P,T)}{\partial P}\right)_T, \qquad P^0 = -\left(\frac{\partial F_0(V,T)}{\partial V}\right)_T \tag{125}$$

The chemical potentials of any particle species x are then found to be related by

$$\tilde{\mu}_x^0 = \frac{\partial G_0(P,T)}{\partial N_x} = \mu_x^0\left(\frac{V}{V^0} n_x, T_x\right) \tag{126}$$

For Boltzmann particles and non-degenerate electrons (APPENDIX A),

$$\mu_x^0(n,T) = T\ln\left(\frac{n}{g_x}\left(\frac{2\pi\hbar^2}{m_x kT}\right)^{3/2}\right) \tag{127}$$

where $g_x$ is the spin degeneracy of the species x . This yields $\tilde{\mu}_K^0 - \mu_K^0 = kT_i \ln\left(V/V^0\right)$ , which does not depend on the species type $K$ , and so, from equations (120) and (122), we obtain the *thermodynamic ionization potential*

$$\phi + \Delta W \equiv -\left(\frac{\partial \Delta F}{\partial z}\right)_{V,T} = \tilde{\mu}_e^0 - \mu_e^0 - \left(\frac{\partial \Delta G}{\partial z}\right)_{P,T} \tag{128}$$

which is the change in the free energy associated with the hypothetical removal of an electron from a bound state within the ion and which defines $\Delta W$ as the thermodynamic IPD (TIPD). Substituting (128) into the equilibrium condition (120) yields

$$\mu_K^0 - \mu_J^0 - \mu_e^0 = \phi_J + \Delta W_J \tag{129}$$

which, with the aid of (127) (for non-degenerate electrons) with $T_e = T_i = T$ , becomes,

$$\frac{n_J n_e}{n_K} = \frac{2g_J}{g_K}\left(\frac{m_e kT}{2\pi\hbar^2}\right)^{3/2} \exp\left(-\frac{\phi_J + \Delta W_J}{kT}\right) \tag{130}$$

Equation (130) is the Saha equation and, importantly, demonstrates that it is the thermodynamic IPD, $\Delta W$ that features in this particular equation of state [11], rather than any of the other forms of the IPD. The SCL has a different role in the equation of state, as discussed in APPENDIX B.

The contribution $\Delta F$ to the free energy is associated with the effective interaction energy $U$ [10], in which case

$$\Delta F = U - T\Delta S - \Phi \tag{131}$$

where

$$\Delta S = -\frac{\partial \Delta F}{\partial T} \tag{132}$$

whereupon

$$\Delta F = T\int_T^\infty \frac{U - \Phi}{T^2}\,\mathrm{d}T \tag{133}$$

.

In the case of pure Coulomb interactions, the scaling laws arising from the virial theorem etc, imply that the interaction free energy can be expressed in terms of some function $f(\Lambda)$ of the coupling parameters $\Lambda_j$ defined by equation (13), in the manner of

$$\begin{aligned} \Delta F_i &= -\frac{kT_i}{3Z_p}\sum_j Z_j f(\Lambda_j) \\ \Delta F_e &= -\Phi \end{aligned} \tag{134}$$

in terms of which, making use of (25) ,

$$U_i = \Delta F_i - T_i \frac{\partial}{\partial T_i}\Delta F = -\frac{kT_i}{2Z_p}\sum_j Z_j \Lambda_j f'(\Lambda_j) \tag{135}$$

$$V\Delta P_{\mathrm{i}} = -V\frac{\partial}{\partial V}\Delta F_{\mathrm{i}} = -\frac{kT_{\mathrm{i}}}{6Z_{\mathrm{p}}}\sum_j Z_j \Lambda_j f'\left(\Lambda_j\right) = \tfrac{1}{3}U_{\mathrm{i}} \tag{136}$$

Comparison of (135) with (27) then yields

$$g\left(\Lambda\right) = \Lambda f'\left(\Lambda\right) \tag{137}$$

Hence, using (26), the thermodynamic ionization potential is

$$-\left(\frac{\partial \Delta F}{\partial z}\right)_{V,T} = \phi_J - \frac{kT_{\mathrm{i}}}{3Z_{\mathrm{p}}}\left(f\left(\Lambda_j\right) + g\left(\Lambda_j\right)\right) = \phi_J + \Delta W_J \tag{138}$$

in which $\Delta W_J = \Delta W_j$ where

$$\frac{\Delta W_j}{kT_{\mathrm{i}}} = -\frac{1}{3Z_{\mathrm{p}}}\left(f\left(\Lambda_j\right) + \Lambda_j f'\left(\Lambda_j\right)\right) \tag{139}$$

is the Thermodynamic Ionization Potential Depression (TIPD), where

$$f\left(\Lambda\right) = \int_0^{\Lambda} \frac{g\left(\lambda\right)}{\lambda}\mathrm{d}\lambda \tag{140}$$

Equation (139) resembles, but is distinct from, the corresponding formula (28) for the static continuum lowering. Taking $g\left(\Lambda\right)$ to be given by (30), yields [12]

$$\begin{aligned} f\left(\Lambda\right) &= \tfrac{9}{10}s^2 - \tfrac{3}{5}\frac{1+s}{1+s+s^2} + \sqrt{3}\arctan\left(\frac{1}{\sqrt{3}}\frac{s-1}{s+1}\right) - \tfrac{3}{2}\ln\left(\frac{1+s+s^2}{3}\right) - \tfrac{1}{2} \\ s\left(\Lambda\right) &= \left(1+\Lambda\right)^{1/3} \end{aligned} \tag{141}$$

In the strong coupling limit, $f\left(\Lambda\right) \sim \tfrac{9}{10}\Lambda^{2/3}$ , $g\left(\Lambda\right) \sim \tfrac{3}{5}\Lambda^{2/3}$ and the TIPD reduces to the static continuum lowering $\Delta U_j^{\mathrm{is}}$, which is as given by (28) in the limit of large $\Lambda$, otherwise there are differences due to temperature-dependent terms associated with the change of entropy.

In the limit of weak coupling $f\left(\Lambda\right) \sim g\left(\Lambda\right) \sim \tfrac{1}{3}\Lambda$, and the TIPD becomes $\Delta W_j = -\tfrac{2}{9}\frac{kT_{\mathrm{i}}}{Z_{\mathrm{p}}}\Lambda_j = \tfrac{2}{3}\Delta U_j$ , which is two thirds of the static value. Equations (24), (29), (137) - (140) imply the following direct relation between the TIPD, $\Delta W(\Lambda)$, and the static continuum lowering, $\Delta U(\Lambda)$,

$$\Delta W = \tfrac{2}{3}\int_0^{\Lambda} \frac{\Delta U\left(\lambda\right)}{\lambda}\mathrm{d}\lambda \tag{142}$$

Let $f$ be any real function of $\lambda$ in $(0,\infty)$ with the property that $f(0)=0$ and suppose that, for some value of $\nu$, $\mathrm{d}\left(\lambda^{-\nu} f(\lambda)\right)/\mathrm{d}\lambda \geq 0 \ \ \forall \lambda \in (0,\infty)$. Integration of the non-negative definite function $\lambda^{\nu}\,\mathrm{d}\left(\lambda^{-\nu} f(\lambda)\right)/\mathrm{d}\lambda$ by parts from zero to $\Lambda > 0$, then implies that $f(\Lambda) \geq \nu \int_0^{\Lambda} \frac{f(\lambda)}{\lambda}\mathrm{d}\lambda$. Application of this lemma to (142) with $f \to \Delta U$ and $\nu = \frac{2}{3}$, implies that

$$|\Delta U| \geq |\Delta W| \tag{143}$$

Moreover, by application of the inequalities $(1+\lambda)^{\nu} \geq 1+\nu\lambda$, $1+\lambda^{\nu} \geq (1+\lambda)^{\nu}$, which hold generally for $\lambda \geq 0,\ 1 \geq \nu \geq 0$, and making use of (24), equation (142), yields that

$$|\Delta W| \leq \tfrac{2}{9}\Lambda \frac{kT_{\mathrm{i}}}{Z_{\mathrm{p}}} \tag{144}$$

to which $\Delta W$ is asymptotic at $\Lambda = 0$, and

$$|\Delta W| \leq \tfrac{1}{2}\Lambda^{2/3} \frac{kT_{\mathrm{i}}}{Z_{\mathrm{p}}} \tag{145}$$

to which it is asymptotic at $\Lambda = \infty$.The equalities in (144) and (145) correspond respectively to the weak and strong coupling limits, as given above. The TIPD is thus distinct from the static continuum lowering, except in the strong-coupling limit, and is generally smaller in the sense of less lowering.

Nor is $\Delta W_j$ the same as the averaged self-energy [10], which is given by $U_j/kT_{\mathrm{i}} = -\left(Z_j/2Z_{\mathrm{p}}\right) g\left(\Lambda_j\right)$, from which it differs, in the weak-coupling limit, by a factor of $4/3$.

### *5.4 The adiabatic ionization potential*

It should by now be clear that neither the thermodynamic nor static IPDs apply directly to "fast" processes, such as photoionization and collisional ionization, which are more reasonably considered to be adiabatic, constant volume processes. Accordingly, we define the adiabatic ionization potential be the energy $\Delta E$ that must be provided to the system in order to increase the ionization of one atom by one unit of charge ($\mathrm{d}z = -1$) while maintaining the volume and entropy of the system. According to (83) or (97), while recalling that $z = z_{j\alpha}$,

$$\Delta E_{j\alpha} = \phi_{j\alpha} - \left(\frac{\partial E}{\partial z_{j\alpha}}\right)_{S,V} \tag{146}$$

The ionization potential for a quasistatic (isobaric) isothermal process (equation (1)), on the other hand, with the aid of (71) and (114), is

$$\phi_{j\alpha} + \Delta U_j + \bar{\varepsilon} = \phi_{j\alpha} - \left(\frac{\partial E}{\partial z}\right)_{n_{\rm e},T} \tag{147}$$

Expressing the energy as a function of $S_{\rm i}, S_{\rm e}, V, T, z$, the two derivatives are related by the chain rule,

$$\left(\frac{\partial E}{\partial z}\right)_{n_{\rm e},T} = \left(\frac{\partial E}{\partial S_{\rm e}}\right)_{V,z}\left(\frac{\partial S_{\rm e}}{\partial z}\right)_{n_{\rm e},T} + \left(\frac{\partial E}{\partial S_{\rm i}}\right)_{V,z}\left(\frac{\partial S_{\rm i}}{\partial z}\right)_{n_{\rm e},T} + \left(\frac{\partial E}{\partial V}\right)_{S,z}\left(\frac{\partial V}{\partial z}\right)_{n_{\rm e},T} + \left(\frac{\partial E}{\partial z}\right)_{S,V} \tag{148}$$

in which, making use Maxwell's relations, while referring to (98) ,

$$\left(\frac{\partial S_{\rm e}}{\partial z}\right)_{n_{\rm e},T} = -\frac{1}{n_{\rm e}}\left(\frac{\partial S_{\rm e}}{\partial V}\right)_{z,T} + \left(\frac{\partial S_{\rm e}}{\partial z}\right)_{V,T} = -\frac{1}{n_{\rm e}}\left(\frac{\partial P_{\rm e}}{\partial T_{\rm e}}\right)_{V} + \left(\frac{\partial S_{\rm e}}{\partial z}\right)_{V,T}$$

$$\left(\frac{\partial S_{\rm i}}{\partial z}\right)_{n_{\rm e},T} = -\frac{1}{n_{\rm e}}\left(\frac{\partial S_{\rm i}}{\partial V}\right)_{z,T} + \left(\frac{\partial S_{\rm i}}{\partial z}\right)_{V,T} = -\frac{1}{n_{\rm e}}\left(\frac{\partial P_{\rm i}}{\partial T_{\rm i}}\right)_{V} + \left(\frac{\partial S_{\rm i}}{\partial z}\right)_{V,T}$$

$$\left(\frac{\partial E}{\partial V}\right)_{S,z} = -P \tag{149}$$

$$\left(\frac{\partial V}{\partial z}\right)_{n_{\rm e},T} = -\frac{1}{n_{\rm e}}$$

where $P = P_{\rm e} + P_{\rm i}$ is the total pressure. Hence, for ionization about the equilibrium state, making use of (109)

$$\left(\frac{\partial E}{\partial z}\right)_{n_{\rm e},T} = \frac{1}{n_{\rm e}}\left(P - T\left(\frac{\partial P}{\partial T}\right)_{V}\right) + \left(\frac{\partial E}{\partial z}\right)_{S,V} = -\Delta\chi + \left(\frac{\partial E}{\partial z}\right)_{S,V} \tag{150}$$

which, when combined with (147), yields

$$-\left(\frac{\partial E}{\partial z}\right)_{S,V} = \Delta U_j - \Delta\chi + \bar{\varepsilon} \tag{151}$$

according to which, the term $\Delta\chi - \Delta U_j - \bar{\varepsilon}$ in the SIPD (45) is the adiabatic ionization potential depression.

The adiabatic IPD applies to discrete ionization processes that occur locally in such a manner that the surrounding plasma is unable to respond, eg photoionization and (fast) collisional ionization. The entropy of the surrounding plasma therefore remains unchanged during the initial process. A prevailing assumption is that the system as a whole remains reasonably near to thermodynamic equilibrium, an assumption which holds reasonably well for the experiments considered in section 8. However, any extrapolation of the results that follow to systems that are strongly driven out of equilibrium, such as when the intensity is sufficient to ionize a significant proportion of the atoms at the same time, or within the same equilibration time frame, would not be justified.

## 6 Phenomenological interpretations

### *6.1 Threshold states and plasma relaxation*

A question that arises from this concerns the nature of those states, which we shall refer to *as threshold states*, that are bound in the static potential by energies less than $\sim \Delta\chi$. In what sense can these states be described as either bound or free, and how should they be treated in model-based calculations?

First of all, if we consider only *excitation* of an electron from an initial ambient state *to* a threshold state, there are no inconsistencies arising from making an *a priori* assumption as to whether the states are bound or free. It would then seem to be an open choice whether the states are treated as bound, according to the static continuum lowering, or free, according to the spectroscopic IPD. Inconsistencies do arise however when we try to consider photoionization *from* such states. Koopman's theorem, and energy conservation, imply that the transition energy for photoexcitation between bound levels must be given by the difference in the photoionization potentials. This only makes sense if the threshold states are deemed always to lie in the spectroscopic continuum, in the sense that any spectroscopic measurement will determine these states to be free continuum states. Threshold states are therefore seriously problematic only if they contain electrons in the ambient state.

The apparent dichotomy about whether or not the states are bound can be resolved by observing that the static equilibrium potential is a fiction in that it represents some equilibrium average of the potential in the vicinity of a fixed charge, and, in particular, applies only when the level corresponding to the threshold state is empty. The fact that a threshold state is apparently represented as being bound actually means only that an infinitesimal test charge would be bound in it. When the level is occupied by a discrete electron with finite charge, as in the immediate post-ionization phase, the ion is maximally screened resulting in the positive charges in the surrounding plasma being less repelled, resulting in increased continuum lowering, compared with the ultimate final equilibrium state, when the electron is "absorbed" into the surrounding plasma. The electron is thus capable of being free while the local state is occupied, while leaving behind an ostensibly bound level when it moves into the surroundings. Anyone used to doing self-consistent atomic physics calculations will be aware of this phenomenon: that the energy of a level generally varies according to its occupancy and that indeed a level can be bound or free depending on whether or not is occupied. This is similar, except that the effect is due to polarisation of the surrounding plasma ("plasma relaxation") rather than of the other bound electrons ("orbital relaxation" [21]).

Ultimately, a complete resolution of this problem has to address the fundamental limitations of the standard picture of ionization, at least for strongly-coupled many-body systems. For modelling purposes, while it is a convenient notion, to consider that the electrons in a closely-coupled many-body system fall into one of just two categories: those that are bound and thereby localised in the vicinity of individual atomic nuclei and those that are free in the sense of being entirely delocalised and virtually decoupled from the ions, is certainly naïve. That there might be electronic states that fall, even approximately, into neither (or both) categories is not only possible, but necessarily so in the pressure ionization regime when bound electrons are evidently interacting with the boundaries of the system. Moreover, the division of any closely-coupled dynamical system into subsystems according to the energy of those systems does not accord with a proper Hamiltonian description. However, if there are sufficiently few electrons occupying threshold states, then the subsystems can be considered to be approximately *dynamically separable*. This approximation is generally applicable to weakly coupled systems in thermodynamic equilibrium. It is also applicable in some dense strongly-coupled regimes when there is a large energy gap between the highest bound state and the continuum – a situation that prevails in typical metals. In the presence of occupied threshold states, the system is not dynamically separable into bound and free electron subsystems. This underlies many of the problems often encountered, including discontinuous behaviour and thermodynamic inconsistencies, in treating plasmas in the high-density pressure ionization regime. The explanation of threshold states given above fails in the pressure ionisation regime, since these states are likely to be already occupied in the ambient system. A different resolution of the dichotomy has therefore to be sought and it is likely that these ambient threshold states possess properties characteristic of both bound and free states, such as being semi-localised and contributing partially to the pressure, as would be implied by a smooth equation of state. As already noted, the pressure ionisation regime is beyond the scope of any theory, like the one given here, that attempts to treat bound and free electrons entirely separately. However, treating the ambient threshold electron states as a separate intermediate group is suggestive of a possible *ad hoc* approach to bridging the pressure ionisation discontinuity within the context of such a picture.

In non-self-confining systems, $P>0$ , the relaxation energy that accounts for the threshold states is due to deviations of the equation of state from perfect gas, which may be due to repulsive atomic cores, and inter-particle forces (Coulomb and exchange) and significantly, the presence of electrons in the threshold states themselves. The Coulomb contributions have already been considered. It is straightforward to show that a hard repulsive core does not contribute to the relaxation energy, by writing the equation of state in the Van der Waals form,

$$\left(P-\frac{U}{3V}\right)\left(V-4Nv_{\mathrm{c}}\right)=NkT \tag{152}$$

where $v_c$ is the core volume, which yields

$$\frac{PV}{NkT}=\frac{V}{V-4Nv_c}+\frac{U}{3NkT} \tag{153}$$

Referring to equation (47), it is clear that the volume-related term involving $v_c$ does not contribute to $\Delta\chi$ , provided that $v_c$ is temperature-independent. Thus the relaxation energy, which is a measure of the width of the threshold band, is essentially determined by the finite-range interatomic forces both Coulomb (which acts to lower the continuum threshold) and exchange (which acts so as to raise it).

The other issue, the difference between the static and thermodynamic IPD's is associated with the change in entropy, as is apparent from the equation $\Delta U-\Delta W=\frac{\partial}{\partial Z}\left(U-\Delta F\right)=T\frac{\partial S}{\partial Z}$. In regimes of moderate to weak coupling, increasing ionisation reduces the entropy, with the result that $\Delta W>\Delta U$ , tending to equality only in the strong coupling limit. In APPENDIX B, a clear link is established between the average static continuum lowering and the equation of state in non-pressure-ionising regimes (In a pressure-ionising regime, the situation is less clear.) while the Saha equation, which holds for weak coupling, depends only on the thermodynamic IPD. The thermodynamic and static IPDs become equal in the strong coupling limit, while the additional entropy-related thermodynamic lowering increases with decreasing plasma coupling.

The entropy connection is motivation for seeking an explanation of the IPD dichotomy in terms of the plasma microfield. This is considered in the following section.

### *6.2 Transient states and the microfield*

A property of a system of charges that might be expected to have a bearing on the IPD and threshold states is the *microfield* [22]. The microfield, expressed in terms of the electric field fluctuation $\Delta\mathbf{E}$, can give rise to *transient states*, or hopping states [23], in which electrons would be only transiently bound to, or localised within, the vicinity of a particular ion. Such states are bound, by virtue of being at negative energies in the average potential, but would be spatially delocalised. By this mechanism, the microfield might be considered to give rise to a reduction in the ionization potential, from the average, by an amount $\sim e\left|\Delta\mathbf{E}\right|R_j$ , which would then appear as an apparent contribution to the observed spectroscopic IPD. The microfield is due to spontaneous *fluctuations* in the charge states in the surrounding plasma, while the spectroscopic IPD, as argued above, depends upon the *response* of the plasma to the changed charge state of the ion. The two processes, while apparently quite separate, are in fact connected through the *fluctuation dissipation theorem* [24], [25], [26], which relates the charge-density correlation function, which characterises the plasma fluctuations, to the imaginary part of the response function, which is expressed, in the spectral representation $(\mathbf{k},\omega)$, by the dielectric function $\epsilon(\mathbf{k},\omega)$ . The variance of the electric microfield, $\left\langle\Delta\mathbf{E}^2\right\rangle$, at an arbitrary location in the plasma, equivalent to the spatially averaged mean microfield, is given by, [10]

$$\left\langle \Delta \mathbf{E}^2 \right\rangle = \frac{\hbar}{(2\pi)^4 \epsilon_0} \int_{-\infty}^{+\infty} \mathrm{d}\omega \int \mathrm{d}^3\mathbf{k}\, \mathrm{Im}\left( \frac{-1}{\epsilon(\mathbf{k},\omega)} \right) \coth\left( \frac{\hbar\omega}{2kT} \right) \tag{154}$$

Applying this formula to the 'slow' ion component of the microfield through the classical approximation $\coth(\hbar\omega/2kT) \simeq 2kT/\hbar\omega$, which requires that the ion temperature be much greater than the ion plasma frequency, and carrying out the integral over $\omega$ using the screening sum rule, then yields, for the quasistatic microfield,

$$\left\langle \Delta \mathbf{E}^2 \right\rangle \simeq \frac{kT}{\pi\epsilon_0} \int_{-\infty}^{\infty} \frac{\mathrm{d}\omega}{\omega} \int \frac{\mathrm{d}^3\mathbf{k}}{(2\pi)^3} \mathrm{Im}\left( \frac{-1}{\epsilon(\mathbf{k},\omega)} \right) = \frac{kT}{\epsilon_0} \int \frac{\mathrm{d}^3\mathbf{k}}{(2\pi)^3} \left( 1 - \frac{1}{\epsilon(\mathbf{k},0)} \right) \tag{155}$$

On the other hand, the dielectric function describes the response of the plasma to a change in the charge state of an ion within it (equivalent to the introduction of a test charge). The change in the self energy of a stationary ion, due to the removal of one electron, is, [10]

$$\Delta U = (2Z+1)\frac{e^2}{\epsilon_0} \int \frac{\mathrm{d}^3\mathbf{k}}{(2\pi)^3} \frac{1}{\mathbf{k}^2} \left( \frac{1}{\epsilon(\mathbf{k},0)} - 1 \right) \tag{156}$$

which is equivalent to the static continuum lowering. Equations (154), (155) and (156) reveal a connection between the microfield and the continuum lowering in terms of a more general underlying theory.

For the classical one-component plasma ( $D_\mathrm{e} \gg D_\mathrm{i}$ ) the static dielectric function $\epsilon(\mathbf{k},0)$ is the reciprocal of the static structure factor $S_\mathrm{ii}(\mathbf{k})$, which is deemed to satisfy $\int (1 - S_\mathrm{ii}(\mathbf{k}))\mathrm{d}^3\mathbf{k} = (2\pi)^3 n_\mathrm{i}$ by virtue of the ion-ion pair distribution, for charged particles of the same sign, vanishing at zero separation. Equation (155) then yields,

$$\left\langle \Delta \mathbf{E}^2 \right\rangle \simeq \frac{n_\mathrm{i} k T_\mathrm{i}}{\epsilon_0} \tag{157}$$

according to which, the classical electric microfield fluctuations in a weakly coupled system of charged particles are equivalent to a single classical normal mode per particle, independently of the actual charges. This gives the energy associated with the microfield as $\frac{1}{2} N_\mathrm{i} k T_\mathrm{i}$ , corresponding to the free energy,

$$F_\mathrm{mf} = \tfrac{1}{2} N_\mathrm{i} k T_\mathrm{i} \ln(\Gamma) \tag{158}$$

where $\Gamma = Z^2 e^2 / 4\pi\epsilon_0 R_\mathrm{WS} k T_\mathrm{i} \equiv \Lambda^2/3$ is the ion coupling parameter. The microfield thus makes a contribution to the equation of state that is different in character and therefore supplementary to the normal quasi-static Coulomb part, such as described in APPENDIX B (cf equation (141). (Moreover, since the resulting pressure satisfies $P_\mathrm{mf} V / N_\mathrm{i}\, k T_\mathrm{i} = \frac{1}{6}$ , the microfield (157) makes no contribution to the relaxation energy (47).)

To understand the strongly coupled limit, we consider a solid state plasma where pointlike ions of charge $Z$ are confined close to specific locations $\{r_i\}$ about which they collectively undergo small harmonic oscillations whereby the displacement of the $i^{\text{th}}$ ion is $\Delta\mathbf{r}_i(t) = \sum_{\mathbf{k}} \mathbf{x}_{\mathbf{k}} \cos(\mathbf{k}\cdot\mathbf{r}_i - \Omega_{\mathbf{k}} t)$ . The resulting microfield is

$$\Delta\mathbf{E}(t) = \frac{Ze}{4\pi\epsilon_0} \sum_i \frac{2}{r_i^3} \sum_{\mathbf{k}} \mathbf{x}_{\mathbf{k}} \cos(\mathbf{k}\cdot\mathbf{r}_i - \Omega_{\mathbf{k}} t) \tag{159}$$

from which an estimate of the mean square microfield at an ion site is

$$\langle \Delta\mathbf{E}^2 \rangle = \left( \frac{Ze}{4\pi\epsilon_0} \right)^2 \frac{4{\nu_{\text{nn}}}^2}{(2R_{\text{WS}})^6} \sum_{\mathbf{k}} \langle {\mathbf{x}_{\mathbf{k}}}^2 \rangle \tag{160}$$

where $\nu_{\text{nn}}$ is the effective number of nearest neighbours, $\langle {\mathbf{x}_{\mathbf{k}}}^2 \rangle$ is the mean square displacement in the mode $\mathbf{k}$ . For a system of classical oscillators of total mass $M$ , in equilibrium at temperature $T_{\text{i}}$ , $\langle {\mathbf{x}_{\mathbf{k}}}^2 \rangle = kT_{\text{i}} / M {\Omega_{\mathbf{k}}}^2$ ; while, for acoustic modes, $\sum_{\mathbf{k}} \frac{1}{{\Omega_{\mathbf{k}}}^2} = \frac{3N_{\text{i}}}{{\Omega_0}^2}$ where $\Omega_0$ is the upper limiting frequency, which we identify with the ion plasma frequency. Hence, combining these formulae with (160)

$$\langle \Delta\mathbf{E}^2 \rangle = \left( \frac{\nu_{\text{nn}}}{12} \right)^2 \frac{3n_{\text{i}}kT_{\text{i}}}{\epsilon_0} \tag{161}$$

which, apart from $O(1)$ numerical factors, is the same as (157). Indeed, $\nu_{\text{nn}} = 12$ is a reasonable choice. meaning that the microfield is now associated with 3 degrees of freedom per particle, and moreover, these should be the same three degrees of freedom as are associated with the potential energy of the oscillators. The fact that two such similar equations as (157) and (161) arise in such different limits suggests that

$$\langle \Delta\mathbf{E}^2 \rangle = \lambda \frac{n_{\text{i}}kT_{\text{i}}}{\epsilon_0} \tag{162}$$

with $\lambda = 1$ in the case of a classical Coulomb fluid and $\lambda = 3$ for a classical solid-state Coulomb plasma, describes the general case. In the solid, the microfield is just the effective oscillator field, and the equation-of-state is adequately described by that for a system comprising a collection of $N_{\text{i}}$ classical oscillators (as per the classical phonon model) with no additional field-related terms. The transition from $\lambda = 1$ to $\lambda = 3$ then corresponds to a discrete phase transition, thus avoiding any need for $\lambda$ , along with the implied scalings, to take on intermediate values.

Expressing the result (162) in terms of the normal field $\mathrm{E}_0 = Ze/4\pi\epsilon_0 R_{\mathrm{WS}}{}^2$ gives

$$\left\langle \Delta\mathbf{E}^2 \right\rangle = 3\lambda \mathrm{E}_0{}^2/\Gamma = 9\lambda \mathrm{E}_0{}^2/\Lambda^2 \tag{163}$$

Note that this gives the spatially-averaged mean microfield. The (time-averaged) mean microfield at the centre of a particular ion is generally what is considered appropriate in line-broadening theory, and is modified from the spatial-average by the correlations with neighbouring ions. This is not necessarily what is relevant to the continuum lowering.

Let the notional microfield "contribution" $\Delta\varepsilon_{\mathrm{mf}}$ to the continuum lowering be the extra energy that an electron can gain from the microfield in moving from an initial location $\mathbf{r}_0$ within the bound-state orbital to the surface of the ion-sphere, where it is deemed to be ionized, ie

$$\Delta\varepsilon_{\mathrm{mf}} = e\int_{\mathbf{r}_0}^{\mathbf{R}_0} \Delta\mathbf{E}\cdot \mathrm{d}\,\mathbf{r} \tag{164}$$

Schwarz's inequality for a randomly directed microfield , in conjunction with (157), then yields

$$\left(\Delta\varepsilon_{\mathrm{mf}}\right)^2 \lesssim 2\frac{\phi k T_{\mathrm{i}}}{Z+1} \tag{165}$$

for $r_0 \ll R_0$ , where $\phi = (Z+1)e^2/2r_0$ , which is a measure of the initial binding energy of the electron. The inequality (165) is perhaps over-strict. In a weakly-coupled system, the mean microfield is relatively weakly correlated with any particular ion, in which case

$$\Delta\varepsilon_{\mathrm{mf}} \simeq kT_{\mathrm{i}}\sqrt{3\Gamma}/Z \tag{166}$$

is a better estimate. A measure of the importance of the microfield in this regime is therefore

$$\frac{\Delta\varepsilon_{\mathrm{mf}}}{\Delta U_{\mathrm{i}}} = \frac{4\pi\epsilon_0 D_{\mathrm{i}}\Delta\varepsilon_{\mathrm{mf}}}{Ze^2} = \frac{1}{\Gamma\sqrt{3}} = \frac{\sqrt{3}}{\Lambda^2} \tag{167}$$

which demonstrates that the microfield is the dominant influence in weakly coupled plasmas. A similar measure of the relative importance of the microfield in the strongly coupled regime, referring to (163), is

$$\frac{2\sqrt{\left\langle \Delta\mathbf{E}^2 \right\rangle}}{3\mathrm{E}_0} = \sqrt{\frac{4\lambda}{3\Gamma}} \tag{168}$$

which shows that the microfield can be expected to cease to dominate the continuum lowering for $\Gamma \gtrsim \frac{4}{3}$. However we should bear in mind that the microfield does not ionize but rather perturbs with the possibility of creating transient states, which spectroscopically and thermodynamically have more in common with bound states.

Unlike threshold states, transient states are, by definition, at negative energies and so are not considered to be within the continuum. While they do represent a possible mechanism whereby an electron can be removed from an ion, the process resembles collisional charge exchange in which increased ionization of one ion is accompanied by an equal reduction in that of another with little or no change in the plasma potential energy. Moreover, since there is no change in the overall particle number, if the electron remains bound, there is no direct contribution to the pressure. This strongly suggests that this process is therefore better regarded as being separate from normal photoionization, while transient states and threshold states are evidently not the same things. Transient states are one-body bound states that are delocalised by interaction with neighbouring ion(s) while threshold states are an emergent property of the many-body (electron + plasma) system. In moderately-coupled or weakly-coupled plasmas, transitions into transient states will merge with the continuum via the Inglis-Teller effect and can be properly described in those terms without invoking an additional continuum-lowering effect. In particular, transient states are spectroscopically bound while threshold states are spectroscopically in the continuum.

## 7 Equation of State

The treatment of continuum lowering in multicomponent Coulomb system comprising ions and electrons in both free and bound states is intrinsically linked to a non-trivial equation of state model, which is developed in APPENDIX B. It is found that the (approximate) applicability of this model to real plasmas is apparently limited only by the inability of the model to treat pressure ionization ( $\theta_V \lesssim 0$ ) which is attributed to the approximation whereby the ions are treated as structureless point charges, and the lack of dynamical separability between electrons bound in threshold states and those in the true continuum. Nevertheless it represents an important enhancement over models that treat the component charges as being inert, one which can be considered to be approximately applicable in all regimes where pressure ionization is not an issue. The deficiency in the pressure ionization regime appears to reside in the equation of state rather than in the model of the IPD. Nevertheless this raises doubts about the general applicability of the Coulomb model and this is something that needs more careful examination, theoretically, using a model that treats pressure ionization, or by direct experimental observation.

## 8 Comparisons with experiment

The experiments that we have modelled fall into two categories: direct measurements using a tuneable FEL [2], [3]; and measurements of 1-3 lines in shocked warm dense aluminium plasmas created using a high power laser [4]. The former provide direct measurements of the ionization thresholds, and hence IPDs, that are virtually model independent but only explore ion configurations, at a fixed density, at the limit of strong coupling ($T_{\mathrm{i}} \simeq 0$). The results of these experiments are found to be remarkably consistent with the Ecker-Kröll formula,

$$\hbar\Delta\omega_j = -\left(Z_j + 1\right)\left(1 + \bar{Z}\right)^{1/3} u_{\mathrm{WS}}^{0} \tag{169}$$

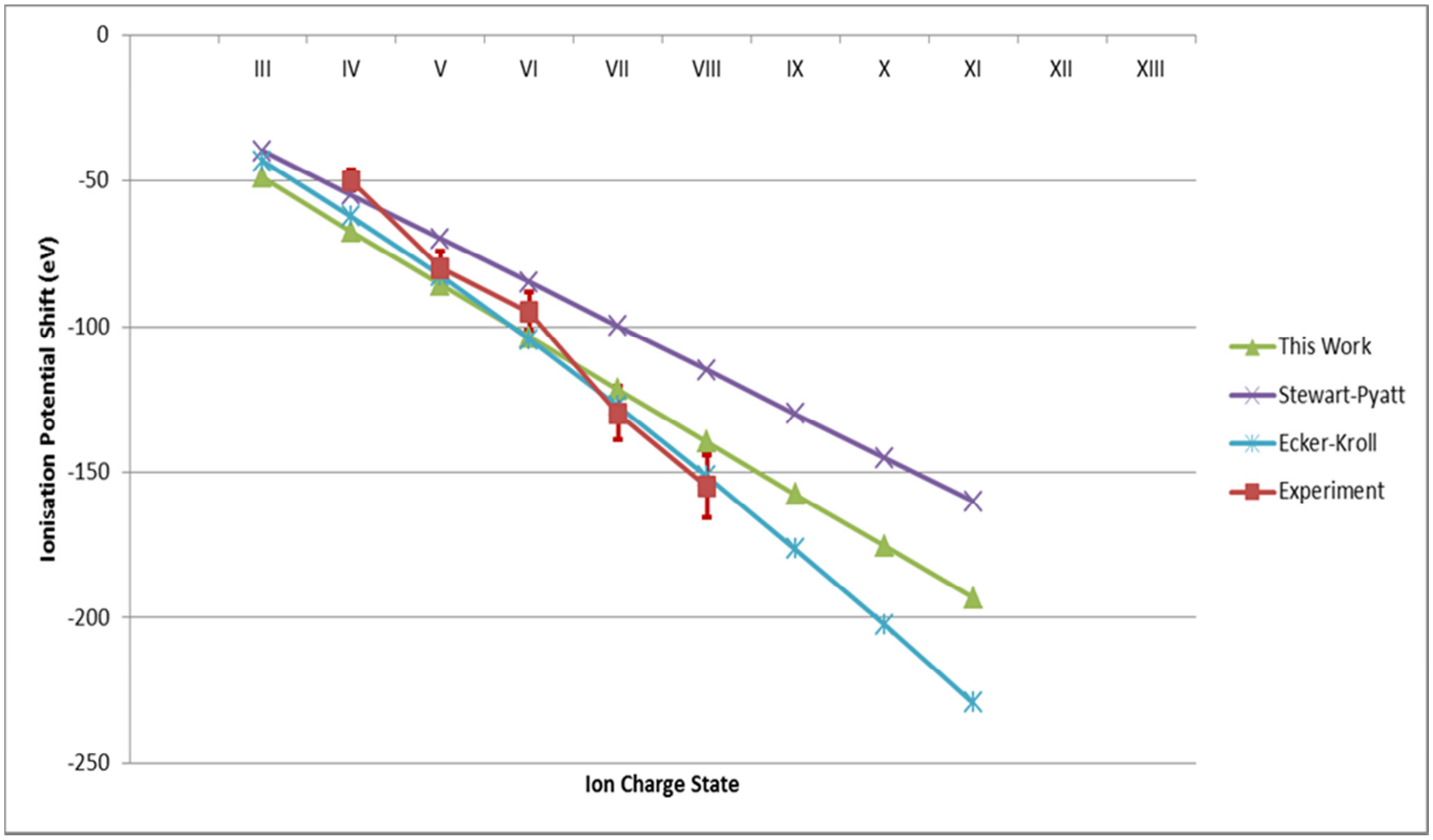

**Figure 2** Calculations of the ionization potential depression for various ion charge states in solid density aluminium compared with the measurements of Ciricosta et al [2]. Stewart-Pyatt is equation (12), Ecker Kröll is (169) and "This Work" is equation (81) .

The experimental results along with the results of various calculations are shown in **Figure 2**, which clearly shows the inadequacy of the Stewart-Pyatt equilibrium model. Most importantly, the experiment is also found to be reasonably well explained by the theory described in this work. In these experiments, the ion plasma coupling parameter is estimated to be in the range 3,000 to 50,000 putting these plasmas clearly in the solid-state regime where the microfield can have no effect on the ionisation potential depression.

An alternative technique involves measuring the strengths of the 1-3 lines as a function of temperature and density and to determine whether the IPD encompasses the 3p states, for example. This is able to explore higher ion temperatures as well as a range of densities, but depends upon some modelling, to determine where the $n = 3$ levels are expected to lie in high charge states as well as to infer the plasma conditions.

Recent experimental results for aluminium from the Orion high-power laser [3], along with predictions of the various models for the putative plasma conditions are given in **Table 2**.

| Plasma state | | Experiment | Model predictions for presence of 1-3 lines. | | | |
|---|---|---|---|---|---|---|
| Density (g/cc) | Temperature (eV) | 1-3 lines observed? | This work Eq (81) | Ion sphere Eq (170) | Stewart-Pyatt Eq (14) | Ecker- Kröll Eq (169) |
| 1.2 | 550 | Yes | Yes | Yes | Yes | Yes |
| 2.5 | 650 | Yes | Yes | Yes | Yes | No |
| 4.0 | 700 | Yes | Yes | Yes | Yes | No |
| 5.5 | 550 | Yes | Yes (Ly$\beta$) | Yes | No | No |
| 9.0 | 700 | No | No | Marginal | Yes | No |
| 11.6 | 700 | NA | No | No | No | No |

**Table 2** Results from the experimental observations of shocked aluminium plasmas at Orion compared with various models for the putative plasma conditions given in [4].

The temperature-density grid is quite coarse and the experiment is thus only able to bracket the IPD along the track of the measurements. The uncertainty in the density at the critical densities, 5.5g/cc and 9g/cc is quoted as being around 10%, which translates to a 3% error in $u_{\mathrm{WS}}^{0}$. Nevertheless the measurements are able to discriminate between various models to the extent that it can be said that the results, shown in **Figure 3**, are consistent with the model derived in this work, equation (81), as well as, as claimed by the authors of the experiment, to the modified ion-sphere (IS) formula,

$$\hbar\Delta\omega_j = -\tfrac{3}{2}\left(Z_j+1\right)\left(\frac{\bar{Z}}{Z_j}\right)^{1/3} u_{\mathrm{WS}}^{0} \tag{170}$$

(cf equation (72) with $C=\frac{9}{10}$) albeit applied in a regime of moderate coupling. On the other hand, they are inconsistent with both Stewart-Pyatt (14) and Ecker-Kröll (169), with the former under-estimating the IPD and the latter considerably over-estimating it.

In this experiment, the plasma coupling strength $\Gamma$ is in the range 2 to 3, indicating a moderately-coupled fluid plasma and that the n=3 bound states lying close to the continuum can be expected to be perturbed by the microfield. The observations are however not consistent with an additional microfield continuum lowering of the magnitude predicted by (168) confirming that affected states, whether transient or not, manifest themselves spectroscopically as bound states.

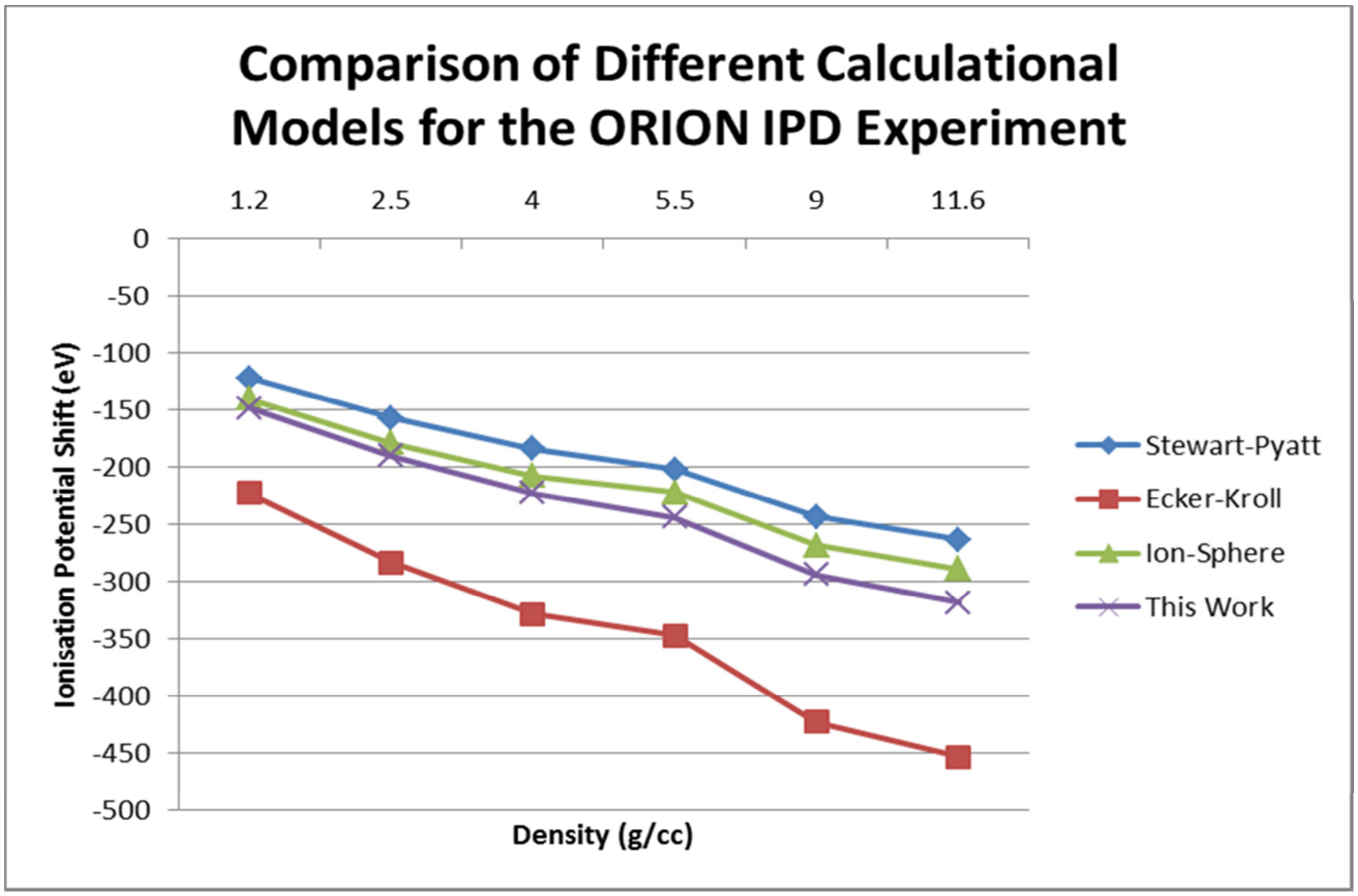

**Figure 3**: comparison of different average-ion IPD calculations for the ORION IPD experiment of [4]. The legend is the same as Figure 2 with the addition of Ion-Sphere, which is as per equation (170), which represents the experiment's authors' considered match to the results. Unlike the FEL experiment, there are no quantitative data. The experiment observes n=3 lines up to and including 5.5 g/cc but not at the higher densities. Stewart Pyatt predicts the presence of n=3 lines up to 9 g/cc while Ecker-Kröll predicts their absence at 2.5 g/cc and above.

The critical measurements are those for 5.5g/cc, from which n=1-3 emission lines are observed, and 9g/cc, which is characterised by an absence of n=1-3 lines. The plasma is determined to be predominantly mixtures of He-like, H-like and fully stripped ions under these conditions. Calculations, employing a simple screened hydrogenic model, without $\ell$ -splitting, differ in that the IS model predicts that n=3 levels should be spectroscopically bound in He-like Al at 5.5 g/cc, whereas the relaxation model proposed here does not. However n=3 bound levels are found to be present in H-like Al at 5.5g/cc using both models. At 9g/cc, the relaxation model predicts that the n=3 levels should be unbound in all ion states, and that no n=1-3 lines should be seen. Taking a realistic view, the experiment is probably unable to discriminate between the predictions for the SIPD given by equations (81) and (170) in this regime, on account of the uncertainties in the plasma conditions and those inherent in the atomic calculations upon which the interpretation may depend. Nevertheless, the new model proposed here does produce a slightly better fit by unequivocally removing the n=1-3 lines from the 9g/cc case. On this basis we conclude that the new IPD model presented in this paper is fully consistent with the observations of this experiment, unlike any of the proposed alternatives. Stewart-Pyatt, as per (14), for example, predicts n=1-3 lines at 9g/cc, while the IS model is marginal under these conditions.

The fact that the new model is able to give a reasonably good account of *both* experiments is compelling. The alternative models, Ecker-Kröll and Ion-Sphere, fit the data only in the regimes of the FEL and laser-shock experiments respectively, and neither fits both experiments. Without an underlying explanation, these alternative formulae should be considered as being no more than fits to the data.

## 9 Conclusions

On the basis of a theoretical re-examination of the IPD problem, motivated and supported by observational evidence from recent experiments on very dense plasmas, we conclude that the Stewart-Pyatt (SP) model, provides only an incomplete description of the ionization potential depression (IPD) as one would define it in terms of a spectroscopic measurement, or even in some equation of state contexts. The SP model and its close derivatives provide only the static continuum lowering (SCL, here denoted by $\Delta U$ ), which represents the average effect of the electrostatic field of the surrounding plasma on the electronic states. Closely related to the static continuum lowering, but distinct from it, is the thermodynamic ionization potential depression (TIPD, here denoted by $\Delta W$ ), which is the change in the thermodynamic free energy associated with ionization. This accounts additionally for the entropy-related terms in the free energy, which arise when the average potential energy becomes temperature dependent. The TIPD is shown to be that which appears in the Saha equation thus demonstrating direct relevance of the TIPD to equation of state modelling. In the limit of strong coupling (high densities and/or low temperatures) when the electrostatic energy becomes independent of temperature (eg the ion sphere approximation) the SCL and the TIPD become synonymous. In general the TIPD is less, in the sense of less depression, than the SCL.

Spectroscopic and other dynamical processes may occur on timescales too fast for the surrounding plasma to respond or come into equilibrium, when they cannot be considered to be transitions between equilibrium states of the plasma. In the case of near-threshold ionization, the electron is deposited close to the parent ion and has to move away before the surrounding plasma has anything to respond to. Neither the TIPD nor the SCL is then a good measure of the ionization potential. The spectroscopic ionization potential depression (SIPD, denoted by $\hbar\Delta\omega$) applies to such adiabatic processes in which energy is exchanged locally by the atomic system interacting only with the photon. The absence of any energy exchange with the plasma surroundings is accounted for by subtracting out an additional relaxation effect. In general, $\hbar|\Delta\omega| > |\Delta U| \geq |\Delta W|$.

The SIPD model proposed in this paper, as represented most generally by equation (79) above, and by the approximate and limiting formulae, (81), (78) and (82), accounts reasonably well for the published observational data, over a wider range of conditions than any of the other simple models on offer. However, there is some doubt over the validity of the model in regimes of pressure ionization due to the underlying equation of state model lacking validity.

A closer look at the plasma equation of state reveals not only close links with the TIPD in the context of the Saha equation, but also a more formal link between the static continuum lowering $\Delta U$ and the correlations between the internal state of the ion and the potential energy due to the surrounding plasma (as expressed by equations (214) and (215)). The fundamental quantity linking the various quasi-static IPDs (SCL and TIPD) is the shift in the Helmholtz free energy, $\Delta F$ , use of which helps maintain thermodynamic consistency . Accordingly, it is possible to model the equation of state of a Coulomb system and derive equations describing the underlying charge-state distribution that are posed as being applicable in all regimes apart from those where pressure ionization is occurring.

## 10 Acknowledgements

The author acknowledges discussions with Justin Wark, Orlando Ciricosta and Dave Hoarty and is grateful to Dave Hoarty and Justin Wark for making their data available. The author is indebted to Stephanie Hansen for a critical reading of the manuscript and for drawing attention to some early mistakes.

**APPENDIX A Equation of state of non-interacting fermion gas**

The following summarises the equation- of state of a gas comprising $N$ identical point-like fermions of mass $m$ in thermodynamic equilibrium at temperature $T$ and chemical potential $\mu = \eta T$ . These equations are given in units such that $\hbar = 1$ , $k = 1$ .

| **Quantity** | **Formula** | **Limiting values** | |
|---|---|---|---|
| | | Degenerate $(\eta \gg 1)$ | Maxwellian $(\eta \ll 0)$ |
| Thermal wavelength | $\Lambda \equiv \left(\frac{2\pi}{mT}\right)^{1/2}$ | | |
| Fermi integral | $I_j(\eta) \equiv \int_0^\infty \frac{y^j}{1+\exp(y-\eta)} \mathrm{d}y$ | $\frac{\eta^{j+1}}{j+1}$ | $\Gamma(j+1)\mathrm{e}^\eta$ |
| Density | $n \equiv N/V = \frac{4}{\Lambda^3\sqrt{\pi}} I_{1/2}(\eta)$ | $\frac{1}{3\pi^2}(2m\mu)^{3/2}$ | $\frac{2}{\Lambda^3}\mathrm{e}^\eta$ |
| Fermi temperature | $T_\mathrm{F} = T\left(\tfrac{3}{2} I_{1/2}(\eta)\right)^{2/3} = \frac{1}{2m}\left(3\pi^2 n\right)^{2/3}$ | $\mu$ | $T\left(\frac{3\sqrt{\pi}}{4}\mathrm{e}^\eta\right)^{2/3}$ |
| Specific energy | $\varepsilon \equiv \frac{U}{N} = \frac{I_{3/2}(\eta)}{I_{1/2}(\eta)} T$ | $\tfrac{3}{5} T_\mathrm{F}\left(1+\frac{5\pi^2}{12}\left(\frac{T}{T_\mathrm{F}}\right)^2\right)$ | $\tfrac{3}{2}T$ |
| Pressure | $P = \tfrac{2}{3} n\varepsilon = \tfrac{2}{3} nT \frac{I_{3/2}(\eta)}{I_{1/2}(\eta)}$ | $\tfrac{2}{5} nT_\mathrm{F}$ | $nT$ |
| Isothermal bulk modulus | $\kappa_\mathrm{T} \equiv n\left(\frac{\partial P}{\partial n}\right)_T = nT_\mathrm{B} \equiv nT \frac{I_{1/2}(\eta)}{I'_{1/2}(\eta)}$ | $\tfrac{2}{3} nT_\mathrm{F}\left(1+\frac{\pi^2}{12}\left(\frac{T}{T_\mathrm{F}}\right)^2\right)$ | $nT$ |
| Specific heat at constant volume | $c_V = \frac{1}{T}\left(\tfrac{5}{2}\varepsilon - \tfrac{9}{4}T_\mathrm{B}\right) = \frac{9}{4}\frac{T_\mathrm{B}}{T}(\gamma-1)$ | $\frac{\pi^2}{2}\frac{T}{T_\mathrm{F}}$ | $\tfrac{3}{2}$ |

| Quantity | Formula | Limiting values | |
|---|---|---|---|
| Volume expansivity | $\beta \equiv \frac{1}{V}\left(\frac{\partial V}{\partial T}\right)_P = \frac{3}{2T}\left(\frac{5}{3}\frac{P}{nT_\mathrm{B}} - 1\right) = \frac{3}{2T}(\gamma - 1)$ | $\frac{\pi^2}{2}\frac{T}{T_\mathrm{F}^2}$ | $\frac{1}{T}$ |
| Ratio of specific heats | $\gamma \equiv \frac{c_P}{c_V} = \frac{10}{9}\frac{I_{3/2}(\eta) I'_{1/2}(\eta)}{\left(I_{1/2}(\eta)\right)^2} = \frac{5}{3}\frac{P}{\kappa_T} = \frac{10}{9}\frac{\varepsilon}{T_\mathrm{B}}$ | $1 + \frac{\pi^2}{3}\left(\frac{T}{T_\mathrm{F}}\right)^2$ | $\frac{5}{3}$ |
| Specific heat at constant pressure | $c_P = \gamma c_V = \frac{5}{2}\frac{\varepsilon}{T}(\gamma - 1)$ | $\frac{\pi^2}{2}\frac{T}{T_\mathrm{F}}$ | $\frac{5}{2}$ |
| Adiabatic bulk modulus | $\kappa_S = \gamma \kappa_T = \frac{5}{3}P$ | $\frac{2}{3}nT_\mathrm{F}$ | $\frac{5}{3}nT$ |
| Entropy | $S = N\left(\frac{5}{2}\frac{P}{nT} - \eta\right)$ | $0$ | $\left(\frac{5}{2} - \eta\right)N$ |
| Helmholtz free energy | $F = N\left(\eta T - \frac{2}{3}\varepsilon\right)$ | $\frac{3}{5}NT_\mathrm{F}$ | $(\eta - 1)NT$ |
| Gibbs free energy | $G = F + PV = \eta NT = \mu N$ | $NT_\mathrm{F}$ | $\eta NT$ |
| Enthalpy | $H = G + TS = \frac{5}{2}N\frac{P}{n} = \frac{5}{3}N\varepsilon$ | $NT_\mathrm{F}\left(1 + \frac{5\pi^2}{12}\left(\frac{T}{T_\mathrm{F}}\right)^2\right)$ | $\frac{5}{2}NT$ |
| | $T\left(\frac{\partial \eta}{\partial T}\right)_n = -\frac{3}{2}\frac{I_{1/2}(\eta)}{I'_{1/2}(\eta)} = -\frac{3}{2}\frac{T_\mathrm{B}}{T}$ | $-\frac{T_\mathrm{F}}{T}$ | $-\frac{3}{2}$ |
| | $n\left(\frac{\partial \eta}{\partial n}\right)_T = \frac{I_{1/2}(\eta)}{I'_{1/2}(\eta)} = \frac{T_\mathrm{B}}{T}$ | $\frac{2}{3}\frac{T_\mathrm{F}}{T}$ | $1$ |
| | $\left(\frac{\partial P}{\partial T}\right)_V = \left(\frac{\partial S}{\partial V}\right)_T = \frac{3}{2}n\frac{T_\mathrm{B}}{T}(\gamma - 1)$ | $\frac{\pi^2}{3}\frac{nT}{T_\mathrm{F}}$ | $n$ |

| Quantity | Formula | Limiting values | |
|---|---|---|---|
| | $\dfrac{\mathrm{d}\gamma}{\mathrm{d}\eta}=\dfrac{T}{T_{\mathrm{B}}}\left(\dfrac{5}{3}-2\gamma\right)+\gamma\dfrac{I''_{1/2}(\eta)}{I'_{1/2}(\eta)}=\dfrac{T}{T_{\mathrm{B}}}\left(\dfrac{5}{3}-\gamma\lambda_T\right)$ | 0 | 0 |
| Adiabatic bulk modulus pressure coefficient | $\lambda_S\equiv\left(\dfrac{\partial\kappa_S}{\partial P}\right)_S=\left(\dfrac{\partial\kappa_S}{\partial P}\right)_T=\tfrac{5}{3}$ | $\tfrac{5}{3}$ | $\tfrac{5}{3}$ |
| Isothermal bulk modulus pressure coefficient | $\lambda_T\equiv\left(\dfrac{\partial\kappa_T}{\partial P}\right)_T=\dfrac{1}{\gamma}\left(\dfrac{5}{3}-\dfrac{T_{\mathrm{B}}}{T}\dfrac{\mathrm{d}\gamma}{\mathrm{d}\eta}\right)$ $=2-\dfrac{T_{\mathrm{B}}}{T}\dfrac{I''_{1/2}(\eta)}{I'_{1/2}(\eta)}$ | $\tfrac{5}{3}$ | 1 |
| Isothermal bulk modulus logarithmic temperature derivative | $\dfrac{T}{\kappa_T}\left(\dfrac{\partial\kappa_T}{\partial T}\right)_V=\tfrac{1}{2}(5-3\lambda_T)$ | 0 | 1 |
| Degeneracy parameter | $\eta=\mu/T$ | $\dfrac{T_{\mathrm{F}}}{T}\left(1-\dfrac{\pi^2}{12}\left(\dfrac{T}{T_{\mathrm{F}}}\right)^2\right)$ | $\tfrac{1}{2}\ln\left(\dfrac{16}{9\pi}\left(\dfrac{T_{\mathrm{F}}}{T}\right)^3\right)+\tfrac{1}{3}\left(\dfrac{2}{\pi}\right)^{1/2}\left(\dfrac{T_{\mathrm{F}}}{T}\right)^{3/2}$ |

**Table 3** Equation of state of a non-interacting fermion gas.

**APPENDIX B Equation of state of a Coulomb system**

## B.1 Basic formulae

Some insight may gained by looking more closely at the implications (of the equations of section 5 ) for the equation-of-state. The thermodynamic properties of a Coulomb system are demonstrated as being describable in terms of the ion parameter $\tilde{\Lambda}_J$, which is a function of the charge state $Z_J$ and the macroscopic thermodynamic coordinates of the surrounding plasma, and which is deemed, as per (25) and (26), to satisfy the Coulomb scaling laws determined by the logarithmic derivatives with respect to volume, temperature and charge according to

$$\frac{\partial \ln \tilde{\Lambda}_J}{\partial \ln V} = -\tfrac{1}{2}$$

$$\frac{\partial \ln \tilde{\Lambda}_J}{\partial \ln T_\mathrm{i}} = -\tfrac{3}{2} \qquad (171)$$

$$\frac{\partial \ln \tilde{\Lambda}_J}{\partial \ln Z_J} = 1$$

and a set of functions $f(\Lambda), g(\Lambda), h(\Lambda)$ that are related through the hierarchy

$$g(\Lambda) = \frac{1}{\Lambda}\int_0^\Lambda h(\lambda)\,\mathrm{d}\lambda$$

$$f(\Lambda) = \int_0^\Lambda g(\lambda)\frac{\mathrm{d}\lambda}{\lambda} \qquad (172)$$

an example of which is the function set based on $h(\lambda)$ given by (24). The equation of state is then given as follows.

*(a) Helmholtz free energy*

$$F = F^0 - \Phi + N_\mathrm{i}\overline{\Delta F}$$

$$\Phi = \sum_J \left(\bar{N}_J - N_J\right) E_J$$

$$\overline{\Delta F} = \frac{1}{N_\mathrm{i}} \sum_J N_J \Delta F_J \tag{173}$$

$$\Delta F_J = -\tfrac{1}{3} kT_\mathrm{i} \frac{Z_J}{Z_\mathrm{p}} f\left(\tilde{\Lambda}_J\right)$$

where $F^0$ is the free energy of the non-interacting system, $E_J = E_\mathrm{c}\left(\mathbf{z}_J\right)$ is the internal energy of the ion bound configuration $J$, and $\bar{N}_J \equiv p_J N_\mathrm{i}$ is the ensemble-average number of electrons in the configuration.

*(b) Internal energy*

$$E_\mathrm{int} = N_\mathrm{i}\left(\tfrac{3}{2} kT_\mathrm{i} + \bar{U}\right) + N_\mathrm{e}\bar{\varepsilon} + E_\mathrm{b}$$

$$\bar{U} = \frac{1}{N_\mathrm{i}} \sum_J N_J U_J$$

$$U_J = -\tfrac{1}{2} kT_\mathrm{i} \frac{Z_J}{Z_\mathrm{p}} g\left(\tilde{\Lambda}_J\right) \tag{174}$$

$$E_\mathrm{b} = \sum_J N_J E_J$$

$\bar{U}$ represents the potential energy per ion and $\bar{\varepsilon} = \tfrac{3}{2} kT_\mathrm{e}\, \zeta\left(n_\mathrm{e}, T_\mathrm{e}\right)$ is the average kinetic energy of an electron ( $\zeta\left(n_\mathrm{e}, T_\mathrm{e}\right) = \tfrac{2}{3}\left(I_{3/2}\left(\mu_\mathrm{e}^0 / kT_\mathrm{e}\right) \big/ I_{1/2}\left(\mu_\mathrm{e}^0 / kT_\mathrm{e}\right)\right)$ , see APPENDIX A).

In addition, we can construct thermodynamic averages of the IPDs, the natural weighting of which is according to the charges of the ions, as follows:

*(c) Charge-weighted average static continuum lowering*

$$\overline{\Delta U} = \frac{1}{N_{\mathrm{i}}\bar{Z}} \sum_J N_J Z_J \Delta U_J$$

$$\Delta U_J = -\tfrac{1}{2} kT_{\mathrm{i}} \frac{1}{Z_{\mathrm{p}}} h\left(\tilde{\Lambda}_J\right) \tag{175}$$

*(d) Charge-weighted average thermodynamic IPD*

$$\overline{\Delta W} = \frac{1}{N_{\mathrm{i}}\bar{Z}} \sum_J N_J Z_J \Delta W_J$$

$$\Delta W_J = -\tfrac{1}{3} kT_{\mathrm{i}} \frac{1}{Z_{\mathrm{p}}} \left( f\left(\tilde{\Lambda}_J\right) + g\left(\tilde{\Lambda}_J\right)\right) \tag{176}$$

where

$$N_{\mathrm{i}} = \sum_J N_J$$

$$\bar{Z} = \frac{1}{N_{\mathrm{i}}} \sum_J N_J Z_J \tag{177}$$

$$Z_{\mathrm{p}} = \frac{1}{N_{\mathrm{i}}\bar{Z}} \sum_J N_J Z_J^{\,2}$$

and where $\tilde{\Lambda}_J$ is provided by

$$\tilde{\Lambda}_J = \xi_J \left( \frac{R_J}{D_{\mathrm{i}}} \right)^3 \tag{178}$$

with $\xi_J$ defined by (42) in terms of the ratio $\alpha = D_{\mathrm{i}}/D$ of the screening lengths, which are taken to be given by the standard formulae

$$\frac{1}{D^2}=\frac{1}{D_\mathrm{i}^2}+\frac{1}{D_\mathrm{e}^{\ 2}};\qquad \frac{1}{D_\mathrm{i}^2}=\sum_K\frac{1}{D_K^{\ 2}}=\frac{N_\mathrm{i}\bar{Z}Z_\mathrm{p}e^2}{V\epsilon_0 kT_\mathrm{i}}$$

$$\frac{1}{D_K^{\ 2}}=\frac{N_K Z_K^{\ 2}e^2}{V\epsilon_0 kT_\mathrm{i}};\qquad \frac{1}{D_\mathrm{e}^2}=\frac{N_\mathrm{e}e^2}{V\epsilon_0 kT_\mathrm{B}} \tag{179}$$

and

$$\tfrac{4}{3}\pi R_J^{\ 3}\sum_K N_K Z_K = Z_J V \tag{180}$$

which defines the volume occupied by each ion as being proportional to its charge independently of the electron density. The degree of approximation to which the proposed function (178) satisfies the relations (171) is asserted to be sufficient.

*(e) Partial Coulomb pressure and entropy*

These equations yield the partial Coulomb pressure and entropy contributions through the potentials

$$\Pi_J \equiv V\Delta P_J = -V\,\partial\Delta F_J/\partial V = \tfrac{1}{3}U_J$$

$$Q_J \equiv T_\mathrm{i}\Delta S_J = -T_\mathrm{i}\,\partial\Delta F_J/\partial T_\mathrm{i} = U_J - \Delta F_J \tag{181}$$

and the following definitive relationships between the thermodynamic properties and the continuum lowering and TIPD:

$$\Delta U_J = 3\frac{\partial \Pi_J}{\partial Z_J} \tag{182}$$

$$\Delta W_J = \frac{\partial \Delta F_J}{\partial Z_J} \tag{183}$$

*(f) Chemical potentials and the Gibbs free energy*

The chemical potentials are given by

$$\mu_K = \mu_K^0 + \Delta\mu_K,\qquad \mu_\mathrm{e} = \mu_\mathrm{e}^0 + \Delta\mu_\mathrm{e}$$

$$\Delta\mu_K = \frac{\partial \Delta F_\mathrm{i}}{\partial N_K} + E_K,\qquad \Delta\mu_\mathrm{e} = \frac{\partial \Delta F_\mathrm{i}}{\partial N_\mathrm{e}} \tag{184}$$

where $\Delta F_{\mathrm{i}} = N_{\mathrm{i}}\overline{\Delta F}$ (using that $\partial\Phi/\partial N_K = -E_K$, $\partial\Phi/\partial N_{\mathrm{e}} = 0$). The Gibbs free energy is then

$$G = \sum_K N_K \mu_K + N_{\mathrm{e}}\mu_{\mathrm{e}} = F^0 + P^0 V + \Delta G$$
$$\Delta G = \sum_K N_K \Delta\mu_K + N_{\mathrm{e}}\Delta\mu_{\mathrm{e}} \tag{185}$$

*(g) Average-atom*

It is often appropriate to represent some of these quantities in terms of their average-atom values, where the function arguments are replaced by their average values, thereby effectively treating the system as if all the atoms were in the average state. Such quantities shall be represented by the notation $J = *$ elevated to superscript, eg

$$Z^* = \sum_J p_J Z_J$$
$$\tilde{\Lambda}^* = \tilde{\Lambda}\left(Z^*, V, T_{\mathrm{i}}, \ldots\right) \tag{186}$$
$$\Delta U^* = -\tfrac{1}{2} k T_{\mathrm{i}} \frac{1}{Z_{\mathrm{p}}^*} h\left(\tilde{\Lambda}^*\right)$$

where $Z_{\mathrm{p}}^* = \frac{1}{Z^*}\sum_J p_J Z_J{}^2$ , which is quantitatively indistinct from $Z_{\mathrm{p}}$ . In the case of the charge-weighted averages, $\overline{\Delta W}$, $\overline{\Delta U}$ , the approximations

$$\overline{\Delta W} \simeq \frac{Z_{\mathrm{p}}^*}{Z^*}\Delta W^* \cong \frac{Z_{\mathrm{p}}}{\overline{Z}}\Delta W^*$$
$$\overline{\Delta U} \simeq \frac{Z_{\mathrm{p}}^*}{Z^*}\Delta U^* \cong \frac{Z_{\mathrm{p}}}{\overline{Z}}\Delta U^* \tag{187}$$

(which are quantitatively exact in the weak coupling limit) are typically appropriate.

The equilibrium ion and electron densities are provided by solution of the Saha equation, for example. The final step in the process is the determination of the pressure, which is complicated by the ion-electron reactions (recombination and ionization) and the interdependence of the pressure and the ionization state, which can lead to significant departures from ideality. The special forms of these equations in the limits of weak and strong coupling are given as follows.

## B.2 Weak Coupling Limit

In the limit of weak coupling, $\Gamma_{\mathrm{i}} \ll 1$ , the above formulae take the simple linear forms

$$f(\Lambda) = g(\Lambda) = \tfrac{1}{3}\Lambda, \quad h(\Lambda) = \tfrac{2}{3}\Lambda \tag{188}$$

with

$$\tilde{\Lambda}_J = \frac{1}{D}\frac{R_J^{\,3}}{D_{\mathrm{i}}^{\,2}} = \frac{3Z_{\mathrm{p}}Z_J e^2}{4\pi\epsilon_0 D k T_{\mathrm{i}}} \tag{189}$$

which give the shifts from the free particle equation of state at the given volume as follows:

*(a) Helmholtz free energy*

$$\Delta F_J = -\frac{Z_J^{\,2} e^2}{12\pi\epsilon_0 D}$$

$$\overline{\Delta F} = -\tfrac{1}{3}\frac{\bar{Z} Z_{\mathrm{p}} e^2}{4\pi\epsilon_0 D} \tag{190}$$

*(b) Internal (potential) energy*

$$U_J = -\frac{Z_J^{\,2} e^2}{8\pi\epsilon_0 D}$$

$$\bar{U} = \tfrac{3}{2}\overline{\Delta F} \tag{191}$$

where the quantity $\bar{U}$ is the averaged self-energy.

*(c) Chemical potential shifts*

$$\Delta\mu_J = -\frac{Z_J^{\,2} e^2}{12\pi\epsilon_0 D}\left(1 + \tfrac{1}{2}\frac{D^2}{D_{\mathrm{i}}^{\,2}}\right)$$

$$\Delta\mu_{\mathrm{e}} = -\frac{Z_{\mathrm{p}} e^2}{12\pi\epsilon_0 D}\left(\tfrac{1}{2}\frac{D^2}{D_{\mathrm{e}}^{\,2}}\right) \tag{192}$$

$$\Delta\mu_{\mathrm{i}} = -\frac{Z_{\mathrm{p}}\bar{Z} e^2}{12\pi\epsilon_0 D}\left(1 + \tfrac{1}{2}\frac{D^2}{D_{\mathrm{i}}^{\,2}}\right)$$

*(d) Gibbs Free energy*

$$\Delta G = \sum_J N_J \Delta\mu_J + N_\mathrm{e}\Delta\mu_\mathrm{e} = N_\mathrm{i}\Delta\mu_\mathrm{i} + N_\mathrm{e}\Delta\mu_\mathrm{e} = -N_\mathrm{i}\frac{Z_\mathrm{p}\bar{Z}e^2}{8\pi\epsilon_0 D} = \tfrac{3}{2}N_\mathrm{i}\overline{\Delta F} \tag{193}$$

The quantity $\Delta G/N_\mathrm{i}$ is known as the *rigid shift*, and is equal to the *averaged self-energy* in the weak-coupling limit [10] .

*(e) Grand potential etc*

$$\begin{aligned} \Pi_J &= \tfrac{1}{3}U_J = \tfrac{1}{2}\Delta F_J \\ Q_J &= U_J - \Delta F_J = \tfrac{1}{2}\Delta F_J \\ \bar{\Pi} &= \bar{Q} = \tfrac{1}{2}N_\mathrm{i}\overline{\Delta F} \end{aligned} \tag{194}$$

*(f) Static continuum lowering*

$$\begin{aligned} \Delta U_J &= 3\frac{\Delta F_J}{Z_J} \\ \overline{\Delta U} &= 3\frac{\overline{\Delta F}}{\bar{Z}} \qquad \left(\cong \frac{Z_\mathrm{p}^*}{Z^*}\Delta U^*\right) \end{aligned} \tag{195}$$

*(g) Thermodynamic continuum lowering*

$$\begin{aligned} \Delta W_J &= 2\frac{\Delta F_J}{Z_J} \\ \overline{\Delta W} &= 2\frac{\overline{\Delta F}}{\bar{Z}} \qquad \left(\cong \frac{Z_\mathrm{p}^*}{Z^*}\Delta W^*\right) \end{aligned} \tag{196}$$

## B.3 Strong Coupling Limit

In the limit of strong coupling, $\Gamma \gg 1$ , we have

$$f(\Lambda) = \tfrac{9}{10}\Lambda^{2/3}, \quad g(\Lambda) = \tfrac{3}{5}\Lambda^{2/3}, \quad h(\Lambda) = \Lambda^{2/3} \tag{197}$$

$$\tilde{\Lambda}_J^{\,2/3} = \tfrac{10}{9} C \left( \frac{R_J}{D} \right)^2 \tag{198}$$

which give the shifts from the free particle equation of state at the given volume as follows:

*(a) Helmholtz free energy*

$$\begin{aligned} \Delta F_J &= -C \frac{D_\mathrm{i}^2}{D^2} \frac{Z_J^{\,2} e^2}{4\pi\epsilon_0 R_J} \\ \overline{\Delta F} &= -C \frac{D_\mathrm{i}^2}{D^2} \frac{\bar{Z} Z_\mathrm{p} e^2}{4\pi\epsilon_0 \bar{R}_\mathrm{i}} = -C \frac{VkT_\mathrm{i}}{4\pi N_\mathrm{i} \bar{R}_\mathrm{i} D^2} \end{aligned} \tag{199}$$

where

$$\frac{1}{\bar{R}_\mathrm{i}} = \frac{1}{N_\mathrm{i} \bar{Z} Z_\mathrm{p}} \sum_J \frac{N_J Z_J^{\,2}}{R_J} \tag{200}$$

*(b) Internal (potential) energy and averaged self-energy*

$$\begin{aligned} U_J &= \Delta F_J \\ \bar{U} &= \overline{\Delta F} \end{aligned} \tag{201}$$

*(c) Chemical potential shifts*

$$\begin{aligned} \Delta\mu_J &= \overline{\Delta F} \left( \tfrac{1}{3} \frac{Z_J}{\bar{Z}} + \frac{Z_J^{\,2}}{\bar{Z} Z_\mathrm{p}} \left( \frac{\bar{R}_\mathrm{i}}{R_J} - 1 \right) + \frac{N_\mathrm{i}}{N_J} \frac{D^2}{D_J^{\,2}} \right) \\ \Delta\mu_\mathrm{e} &= \frac{\overline{\Delta F}}{\bar{Z}} \left( \frac{D^2}{D_\mathrm{e}^{\,2}} \right) \\ \Delta\mu_\mathrm{i} &= \overline{\Delta F} \left( \tfrac{1}{3} + \frac{D^2}{D_\mathrm{i}^{\,2}} \right) \end{aligned} \tag{202}$$

*(d) Gibbs free energy and rigid shift*

$$\Delta G = \sum_J N_J \Delta\mu_J + N_\mathrm{e}\Delta\mu_\mathrm{e} = N_\mathrm{i}\Delta\mu_\mathrm{i} + N_\mathrm{e}\Delta\mu_\mathrm{e} = \tfrac{4}{3} N_\mathrm{i}\overline{\Delta F} \tag{203}$$

which now gives the rigid shift as 4/3 of the averaged self-energy.

*(e) Grand potential etc*

$$\begin{aligned} \Pi_J &= \tfrac{1}{3}U_J = \tfrac{1}{3}\Delta F_J \\ Q_J &= U_J - \Delta F_J = 0 \\ \overline{\Pi} &= \tfrac{1}{3} N_\mathrm{i}\overline{\Delta F} \end{aligned} \tag{204}$$

*(f) Static continuum lowering and TIPD*

$$\begin{aligned} \Delta U_J &= \Delta W_J = \tfrac{5}{3}\frac{\Delta F_J}{Z_J} \\ \overline{\Delta U} &= \overline{\Delta W} = \tfrac{5}{3}\frac{\overline{\Delta F}}{\overline{Z}} \end{aligned} \tag{205}$$

## B.4 Total Pressure

The total pressure follows from the Helmholtz free energy (173) according to the standard thermodynamic formula, $P = -\partial F/\partial V$ , which yields

$$P = P_{\mathrm{e}}^0 + P_{\mathrm{i}}^0 + \tfrac{1}{3}\frac{N_{\mathrm{i}}}{V}\bar{U} + \frac{\partial \Phi}{\partial V} \tag{206}$$

All quantities on the right-hand side of this equation are given, in terms of known quantities, by the equations above or those in APPENDIX A, except for $\partial\Phi/\partial V$, which represents the effect of electron-ion binding on the pressure . Making reference to (112), this is

$$\frac{\partial \Phi}{\partial V} = \sum_J \frac{\partial \bar{N}_J}{\partial V} E_{\mathrm{c}}\left(\mathbf{z}_J\right) \equiv N_{\mathrm{i}} \sum_J \frac{\partial p_J}{\partial V} E_J \tag{207}$$

where $p_J = p\left(\mathbf{z}_J\right)$ denotes the probability of an ion being in the configuration $J$ with energy $E_J$ . This is taken to be given, in the first instance, by the Gibbs distribution, $p_J \propto g_J \exp\left(-S_J\left(\mu_{\mathrm{e}}, T_{\mathrm{e}}\right)\right)$ where $S_J$ is the entropy of the ion state $J$ embedded in a plasma whose properties, together with the motion of the ion, are averaged over and are thus representable in terms of the thermodynamic coordinates. The entropy $S_J$ is found by expressing the Helmholtz free energy, of a neutral ion-electron system containing $Z = \Sigma_J + Z_J$ electrons in chemical equilibrium, as follows

$$F_J + E_J - TS_J = \mu_{\mathrm{e}} Z + \mu_{\mathrm{i}} - PV \tag{208}$$

the right-hand side of which does not depend upon the state of the individual ion. Then, substituting $F_J = F_J^0 + \Delta F_J \equiv \left(\mu_{\mathrm{e}}^0 Z_J + \mu_{\mathrm{i}}^0 - P^0 V\right) + \Delta F_J$ and rearranging terms, leads to

$$-TS_J = \mu_{\mathrm{e}}^0 \Sigma_J - \Delta F_J - E_J + \left\{ Z\Delta\mu_{\mathrm{e}} + \Delta\mu_{\mathrm{i}} - \left(P - P^0\right)V \right\} \tag{209}$$

where $\Delta\mu_{\mathrm{e}} = \mu_{\mathrm{e}} - \mu_{\mathrm{e}}^0$ and $\Delta\mu_{\mathrm{i}} = \mu_{\mathrm{i}} - \mu_{\mathrm{i}}^0$ are the chemical potential shifts. In (209), the term in $\{\ \}$ brackets does not depend on the state of the ion, and can thus be incorporated in the grand potential $T_{\mathrm{e}}\mathfrak{Z}$ , which is equivalent to the partition function. Hence the probability of finding the ion in the state $J$ is

$$p_J = g_J \exp\left(\left(\mu_{\mathrm{e}}^0 \Sigma_J - \Delta F_J - E_J\right)/kT_{\mathrm{e}} + \mathfrak{Z}\left(\mu_{\mathrm{e}}^0, T_{\mathrm{e}}\right)\right) \tag{210}$$

which is consistent with (130), for example, and in which $\Sigma_J = Z - Z_J$ is the number of bound electrons in the configuration $J$. The grand potential $T_e\mathfrak{Z}$ is wholly determined by the normalization $\sum_J p_J = 1$. It then follows straightforwardly that

$$V\frac{\partial p_J}{\partial V} = \frac{p_J}{kT_e}\left(\left(\Sigma_J - \langle\Sigma_c\rangle\right)V\left(\frac{\partial \mu_e^0}{\partial V}\right)_T + \left(\Pi_J - \langle\Pi_c\rangle\right)\right)$$

$$= -\frac{p_J}{kT_e}\left(kT_B\left(1-\theta_V\right)\Delta\Sigma_J - \Delta\Pi_J\right) \tag{211}$$

where $\langle X_c\rangle = \sum_J p_J X_J$, $\Delta X_J = X_J - \langle X_c\rangle \;\forall\; X_J = X_c(\mathbf{z}_J)$ and

$$\theta_x = \frac{x}{\langle Z_c\rangle}\frac{\partial\langle Z_c\rangle}{\partial x} = \frac{\partial \ln\langle Z_c\rangle}{\partial \ln x} \tag{212}$$

with $x$ standing for any macroscopic thermodynamic coordinate, such as $V$ or $T$, and $\langle Z_c\rangle$ is the statistical average over an ensemble. While there is little or no quantitative difference between this and $\bar{Z}$, the average ion charge within a microstate of the many-body system, there is nevertheless a subtle, but significant, in the present context, distinction, namely that $\bar{Z}$ is a function of the microstate coordinates, $N_K$, which are mathematically independent of the thermodynamic coordinates; while $\langle Z_c\rangle = Z^*$ (see above) depends only on the averages $\bar{N}_K = p_K N_i$ and is therefore a function of the thermodynamic coordinates alone.

Substituting (211) into (207) yields

$$\frac{\partial\Phi}{\partial V} = -\frac{n_i}{kT_e}\sum_J p_J\left(kT_B\left(1-\theta_V\right)\Delta\Sigma_J - \Delta\Pi_J\right)\Delta E_J \equiv -\frac{n_i}{kT_e}\left(kT_B\left(1-\theta_V\right)\langle\Delta\Sigma_c\Delta E_c\rangle - \langle\Delta\Pi_c\Delta E_c\rangle\right)$$

$$= \frac{n_i}{kT_e}\left(kT_B\left(1-\theta_V\right)\langle\Delta Z_c\Delta E_c\rangle + \langle\Delta\Pi_c\Delta E_c\rangle\right) \tag{213}$$

Now, making use of (182), the first order variation in $\Pi_J$ is given by

$$\Delta\Pi_J = \Delta Z_J \left.\frac{\partial \Pi_J}{\partial Z_J}\right|_{Z_J = Z^*} = \tfrac{1}{3}\Delta Z_J \Delta U^* \tag{214}$$

giving

$$\left\langle \Delta\Pi_{\rm c}\Delta E_{\rm c}\right\rangle = \tfrac{1}{3}\Delta U^{*}\left\langle \Delta Z_{\rm c}\Delta E_{\rm c}\right\rangle = -\tfrac{1}{3}\Delta U^{*}\left\langle \Delta\Sigma_{\rm c}\Delta E_{\rm c}\right\rangle$$

$$\left\langle \Delta\Pi_{\rm c}\Delta Z_{\rm c}\right\rangle = -\left\langle \Delta\Pi_{\rm c}\Delta\Sigma_{\rm c}\right\rangle = \tfrac{1}{3}\Delta U^{*}\left\langle \Delta Z_{\rm c}^{\;2}\right\rangle = \tfrac{1}{3}\Delta U^{*}\left\langle \Delta\Sigma_{\rm c}^{\;2}\right\rangle \tag{215}$$

These equations impart a particular significance to the average-atom static continuum lowering, $\Delta U^{*}$ .

The total pressure is therefore

$$PV = P^{0}V + N_{\rm i}\bar{\Pi} + \frac{N_{\rm i}}{kT_{\rm e}}\left(kT_{\rm B}\left(1-\theta_{V}\right) + \tfrac{1}{3}\Delta U^{*}\right)\left\langle \Delta Z_{\rm c}\Delta E_{\rm c}\right\rangle \tag{216}$$

where the effect of the bound electrons is described by the correlations $\left\langle \Delta\Sigma_{\rm c}\Delta E_{\rm c}\right\rangle$ involving the configuration charge states and energies; and

$$P^{0}V = \zeta\left(n_{\rm e},T_{\rm e}\right)N_{\rm e}kT_{\rm e} + N_{\rm i}kT_{\rm i} \tag{217}$$

(see APPENDIX A). The quantity $PV^{0}$ is, on the other hand,

$$PV^{0} = \zeta\left(\frac{V}{V^{0}}n_{\rm e},T_{\rm e}\right)N_{\rm e}kT_{\rm e} + N_{\rm i}kT_{\rm i} \tag{218}$$

(cf equation (126)) which does not, in general, give $V^{0}$ in closed form. For non-degenerate electrons however, $PV^{0} = P^{0}V$ , while in the limit of high density = extreme electron degeneracy when $T_{\rm F} \gg T_{\rm e},T_{\rm i}$ ,

$$\frac{V^{0}}{V} \simeq \left(\frac{P^{0}}{P}\right)^{3/5} + \tfrac{3}{5}\frac{N_{\rm i}kT_{\rm i}}{PV}\left(1-\left(\frac{P}{P^{0}}\right)^{2/5}\right) \tag{219}$$

## B.5 Charge-state fluctuations and derivatives

Equation (210) and its derivatives, eg (211), are of key importance, since they provide the configuration averages and their derivatives of any quantity, in terms of the macroscopic coordinates of the system.. In particular, the dependence of the CSD on the thermodynamic state is so provided. With $T_{\rm e} = T_{\rm i} = T$ , we can derive the companion equation to (211)

$$T\frac{\partial p_{J}}{\partial T} = \frac{p_{J}}{kT}\left(\left(\Sigma_{J}-\left\langle\Sigma_{\rm c}\right\rangle\right)kT^{2}\left(\frac{\partial\eta_{\rm e}^{0}}{\partial T}\right)_{V} + \left(E_{J}-\left\langle E_{\rm c}\right\rangle\right) + \left(U_{J}-\left\langle U_{\rm c}\right\rangle\right)\right)$$

$$= -\frac{p_{J}}{kT}\left(\left(\tfrac{3}{2}-\theta_{T}\right)kT_{\rm B}\Delta\Sigma_{J} - \Delta E_{J} - 3\Delta\Pi_{J}\right) \tag{220}$$

These equations straightforwardly yield the derivatives of the moments of the CSD, eg,

$$V\frac{\partial\langle Z_{\rm c}\rangle}{\partial V}=V\sum_J\frac{\partial p_J}{\partial V}Z_J=\frac{1}{kT}\left(\left(1-\theta_V\right)kT_{\rm B}+\tfrac{1}{3}\Delta U^*\right)\left\langle\Delta\Sigma_{\rm c}^{\ 2}\right\rangle$$

$$T\frac{\partial\langle Z_{\rm c}\rangle}{\partial T}=T\sum_J\frac{\partial p_J}{\partial T}Z_J=\frac{1}{kT}\left(\left(\left(\tfrac{3}{2}-\theta_T\right)kT_{\rm B}+\Delta U^*\right)\left\langle\Delta\Sigma_{\rm c}^{\ 2}\right\rangle-\left\langle\Delta E_{\rm c}\Delta\Sigma_{\rm c}\right\rangle\right) \tag{221}$$

which imply

$$\theta_V\equiv\frac{V}{\langle Z_{\rm c}\rangle}\frac{\partial\langle Z_{\rm c}\rangle}{\partial V}=\frac{kT_{\rm B}+\tfrac{1}{3}\Delta U^*}{kT_{\rm B}\left\langle\Delta\Sigma_{\rm c}^{\ 2}\right\rangle+\langle Z_{\rm c}\rangle kT}\left\langle\Delta\Sigma_{\rm c}^{\ 2}\right\rangle$$

$$\theta_T\equiv\frac{T}{\langle Z_{\rm c}\rangle}\frac{\partial\langle Z_{\rm c}\rangle}{\partial T}=\frac{\left(\tfrac{3}{2}kT_{\rm B}+\Delta U^*\right)\left\langle\Delta\Sigma_{\rm c}^{\ 2}\right\rangle-\left\langle\Delta E_{\rm c}\Delta\Sigma_{\rm c}\right\rangle}{kT_{\rm B}\left\langle\Delta\Sigma_{\rm c}^{\ 2}\right\rangle+\langle Z_{\rm c}\rangle kT} \tag{222}$$

The corresponding derivatives of $\left\langle\Delta\Sigma_{\rm c}^{\ 2}\right\rangle$ are found to involve only odd moments of the CSD, which can typically be considered to be negligible, which, since

$$\left\langle\Delta\Sigma_{\rm c}^{\ 2}\right\rangle\equiv\langle Z_{\rm c}\rangle\left(\langle Z_{\rm p}\rangle-\langle Z_{\rm c}\rangle\right)=Z^*\left(Z_{\rm p}^*-Z^*\right) \tag{223}$$

then leads to

$$\tilde{\theta}_x\equiv\frac{x}{\langle Z_{\rm p}\rangle}\frac{\partial\langle Z_{\rm p}\rangle}{\partial x}=\frac{2Z^*-Z_{\rm p}^*}{Z_{\rm p}^*}\theta_x \tag{224}$$

If the bound and free electrons possess putatively independent Hamiltonian descriptions, ie, in some approximation, are *dynamically separable*, then they can be treated as separate thermodynamic systems in the context of the Grand Canonical Ensemble vis à vis equation (210). With the constraint of overall electrical neutrality, which requires that $\langle\Sigma_{\rm c}\rangle+\langle Z_{\rm c}\rangle=Z\ \Rightarrow\ \langle\Delta\Sigma_{\rm c}\rangle+\langle\Delta Z_{\rm c}\rangle=0$ , the correlation functions $\left\langle\Delta E_{\rm c}\Delta\Sigma_{\rm c}\right\rangle$, $\left\langle\Delta\Sigma_{\rm c}^{\ 2}\right\rangle$ are related through a general formula that can be found in the standard treatment of fluctuations in a grand canonical ensemble. The probability distribution (210) gives the standard formulae $\left\langle\Delta E_{\rm c}\Delta\Sigma_{\rm c}\right\rangle=-\partial\langle\Sigma_{\rm c}\rangle/\partial\beta$, $\left\langle\Delta\Sigma_{\rm c}^{\ 2}\right\rangle=\partial\langle\Sigma_{\rm c}\rangle/\partial\eta$ , $\beta=1/kT$ , $\eta=\mu_{\rm e}/kT$ , and hence that

$$\frac{\left\langle\Delta E_{\rm c}\Delta\Sigma_{\rm c}\right\rangle}{\left\langle\Delta\Sigma_{\rm c}^{\ 2}\right\rangle}=\mu_{\rm e}-T\left(\frac{\partial\mu_{\rm e}}{\partial T}\right)_{\langle\Sigma_{\rm c}\rangle}=\mu_{\rm e}-T\left(\frac{\partial\mu_{\rm e}}{\partial T}\right)_{Z^*} \tag{225}$$

where

$$\left(\frac{\partial \mu_{\mathrm{e}}}{\partial T}\right)_{Z^*} = \left(\frac{\partial \mu_{\mathrm{e}}}{\partial T}\right)_{n_{\mathrm{e}}} + \frac{n_{\mathrm{e}}}{T}\frac{\theta_T}{\theta_V}\left(\frac{\partial \mu_{\mathrm{e}}}{\partial n_{\mathrm{e}}}\right)_T \tag{226}$$

Application of equations in APPENDIX A then yields

$$T\left(\frac{\partial \mu_{\mathrm{e}}}{\partial T}\right)_{Z^*} = T\left(\frac{\partial \mu_{\mathrm{e}}^0}{\partial T}\right)_{Z^*} + T\left(\frac{\partial \Delta\mu_{\mathrm{e}}}{\partial T}\right)_{Z^*} = \mu_{\mathrm{e}}^0 - kT_{\mathrm{B}}\left(\tfrac{3}{2} - \frac{\theta_T}{\theta_V}\right) + T\left(\frac{\partial \Delta\mu_{\mathrm{e}}}{\partial T}\right)_{Z^*} \tag{227}$$

where, in the weak coupling limit and low degeneracy, for which $n_{\mathrm{e}}\partial\Delta\mu_{\mathrm{e}}/\partial n_{\mathrm{e}} = -T\partial\Delta\mu_{\mathrm{e}}/\partial T = \tfrac{1}{2}\Delta\mu_{\mathrm{e}}$,

$$T\left(\frac{\partial \Delta\mu_{\mathrm{e}}}{\partial T}\right)_{Z^*} = \tfrac{1}{2}\left(\frac{\theta_T}{\theta_V} - 1\right)\Delta\mu_{\mathrm{e}} \tag{228}$$

and hence

$$\frac{\left\langle \Delta E_{\mathrm{c}}\Delta\Sigma_{\mathrm{c}}\right\rangle}{\left\langle \Delta\Sigma_{\mathrm{c}}{}^2\right\rangle} = \left(\tfrac{3}{2} - \frac{\theta_T}{\theta_V}\right)kT + \tfrac{1}{2}\left(3 - \frac{\theta_T}{\theta_V}\right)\Delta\mu_{\mathrm{e}} \tag{229}$$

Substituting into (222) and solving for $\theta_V$ and $\theta_T$ yields the key relation

$$\theta_T = 3\theta_V \tag{230}$$

with $\theta_V$ given by the first of (222), together with

$$\frac{\left\langle \Delta E_{\mathrm{c}}\Delta\Sigma_{\mathrm{c}}\right\rangle}{\left\langle \Delta\Sigma_{\mathrm{c}}{}^2\right\rangle} = -\tfrac{3}{2}kT \tag{231}$$

which, in turn, lead, via (222), to

$$\theta_V = \tfrac{1}{3}\theta_T = \left(1 + \frac{\Delta U^*}{3kT}\right)\frac{\left\langle \Delta\Sigma_{\mathrm{c}}{}^2\right\rangle}{\left\langle \Delta\Sigma_{\mathrm{c}}{}^2\right\rangle + \left\langle Z_{\mathrm{c}}\right\rangle} \tag{232}$$

Equations (230), (231) and (232) are, in principle, exact in the weak coupling (Saha-Boltzmann) limit when degeneracy and pressure ionization are not significant factors. Furthermore, it is found, by considering the details of electron fluctuations within atomic shells in the average-atom picture, with regard for the negative correlations due to the electrons' mutual Coulomb repulsion, that $\left\langle \Delta\Sigma_{\mathrm{c}}{}^2\right\rangle < \left\langle Z_{\mathrm{c}}\right\rangle$ . In many cases, $\left\langle \Delta\Sigma_{\mathrm{c}}{}^2\right\rangle \ll \left\langle Z_{\mathrm{c}}\right\rangle$ , which leads to further simplifications.

In the regime of strong coupling and arbitrary degeneracy, using (199) and (202), we find $n_{\mathrm{e}}\partial\Delta\mu_{\mathrm{e}}/\partial n_{\mathrm{e}} = -\tfrac{2}{3}T\partial\Delta\mu_{\mathrm{e}}/\partial T = \Delta\mu_{\mathrm{e}}\left(1-\lambda_T\right)$ where $\lambda_T$ is the electronic isothermal bulk modulus pressure coefficient, $\left(\partial\kappa_T/\partial P_\mathrm{e}\right)_{T_\mathrm{e}}$, which has the value 1 in the non-degenerate limit and $\frac{5}{3}$ in the degenerate limit and otherwise $1 \le \lambda_T \le \frac{5}{3}$ (see APPENDIX A) . Hence

$$T\left(\frac{\partial\Delta\mu_\mathrm{e}}{\partial T}\right)_{Z^*} = \left(\frac{\theta_T}{\theta_V} - \frac{3}{2}\right)\left(1-\lambda_T\right)\Delta\mu_\mathrm{e} \tag{233}$$

which gives

$$\frac{\left\langle \Delta E_\mathrm{c}\Delta\Sigma_\mathrm{c}\right\rangle}{\left\langle \Delta\Sigma_\mathrm{c}{}^2\right\rangle} = \left(kT_\mathrm{B} + \left(1-\lambda_T\right)\Delta\mu_\mathrm{e}\right)\left(\frac{3}{2} - \frac{\theta_T}{\theta_V}\right) + \Delta\mu_\mathrm{e} \tag{234}$$

leading to

$$\frac{\theta_T}{\theta_V} = 3 + \frac{3}{2}\frac{\left(3\lambda_T - 1\right)\Delta\mu_\mathrm{e}}{3\left(\lambda_T - 1\right)\Delta\mu_\mathrm{e} - \Delta U^*} \tag{235}$$

in place of (230). Since, for degenerate systems, $\left|\Delta\mu_\mathrm{e}\right| \ll \left|\Delta U^*\right|$, this yields that $0 < \theta_T/\theta_V \lesssim 3$ .

While equations (222) can describe *pressure ionization* ( $\theta_V < 0$ ) this is only at the expense of implying $\theta_T < 0$, which is clearly wrong. The model is evidently incomplete in this regime. The principal reason for this is that the conditions for dynamical separability are not met: The electrons in the ionising states are contributing to the properties of both bound and free electron systems, precisely the situation described above in section 6.1. Fundamentally, this is a failure of the Coulomb model, which treats the particles, ions in particular, as being point-like and because, during pressure ionization, the bound and free electron systems are not dynamically separable. The electron bound states within ions have a finite size and, at high enough densities, can start to overlap and thereby contribute to the pressure, while still remaining bound. The resulting degeneracy pressure would introduce a repulsive (positive) element into the interaction pressure, $\frac{1}{3}U$ , which is otherwise wholly attractive. This can also be expected to affect the relaxation energy. Pressure ionization contributes to departures of the equation of state from ideality and will therefore affect the relaxation energy (47) and hence the spectroscopic IPD. This is important because it suggests that that the model of the SIPD presented here may be incomplete in this regime.

Pressure ionization occurs when the continuum threshold is depressed through full or nearly-full electronic levels, and is thus essentially a characteristic of degenerate systems. For such systems, the level energies below the Fermi level are insensitive to temperature and, as the continuum lowering in the strong-coupling limit is temperature independent also, this means that the bound level occupancies are temperature independent, implying $\theta_T \simeq 0$. Clearly therefore, neither of equations (234) and (235) can apply to a system while it is subject to pressure ionization.

## APPENDIX C List of symbols

*C.1 List of symbols used for mathematical and physical quantities*

(Symbols used in APPENDIX A are defined separately therein.)

$a_\infty$ Bohr radius ($= 4\pi\epsilon_0\hbar^2 / m_e e^2$)

$C_j$ Force constant associated with an ion species $j$ whereby the electrostatic energy of an arrangement of ions on a lattice is given by (43).

$C$ Mean (or common) force constant given by (44).

$D$ Total plasma screening length.

$D_e$ Electron (Thomas-Fermi) screening length.

$D_i$ Ion (Debye) screening length.

$E$ Plasma energy $= E_{int} - E_b = E_i + E_e$

$E_i$ Ion component of the plasma energy.

$E_e$ Electron component of the plasma energy.

$E_{int}$ Plasma internal energy $= E + E_b$ .

$E_b$ Component of $E$ representing the total energy of the bound electrons in a plasma.

$E_c(\mathbf{z})$ Energy of the electronic configuration (of an ion) denoted by $\mathbf{z}$ .

$E_J$ $= E_c(\mathbf{z}_J)$ = energy of the electronic configuration $J$ .

$e$ Unit of charge.

e Euler's constant, or, as subscript label denoting "electron".

$F$ Helmholtz free energy $= F_e + F_i$ where $F_e, F_i$ are the Helmholtz free energies for the electron and ion subsystems respectively, as defined by $F_e = G_e - P_e V$ , $F_i = G_i - P_i V$ .

$F^0$ $F_0^e + F_0^i$ = Helmholtz free energy of closed system of *non-interacting* particles with the same temperature(s), volume and concentrations as an interacting system with Helmholtz free energy $F$ .

$F_0$ $= F_0(V,T)$ is the Helmholtz function for a non-interacting particle system having the same particle concentrations as the interacting system.

$f(\Lambda)$ Function that yields the interaction free energy as per (134) and defined by (140) - (141).

$G$ Gibbs free energy $= G_{\mathrm{e}} + G_{\mathrm{i}}$ where $G_{\mathrm{e}}, G_{\mathrm{i}}$ are the Gibbs free energies for the electron and ion subsystems respectively, as defined by (99).

$G^0$ $G^0 = F^0 + P^0 V = G_{\mathrm{i}}^0 + G_{\mathrm{e}}^0$ where $G_{\mathrm{i}}^0$, $G_{\mathrm{e}}^0$ are the Gibbs free energies for the electron and ion subsystems respectively, as defined by $G_{\mathrm{i}}^0 = F_{\mathrm{i}}^0 + P_{\mathrm{i}}^0 V = \sum_K N_K \mu_K^0$ , $G_{\mathrm{e}}^0 = F_{\mathrm{e}}^0 + P_{\mathrm{e}}^0 V = N_{\mathrm{e}} \mu_{\mathrm{e}}^0$ .

$G_0$ $= G_0(P,T)$ is the Gibbs function for a non-interacting particle system having the same particle concentrations as the interacting system.

$g(\Lambda)$ Function that yields the interaction energy as per (27) and defined by (30).

$g_j$ $g_j(r)$ = ion pair correlation function around the *j*th ion.

$g_J$ Spin (or other internal) degeneracy weighting ($= 2s_J + 1$ ) of the atomic state $J$ .

$h(\Lambda)$ Function that yields the static continuum lowering as per (24).

$I_k(x)$ Fermi integral function defined by $I_k(x) = \int_0^\infty \frac{y^k}{1+\exp(y-x)} \mathrm{d}y$ .

$I(z)$ Thermodynamic potential associated with a change in the internal coordinate, $z$ .

i $\sqrt{-1}$, or subscript label denoting ”ion”.

$J, K, ..$ Labels denoting atomic states or configurations present in the plasma.

$j$ Label denoting a particular ion in the plasma.

$k$ Boltzmann’s constant.

$\mathbf{k}$ Wave-number used in the spectral representation, eg of microfield fluctuations and charge-density oscillations..

$M$ Total mass.

$m$ Particle mass.

$m_{\mathrm{e}}$ Electron mass.

$N_{\mathrm{e}}$ Number of free electrons present in the plasma, $= \sum_K Z_K N_K$

$N_{\mathrm{i}}$ Number of atomic ions present in the plasma, $= \sum_K N_K$

$N_K$ Number of ions in the configuration $K$ present in the plasma.

$n_{\mathrm{e}}$ Free electron density $= N_{\mathrm{e}}/V$ .

$n_{\mathrm{i}}$ Total ion density $= N_{\mathrm{i}}/V$ .

$P$ Pressure $= P_{\mathrm{e}} + P_{\mathrm{i}}$ where $P_{\mathrm{e}}$ and $P_{\mathrm{i}}$ are the electron and ion pressures respectively.

$P^0$ $= P_{\mathrm{i}}^0 + P_{\mathrm{e}}^0 =$ pressure of closed system of *non-interacting* particles with the same temperature(s), volume and concentrations as an interacting system of pressure $P$ .

$p(z)$ Probability distribution of $z$ , = probability that a randomly chosen ion has $z$ in $(z,\ z+\mathrm{d}z)$ (continuous distribution) or that $z$ has a particular value (discrete distribution).

$p_J$ $p(\mathbf{z}_J)$

$Q_J$ $= T\Delta S_J$

$R_j$ Ion sphere radius (7).

$\bar{R}_{\mathrm{i}}$ Charge-squared-weighted harmonic mean ion-sphere radius as defined by (200).

$R_{\mathrm{WS}}$ Wigner-Seitz radius (18).

$r_j$ Ion core radius, which is the radius that separates the inner local region surrounding the nucleus of an ion from the "collective" region occupied by the plasma as a whole where no individual ion has a dominant influence.

$r_s$ Brueckner parameter $= R_{\mathrm{WS}}/a_\infty$ .

$S$ Entropy $= S_{\mathrm{e}} + S_{\mathrm{i}}$ where $S_{\mathrm{e}}$ and $S_{\mathrm{i}}$ are the electron and ion entropies respectively.

$S_0$ $= S_0(V,T) =$ macroscopic entropy $= \langle S(V,T,z)\rangle$

$S_J$ Internal entropy associated with ion in state $J$ .

$S_{\mathrm{ii}}$ $S_{\mathrm{ii}}(\mathbf{k})$ = static ion-ion structure factor.

$s$ Denotes the function, $s(\Lambda) = (1+\Lambda)^{1/3}$

$s_J$ Total spin of atomic state $J$ .

$T$ Temperature.

$T_{\mathrm{B}}$ Effective electron temperature, $= \kappa_T/n_{\mathrm{e}}$ , defined, in the first instance, by (16).

$T_{\mathrm{F}}$ Fermi temperature.

$T_{\mathrm{e}}$ Electron temperature.

$T_{\mathrm{i}}$ Ion temperature.

$U$ Coulomb energy of the plasma.

$U_{\mathrm{DH}}$ Coulomb energy in Debye-Hückel approximation (33).

$U_{\mathrm{is}}$ Coulomb energy in ion-sphere approximation (32).

$U_J$ $= -\dfrac{kT_{\mathrm{i}}}{2Z_{\mathrm{p}}} Z_J\, g\left(\Lambda_J\right)$ =plasma Coulomb energy attributed to atom in configuration $J$ .

$u^0_{\mathrm{DH}}$ *Debye-Hückel energy*, as defined by (23).

$u^0_{\mathrm{WS}}$ *Wigner-Seitz energy*, as defined by (21).

$V$ Volume

$V^0$ Volume of closed system of *non-interacting* particles with the same temperature(s), pressure and concentrations as an interacting system of volume $V$ .

$v_{\mathrm{c}}$ Repulsive core volume.

$w(\eta)$ Positive definite monotone function that, when multiplied by the electron temperature, represents the electron degeneracy contribution to the spectroscopic IPD and in terms of which the electron energy offset is given by (62) - (65).

$X_j$ $= r_j / R_j$

$x$ $= 1/\Lambda_j$ , or general variable or thermodynamic coordinate.

$Y_j$ Ion-plasma coupling parameter (9).

$y$ Parameter defined by (38).

$Z$ Ion atomic number.

$Z_j$ Charge on ion $j$ .

$Z_{\mathrm{c}}(\mathbf{z})$ $= Z - \Sigma_{\mathrm{c}}(\mathbf{z})$ = charge of ion in configuration denoted by $\mathbf{z}$

$Z_J$ $= Z_{\mathrm{c}}(\mathbf{z}_J) = Z - \Sigma_J$ = charge of ion in the configuration $K$

$\bar{Z}$ Mean charge state of plasma $= n_{\mathrm{e}}/n_{\mathrm{i}}$ .

$Z_{\mathrm{p}}$ Plasma effective ion perturber charge (4).

$Z^*$ $= \langle Z_{\mathrm{c}} \rangle$ = average-atom ion charge.

$Z_{\mathrm{p}}^{*}$ $=\langle Z_{\mathrm{c}}^{2}\rangle / Z^{*}$ = average effective plasma perturber charge.

$z$ General internal coordinate(s), eg one or more of the components of $\mathbf{z}$, used to represent the internal configuration of an atom or ion.

$\bar{z}$ Average of $z$ as defined by (85).

$z_{\alpha}$ Number of electrons in the bound level $\alpha$ within a configuration or state of an ion.

$\mathbf{z}$ Configuration vector representing the electronic configuration of an atom or ion, $=\sum_{\alpha} z_{\alpha}\hat{\boldsymbol{v}}_{\alpha}$ .

$\mathbf{z}_{J}$ Configuration vector of the configuration $J$

$\mathfrak{Z}$ Lagrange multiplier arising from the requirement that the probabilities $p(z)$ should be normalised, in terms of which $\ln\mathfrak{Z}$ is the partition function and $kT\mathfrak{Z}$ is the grand potential.

$\alpha$ Labels electronic state within a particular configuration ($K$) or ion ($j$).

$\alpha$ Plasma electron screening parameter $= D_{\mathrm{i}}/D$

$\alpha_{\mathrm{M}}$ Madelung constant.

$\Gamma_{j}$ Ion coupling parameter (9) $= Z_{\mathrm{p}} Z_{j} e^{2} / 4\pi\epsilon_{0} R_{j} kT_{\mathrm{i}}$ , relating to specific ion, $j$

$\Gamma$ Plasma ion coupling parameter (80), $= 3Z_{\mathrm{p}}\bar{Z}e^{2} / 4\pi\epsilon_{0} R_{\mathrm{WS}} kT_{\mathrm{i}}$ , relating to plasma as a whole.

$\Delta_{\mathrm{e}}$ Electron energy offset in the SIPD as per equation (45)

$\Delta_{KJ}$ Chemical potential difference defined by (106)

$\Delta E_{j\alpha}$ Ionization potential of state $\alpha$ in ion $j$ .

$\Delta\mathbf{E}$ Electric microfield.

$\Delta\varepsilon_{\mathrm{mf}}$ Postulated microfield contribution to the continuum lowering.

$\Delta F$ $= F - F^{0}$ = interaction free energy.

$\Delta F_{J}$ contribution to $\Delta F$ due to n individual ion in state $J$ .

$\Delta S_{J}$ $= -\partial\Delta F_{J} / \partial T_{\mathrm{i}} = (U_{J} - \Delta F_{J}) / T_{\mathrm{i}}$ = entropy associated with the interaction between an ion in configuration $J$ and the surrounding plasma with which it is in equilibrium.

$\Delta U_j$ Static continuum lowering.

$\Delta W_j$ Thermodynamic ionization potential.

$\Delta z$ $= z - \bar{z}$

$\Delta\chi$ $= \Delta\chi_\mathrm{e} + \Delta\chi_\mathrm{i}$ Relaxation energy (47).

$\Delta\mu_\mathrm{e}$ $= \mu_\mathrm{e} - \mu_\mathrm{e}^0$ is the shift in the electron chemical potential due to Coulomb interactions.

$\Delta\mu_\mathrm{i}$ $= \mu_\mathrm{i} - \mu_\mathrm{i}^0$ is the shift in the electron chemical potential of a bare ion due to Coulomb interactions.

$\Delta\chi_\mathrm{e}$ The specifically electronic component of the relaxation energy given by (51).

$\Delta\chi_\mathrm{i}$ The ionic (Coulomb energy) component of the relaxation energy given by (50).

$\Delta\omega_j$ Spectroscopic ionization potential depression (SIPD) (45).

$\mathrm{E}_0$ Normal electric field strength $\simeq Ze/4\pi\epsilon_0 R_\mathrm{WS}{}^2$ , $Z^2 = \langle Z^2 \rangle = Z_\mathrm{p}\bar{Z}$ .

$\epsilon$ $= \epsilon(\mathbf{k},\omega)$ = longitudinal dielectric function.

$\epsilon_0$ Permittivity of free space.

$\bar{\varepsilon}$ Mean kinetic energy associated with a free electron .

$\zeta$ $= \zeta(n_\mathrm{e}, T_\mathrm{e})$ = ratio of electron pressure to ideal gas pressure of a system of non-interacting electrons, $= \frac{2}{3}\left(I_{3/2}\left(\mu_\mathrm{e}^0/kT_\mathrm{e}\right)/I_{1/2}\left(\mu_\mathrm{e}^0/kT_\mathrm{e}\right)\right)$ .

H Heaviside step-function: $\mathrm{H}(x) = 1, \quad x > 0$ , $\mathrm{H}(x) = 0, \quad x < 0$

$\eta_\mathrm{e}$ $= \mu_\mathrm{e}/T_\mathrm{e}$ or $\mu_\mathrm{e}/kT_\mathrm{e}$

$\eta_K$ $= \mu_K/T_\mathrm{i}$ or $\mu_K/kT_\mathrm{i}$

$\kappa_T$ Electron isothermal bulk modulus $= n_\mathrm{e}\left(\dfrac{\partial P_\mathrm{e}}{\partial n_\mathrm{e}}\right)_{T_\mathrm{e}}$

$\Lambda_j$ $= 1/Y_j^3 = \left(3\Gamma_j\right)^{3/2}$

$\tilde{\Lambda}_j$ $= \xi_j \Lambda_j$

$\Lambda_j^\nu$ $= \dfrac{Z_j + \nu\frac{1}{2}}{\bar{Z}}\Lambda_0^0$ , $\nu = \pm,\ 0$

$\Lambda_0^0$ $=(3\Gamma)^{3/2}$

$\Lambda_J$ $\Lambda_j$ or $\tilde{\Lambda}_j$ for any atom in configuration $J$ .

$\lambda_T$ Electronic isothermal bulk modulus pressure coefficient. $=\left(\dfrac{\partial \kappa_T}{\partial P_e}\right)_{T_e}$ .

$\theta_x$ $=\dfrac{x}{Z^*}\left(\dfrac{\partial Z^*}{\partial x}\right)_y$ where $x, y$ denote the macroscopic thermodynamic coordinate $V, T$ in either order.

$\tilde{\theta}_x$ $=\dfrac{x}{Z_p^*}\left(\dfrac{\partial Z_p^*}{\partial x}\right)_y$ where $x, y$ denote the macroscopic thermodynamic coordinate $V, T$ in either order.

$\mu_e$ Electron chemical potential.

$\mu_e^0$ $=\mu_e^0(n_e, T_e)$ = chemical potential for non-interacting electrons.

$\mu_i$ Chemical potential of a bare ion.

$\mu_i^0$ Chemical potential of a bare ion in the equivalent (same ion density and temperature) non-interacting system.

$\mu_K$ Chemical potential of ion in the configuration $K$ .

$\mu_K^0$ $=\mu_K^0(n_K, T_i)$ = chemical potential of non-interacting ions in the configuration $K$ .

$\mu_0$ Chemical potential that a neutral atom would have to be in equilibrium with the plasma.

$\hat{\boldsymbol{v}}_\alpha$ Unit configuration vector denoting one electron in the level $\alpha$ .

$\Pi_c(\mathbf{z})$ Electrostatic contribution of an atomic configuration to $PV$ .

$\Pi_J$ $=\Pi_c(\mathbf{z}_J)=\frac{1}{3}U_J$

$\rho(z)$ Density of internal states.

$\Sigma_c(\mathbf{z})$ Number of bound electrons in the configuration denoted by $\mathbf{z}$ , $=\sum_\alpha \hat{\boldsymbol{v}}_\alpha \cdot \mathbf{z}$

$\Sigma_J$ $=\Sigma_c(\mathbf{z}_J)=Z-Z_J$

$\Phi$ Energy of the bound electrons in a plasma equal to the sum over the binding energies of all the configurations when the ions are considered to be effectively isolated from each other.

$\phi_{j\alpha}$ $>0$ Ionization potential of electron in state $\alpha$ in an isolated ion $j$ .

$\varphi$ Hard sphere packing fraction of a lattice.

$\xi_j$ Parameter given by (13).

$\Omega_{\mathbf{k}}$ Collective frequency of charge density oscillations corresponding to wavevector $\mathbf{k}$ .

$\omega$ Photon frequency.

$\omega_0$ Unshifted edge frequency given by $\hbar\omega_0 = \phi_{j\alpha}$

Other notations

$$\bar{X} = \frac{1}{N_{\mathrm{i}}}\sum_j X_j = \frac{1}{N_{\mathrm{i}}}\sum_j N_J X_J$$

$$\langle X_{\mathrm{c}} \rangle = \sum_J p_J X_J = \int p(z) X(z) \rho(z) \mathrm{d}z$$

**References**

[1] D.R. Inglis, E. Teller, Ionic Depression of Series Limits of One-Electron Spectra, Astrophys. J. 90 (1939) 439.

[2] O. Ciricosta, S. Vinko, H.-K. Chung, B.-I. Cho, C. Brown, T. Burian, et al., Direct Measurements of the Ionization Potential Depression in a Dense Plasma, Phys. Rev. Lett. 109 (2012) 065002.

[3] T.R. Preston, S.M. Vinko, O. Ciricosta, H.-K. Chung, R.W. Lee, J.S. Wark, The effects of ionization potential depression on the spectra emitted by hot dense aluminium plasmas, High Energy Density Phys. 9 (2013) 258–263.

[4] D.J. Hoarty, P. Allan, S.F. James, C.R.D. Brown, L.M.R. Hobbs, M.P. Hill, et al., Observations of the effect of ionization potential depression in hot dense plasma, Phys. Rev. Lett. 110 (2013) 265003.

[5] J.C. Stewart, K.D. Pyatt, Lowering of Ionization Potentials in Plasmas, Astrophys. J. 144 (1966) 1203.

[6] G. Ecker, W. Kröll, Lowering of the Ionization Energy for a Plasma in Thermodynamic Equilibrium, Phys. Fluids. 6 (1963) 62.

[7] G. Ecker, W. Weizel, Zustandssumme und effektive Ionisierungsspannung eines Atoms im Inneren des Plasmas, Ann. Phys. 452 (1956) 126–140.

[8] A.V. Demura, Physical Models of Plasma Microfield, Int. J. Spectrosc. 2010 (2010) 671073.

[9] B.J.B. Crowley, Average-atom quantum-statistical cell model for hot plasma in local thermodynamic equilibrium over a wide range of densities, Phys. Rev. A. 41 (1990) 2179–2191.

[10] D. Kremp, M. Schlanges, W.-D. Kraeft, T. Bornath, Quantum Statistics of Nonideal Plasmas, Springer, 2005.

[11] R.M. More, Atomic Physics in ICF, LLNL Report UCRL-84991, 1981.

[12] B.J.B. Crowley, J.W. Harris, Modelling of plasmas in an average-atom local density approximation: the CASSANDRA code, J. Quant. Spectrosc. Radiat. Transf. 71 (2001) 257–272.

[13] L.K. Pattison, B.J.B. Crowley, J.W.O. Harris, L.M. Upcraft, The calculation of free electron density in Cassandra, High Energy Density Phys. 6 (2010) 66–69.

[14] G.B. Zimmerman, R.M. More, Pressure ionization in laser-fusion target simulation, J. Quant. Spectrosc. Radiat. Transf. 23 (1980) 517–522.

[15] E.H. Lieb, H.J. Narnhofer, The Thermodynamic Limit for Jellium, J.Stat. Phys. 12 (1975).

[16] E.H. Lieb, H.J. Narnhofer, The Thermodynamic Limit for Jellium: Errata, J.Stat. Phys. 14 (1976).

[17] N. Mermin, Exact Lower Bounds for Some Equilibrium Properties of a Classical. One-Component Plasma, Phys. Rev. 171 (1968).

[18] H. DeWitt, Asymptotic form of the classical one-component plasma fluid equation of state, Phys. Rev. A. 14 (1976) 1290–1293.

[19] R.M. Martin, Electronic Structure: Basic Theory and Practical Methods, Cambridge University Press, 2004.

[20] M. Zemansky, Heat and Thermodynamics, 5th ed., McGraw-Hill, New York and London, 1968.

[21] S.J. Rose, The effect of orbital relaxation in medium and high-Z average-atom calculations, J. Quant. Spectrosc. Radiat. Transf. 53 (1995) 39–44.

[22] A.V. Demura, Physical Models of Plasma Microfield, Int. J. Spectrosc. 2010 (2010).

[23] M. Dharma-Wardana, F. Perrot, Level shifts, continuum lowering, and the mobility edge in dense plasmas, Phys. Rev. A. 45 (1992) 5883–5896.

[24] P.M. Platzmann, P.A. Wolff, Waves and Interactions in Solid State Plasmas, Academic Press, New York and London, 1973.

[25] L.D. Landau, E.M. Lifshitz, Statistical Physics, Part 1, 3rd ed., Pergamon Press, Oxford, 1980.

[26] W. Jones, N.H. March, Theoretical Solid State Physics, Courier Dover Publications, 1973.